\documentclass[a4paper,11pt, titlepage, twoside]{article}

\usepackage{titling}
\title{Report on Stability Analysis of Islanded Power Systems with Grid-Forming Converters}
\author{Authors}
\date{Date}

\usepackage{lipsum}
\usepackage[utf8]{inputenc}
\usepackage[T1]{fontenc}
\usepackage{parskip}
\usepackage[a4paper,top=3cm,bottom=4cm,left=2.5cm,right=2.5cm]{geometry}

\usepackage{graphicx} 
\usepackage{float}
\usepackage[export]{adjustbox}
\usepackage{eurosym}
\usepackage{amsmath,amssymb,amsfonts}
\usepackage{textgreek}
\usepackage{steinmetz}
\usepackage{systeme}
\usepackage{sectsty}
\usepackage{multirow}
\usepackage{caption}
\usepackage{comment}
\usepackage{subcaption}
\usepackage{wrapfig}
\usepackage{listings}
\usepackage{xurl} 
\usepackage[colorlinks=true, allcolors=blue]{hyperref}
\usepackage{color} 
\usepackage[dvipsnames]{xcolor}
\usepackage[section]{placeins} 
\usepackage{longtable}
\usepackage{mathtools}
\usepackage{commath}

\usepackage{enumitem}
\usepackage{scalerel}
\usepackage{tikz}
\usepackage[inkscapelatex=false]{svg}
\usepackage{booktabs}
\usepackage{tabularx}
\usepackage{ragged2e} 
\newcolumntype{Y}{>{\RaggedRight\arraybackslash}X} 

\usetikzlibrary{svg.path}
\usetikzlibrary{shapes,arrows}
\usepackage{pgfplots}
\usepgfplotslibrary{groupplots}
\usepgfplotslibrary{fillbetween}
\usetikzlibrary{pgfplots.statistics}

\usepackage{pdflscape}   
\usepackage{array}       
\usepackage{dtk-logos}
\usepackage{fancyhdr}
\usepackage{titlesec}
\usepackage{transparent}
\usepackage{booktabs}
\usepackage{graphicx}

\makeatletter
\renewcommand*{\equationautorefname}{\@gobble}

\makeatother
\usepackage[export]{adjustbox}
\usepackage{url}

\usepackage{acronym}

\usepackage{listings}
\definecolor{codegreen}{rgb}{0,0.6,0}
\definecolor{codegray}{rgb}{0.5,0.5,0.5}
\definecolor{codepurple}{rgb}{0.58,0,0.82}
\definecolor{light-gray}{gray}{0.95}
\definecolor{ADoredColor}{rgb}{0, 0.2667, 0.5804}

\lstdefinestyle{Python}{
  backgroundcolor=\color{light-gray}, 
  commentstyle=\color{codegreen},
  keywordstyle=\color{violet},
  numberstyle=\tiny\color{codegray},
  stringstyle=\color{codepurple},
  basicstyle=\ttfamily\footnotesize,
  emph={self,__init__},          
  emphstyle=\color{red},
  breakatwhitespace=false,         
  breaklines=true,                 
  captionpos=b,                    
  keepspaces=true,                 
  numbers=left,                    
  numbersep=5pt,                  
  showspaces=false,                
  showstringspaces=false,
  showtabs=false,                  
  tabsize=2
}

\lstdefinestyle{Matlab}{
    backgroundcolor=\color{light-gray}, 
    commentstyle=\color{codegreen},
    breaklines=true,%
    morekeywords={matlab2tikz},
    basicstyle=\ttfamily\footnotesize,%
    keywordstyle=\color{blue},%
    morekeywords=[2]{1}, keywordstyle=[2]{\color{black}},
    identifierstyle=\color{black},%
    stringstyle=\color{codepurple},
    showstringspaces=false,
    numbers=left,%
    numberstyle={\tiny \color{black}},
    numbersep=5pt, 
    emph=[1]{for,end,break},emphstyle=[1]\color{blue}, 
}

\usepackage[sorting = none, backend = biber, style=numeric-comp]{biblatex}
\AtBeginBibliography{\sloppy\setlength{\emergencystretch}{2em}}

\newcommand{\sectionbreak}{\clearpage}

\usepackage{titlesec}
\titleformat{\section}{\clearpage\normalfont\Large\bfseries\color{Blue}}{\thesection}{1em}{}
\titleformat{name=\section,numberless}{\clearpage\normalfont\Large\bfseries\color{Blue}}{}{0pt}{}

\chapterfont{\color{ADoredColor}}  
\sectionfont{\color{ADoredColor}}  
\subsectionfont{\color{MidnightBlue}}
\subsubsectionfont{\color{NavyBlue}}

\usepackage{nomencl}

\makenomenclature

\usepackage[nonumberlist]{glossaries}

\newacronym{VSC}{\textcolor{black}{VSC}}{\textcolor{black}{Voltage Source Converters}}
\newacronym{igbt}{\textcolor{black}{IGBT}}{\textcolor{black}{Insulated-Gate Bipolar Transistor}}

\newacronym{res}{\textcolor{black}{RES}}{\textcolor{black}{Renewable Energy Systems}}
\makeglossaries

\begin{document}

\graphicspath{ {Figures/} } 
\begin{titlepage}

\includegraphics[height=2.6cm,center]{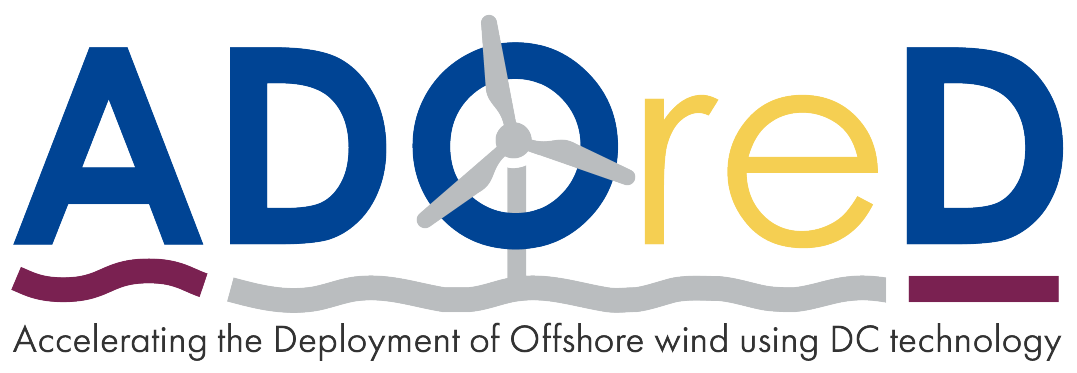}
\vspace{1mm}
\includegraphics[height=1.52cm,center]{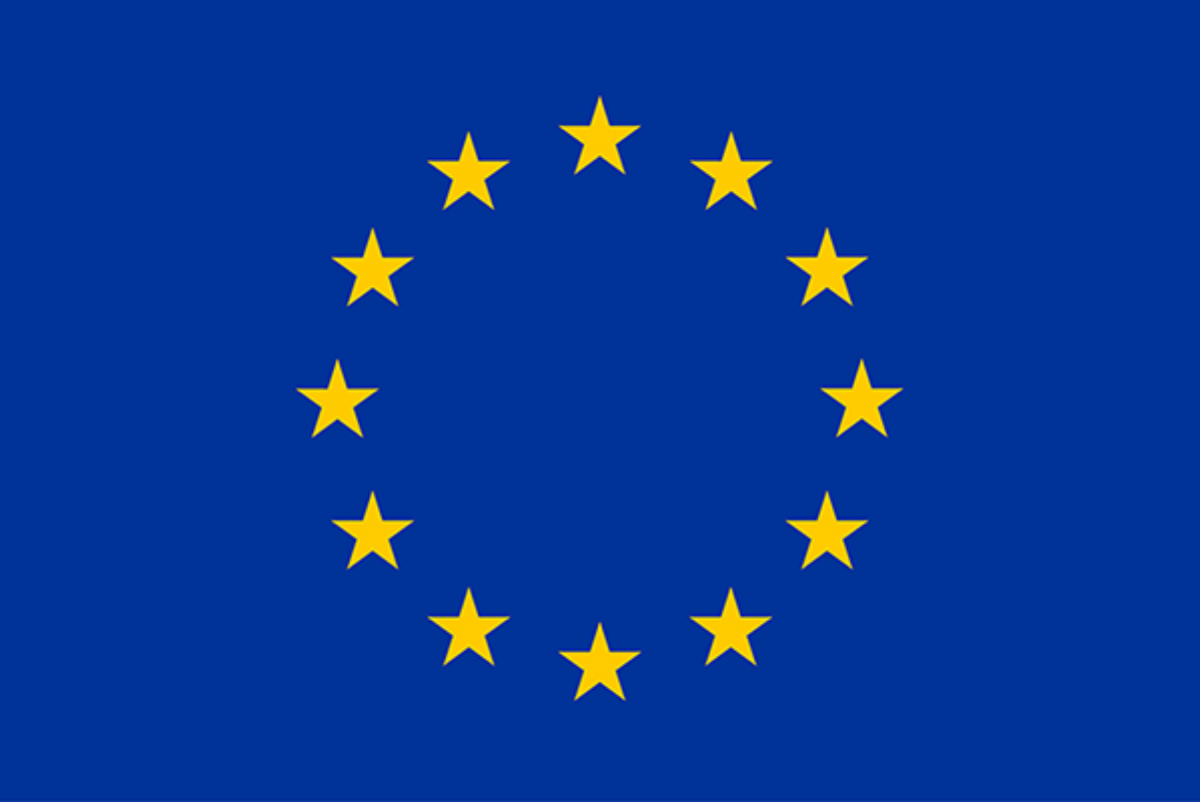} 

  \vspace{25mm}
    \hspace{2mm}
\centering
    {\Large \textbf{Doctoral Network Horizon Europe} \\
    \Large \textbf{Grant Agreement number: 101073554 }}\\

    \vspace{20mm}
    \Large \textcolor{ADoredColor}{Deliverable number : D3.4}\\
    \huge \textcolor{ADoredColor}{\textbf{\thetitle}}\\
    \vspace{10mm}
    \Large \textcolor{ADoredColor}{Description of deliverable:}\\
    \vspace{4mm}
   \vspace{4mm}
    
    \hspace{2mm}
    
\begin{table}[H]
\centering
\begin{tabular}{|l|l|}
\hline
Deliverable number     & D3.4               \\ \hline
Work Package           & WP3 : AC/DC Interactions    \\ \hline
Dissemination level    & Sensitive             \\ \hline
Project period         & From Jun - to Nov \\ \hline
Due date &  30/ 11 / 2025  \\\hline
Actual submission date & 30/ 01  / 2026  \\  \hline
Authors & Anup Joshi,\\ 
        & Luis Angel Garcia Reyes,\\
        & Qingyuan Gong  \\ \hline
Peer reviewed by & Xavier Guillaud   \\
                 & Javier Renedo \\
                 & Dirk Van Hertem \\
\hline
Version & v3.0  \\ \hline
\end{tabular}
\end{table}

    \end{titlepage}
\hypersetup{linkcolor=black}
\tableofcontents
\listoffigures
\listoftables
\renewcommand*{\glspostdescription}{}
\printglossary[type=\acronymtype, title={List of Abbreviations}, toctitle={List of Abbreviations}]
\section*{List of Abbreviations}

\begin{description}[style=multiline,
                    leftmargin=2.8cm,
                    labelwidth=2.3cm,
                    labelsep=0.7cm]

\item[AC]        Alternating Current
\item[AGC]       Automatic Generation Control
\item[AEMO]      Australian Energy Market Operator
\item[BESS]      Battery Energy Storage System
\item[CIG]       Converter-Interfaced Generation
\item[CCT]       Critical Clearing Time
\item[DC]        Direct Current
\item[DFIG]      Doubly-fed induction generators
\item[EMT]       Electromagnetic Transients
\item[ESIG]      Energy Systems Integration Group
\item[ESS]       Energy Storage System
\item[FD]        Frequency-Domain
\item[FRC]       Frequency Restoration Control
\item[FFR]       Fast-Frequency Response
\item[GFL]       Grid-Following
\item[GFM]       Grid-Forming
\item[HIL]       Hardware-in-the-Loop
\item[HVDC]      High-Voltage Direct Current
\item[IBR]       Inverter-Based Resource
\item[IPS]       Islanded Power System
\item[IGP]       Integrated Grid Plan
\item[LCC]       Line-Commutated Converter
\item[LVRT]      Low-Voltage Ride-Through
\item[MPC]       Model Predictive Controller
\item[OWP]       Offshore Wind Plants
\item[PCC]       Point of Common Coupling
\item[PHS]       Pumped Hydro Storage
\item[PLL]       Phase-Locked Loop
\item[PPM]       Power-Pack Module
\item[PSC]       Power-Synchronization Control
\item[PSS]       Power System Stabilizer
\item[PUC]       Public Utilities Commission
\item[PV]        Photovoltaic
\item[QSS]       Quasi Static Simulations
\item[RES]       Renewable Energy Sources
\item[RMS]       Root-Mean-Square
\item[ROCOF]     Rate of Change of Frequency
\item[SCR]       Short-Circuit Ratio
\item[SG]        Synchronous Generator
\item[SS]        State Space
\item[STATCOM]   Static Synchronous Compensator
\item[SSA]      Small-Signal Stability Analysis
\item[SSO]      Sub-synchronous Oscillation
\item[TD]       Time-Domain
\item[TSO]       Transmission System Operator
\item[UC]        Ultracapacitor
\item[UFLS]      Under-Frequency Load Shedding
\item[V2G]       Vehicle-to-Grid
\item[VSC]       Voltage-Source Converter
\item[VOC]       Virtual Oscillator Control
\item[VSM]       Virtual Synchronous Machine

\end{description}

\hypersetup{linkcolor=blue} 

\printnomenclature

\section*{Executive Summary}
\addcontentsline{toc}{section}{\textbf{Executive Summary}}

Islanded power systems are inherently challenging to operate due to their limited interconnection, low inertia, and weak short-circuit strength. As renewable and inverter-based resources replace synchronous machines, these challenges exacerbate, leading to faster frequency deviations and greater voltage volatility which leads to increased risk of instability. In this report, a summary of differences between islanded and interconnected power system is provided leading to a review of key stability issues in islanded systems, focusing on: frequency stability, converter-driven interactions, \color{black} voltage stability, \color{black} large-disturbance behavior, and small-signal dynamics.

Grid-forming (GFM) \color{black} voltage source \color{black} converters have emerged as a promising solution because they can establish voltage and frequency, emulate synchronous machine behavior, and provide \color{black} technical capabilities \color{black} such as \color{black} grid stregth, \color{black} virtual inertia, fast frequency response, voltage regulation, and black-start capability, \color{black} among others. \color{black} The report examines \color{black} some relevant \color{black} GFM control strategies and evaluates their functional capabilities in the context of weak, low-inertia islanded grids.

While GFM technology offers significant benefits, challenges remain—particularly regarding interoperability, control coordination among multiple inverter-based resources, lack of standardized grid code requirements and coordination of protection devices. The report highlights that successful deployment will require advances in interoperability, harmonized grid codes, and grid planning taking into account the unique features of islanded power system and clear GFM functionality requirements.

Finally, a case study is performed to illustrate the stability enhancing capability of GFM converters in weak grids \color{black} by means of \color{black} of small signal stability analysis. Overall, the report finds that GFM technology is a key enabler for reliable, renewable-dominant island power systems, but broader adoption will depend on continued technical and regulatory development.

\section{Introduction}
The energy transition and push for decarbonization of the electric power system has meant moving away from conventional fossil fuel based generation and deployment of more environment friendly renewable energy sources (RES) like solar and wind. Different countries have adopted their own decarbonization roadmap and it will drastically change the generation mix in the future, with the converter-interfaced generation (CIG) such as solar \color{black} photovoltaic (PV) \color{black} and wind turbine generators poised to take a large share. Unlike continental grids, island power systems are characterized by limited interconnection, have very less number of generating units compared to continental grids and provide favorable conditions for renewable energy generation. Traditionally, many of these \color{black} islands \color{black} have been powered by fossil-fuel based thermal units with exported fuel due to their isolation \cite{cross_benchmarking_2017}. However, due to a push towards decarbonization, the share of CIGs such as solar PV and wind energy is increasing in islands. Moreover, some of these island power \color{black} systems \color{black} are interconneted to the mainland grid through high voltage direct current (HVDC) \color{black} systems. \color{black} Thus, the islanded power \color{black} systems \color{black} have distinct characteristics depending upon their generation mix, interconnection status, electricity demand etc. A list of islands with their demands and generation mix is provided in the Table \ref{Island_list}. 

\begin{table}[H]
\centering
\caption{Some island power systems with their max demand and generation mix}
\label{tab:island_power_systems}
\begin{tabular}{p{2.8cm} p{2.5cm} p{8cm}}
\textbf{Island} 
& \textbf{Max demand (MW)} 
& \textbf{Generation mix}  \\[6pt]
\toprule
Balearic Islands 
& 1286
& Combined cycle, coal, LCC-HVDC, gas, waste, \\[4pt]

Canary Islands 
& 1372
& Combined cycle, diesel gen, wind, gas, solar PV \\[4pt]

Hawaii 
& 1102 
& Thermal, solar PV, wind, hydro, geothermal \\[4pt]

Azores 
& 76 
& Thermal, geothermal, wind, hydro \\[4pt]

Crete  
& 650   
& Thermal, wind, solar PV  \\[4pt]

Faroe Island 
& 48 
& Thermal, wind, hydro  \\[4pt]

Guadeloupe 
& 256 
& Fossil fuels, coal, biofuels, wind, solar, hydro  \\[4pt]

Jeju Island
& 1000 
& HVDC links, wind, solar PV \\[4pt]

Madeira
& 150 
& Thermal, hydro, wind, solar PV  \\[4pt]
\bottomrule
\label{Island_list}
\end{tabular}
\vspace{-1cm}
\end{table}

A larger list of islanded power system and information on them is provided in \cite{Island_book}.

\subsection{Characteristics of an Islanded Power System}
Most islanded power \color{black} systems \color{black} share some common characteristics: limited interconnection, limited generating units, favorable conditions for renewable energy generation. Due to the first two, there are already challenges regarding frequency stability of the islanded power system and that in turn impacts higher renewable energy integration. Most renewable energy sources are connected to the grid through power converters and converter behavior differs from conventional synchronous generators (SG) in two major areas which are critical for an electrical grid: lack of rotational inertia and reduced short-circuit capability. The issue with frequency and voltage stability due to reduced inertia and short circuit power with  large-scale deployment of CIGs have been widely known \cite{milano_foundations_2018}. These issues have been studied for their impact on the wide area interconnected network, but islanded power \color{black} systems \color{black} are more susceptible to them due to their unique characteristics as compared to the interconnected grid. Thus, islanded power systems require even more urgent measures to offset the issues that comes with growing share of CIGs. While the challenges with growing CIG penetration in islanded power systems are evident, they also provide an experimental ground to test the technology enablers for systems with very high amount of RES due to their smaller size, before being introduced in the interconnected grid \cite{EASE2022Islands}. In order to develop unique solutions required for islanded power systems, it is important to understand how islanded power systems are different from large area interconnected power system. A summary of how comparison between islanded power system and interconnected power system based on the concepts discussed in \cite{Hoke2021, ochoa-correa_pathways_2025,milano_foundations_2018} is provided in Table \ref{Table_comp}. 

\begin{table}[H]
\vspace{0.6cm}
\centering
\caption{Comparison between islanded and interconnected power systems}
\renewcommand{\arraystretch}{1.15} 
\begin{tabularx}{\linewidth}{l Y Y}
\toprule
\textbf{Aspect} & \textbf{Islanded power system} & \textbf{Interconnected power system} \\
\midrule
System configuration 
  & Geographically isolated
  & Strongly meshed grid  \\
Size \& generation diversity
  & Small generation fleet generally consisting of thermal plants as the dominating source 
  & Large and diverse generation mix consisting of thermal, hydro, nuclear, renewables \\
System Inertia 
  & Low inertia resulting in fast frequency deviations and high RoCoF 
  & High inertia resulting in slower frequency deviations; comparatively smaller ROCOF \\
Short-circuit ratio 
  & Low SCR; weak-grid behavior; limited fault current and voltage stiffness 
  & High SCR; strong voltage stiffness and predictable fault behaviour.  \color{black} Certain areas could have low SCR. \color{black} \\
Renewable integration. 
  & Intermittency strongly affects stability due to small system size; large scale RES integration difficult
  & Intermittency smoothened by large geographical area and diverse generation mix ; large-scale RES integration is easier \\
System restoration 
  & Restoration depends on few local resources often a thermal unit or \color{black} VSC-HVDC interconnections, if any. \color{black} \color{black} If sufficient generation units are available, system restoration is easier than in large-power systems, due to the size of the power system. \color{black}
  & Restoration supported by hydro/thermal units or \color{black} VSC-HVDC interconnections. System restoration of a large-scale power system is more difficult in general, due to the size of the power system.  \color{black} \\
\bottomrule
\label{Table_comp}
\end{tabularx}
\end{table}

\subsection{Issues in Islanded Power System}
The increase in share of CIGs and decrease in share of conventional sources in the energy mix of the island power system  leads to a reduction in system inertia. Inertia of the system, which is normally provided by the SG's present in the system is crucial to keep the system frequency variation within a limit during the early stages of unbalance of generation and demand. A reduction in inertia in a system which is already isolated is detrimental to the frequency stability of the system. Thus, \color{black} traditionally, \color{black} frequency stability has been pointed to as the most challenging \color{black} stability issue \color{black} in islanded \color{black} systems \color{black} with high number of CIGs \cite{Island_book,ochoa-correa_pathways_2025} . Another challenge in power systems with high amount of CIGs is the problem of voltage regulation \cite{ochoa-correa_pathways_2025}. Voltage regulation is the ability of the system to maintain the system voltage within a small band, despite  fluctuation in network conditions. An islanded power system consists of very less number of generating units and with the loss of even a few of those rotating generators the short circuit power provided by them is reduced by a large amount. The CIGs \color{black} may not be \color{black} able to compensate for it as the short circuit contribution by them is inherently limited and thus the grid becomes weaker. In a weak grid, even a small event or disturbance can lead to large voltage fluctuations. \color{black} Notice that voltage stability in island power systems is relevant, and enough reactive-power reserve in the system is needed. However, island power systems are smaller and problems related to voltage control in different areas may be less critical than in large-scale power systems. A summary of the critical issues related to the stability phenomena is presented in the Table \ref{Table_comp} with a detailed comparison against interconnected power systems.

 \begin{table}[H]
 \vspace{0.5cm}
\centering
\caption{Comparison of stability phenomena between islanded and interconnected power systems}
\renewcommand{\arraystretch}{1.15} 
\begin{tabularx}{\linewidth}{Y Y Y}
\toprule
\textbf{Aspect} & \textbf{Islanded power system} & \textbf{Interconnected power system} \\

\midrule
Frequency stability 
  & Highly critical because of low inertia and isolated nature.
  & Less critical; Frequency deviations smaller and \color{black} system inertia is higher. \color{black} \\
Converter-driven stability 
  & Very significant if IBRs present in weak grids; resonance and control interactions.
  & Similarly critical in systems with high IBR penetration.\\
Voltage stability 
  & Voltage control is critical: Enough reactive-power reserve required.  
  & Voltage control is critical: Enough reactive-power reserve required. Additionally, voltage control in different areas of a large-scale power system. \\
Transient stability 
  & Non critical.
  & Critical. \\
 Electromechanical oscillations 
  & Inter-area oscillations: Not critical, only local oscillations. 
  & Inter-area oscillations: Critical, also local oscillations.  \\

\bottomrule
\label{Table_comp}
\end{tabularx}
\end{table}

  \color{black} In \cite{oscillationsmicrogrid} converter-driven oscillations have been investigated in islanded microgrid with high amount of converters. In the literature the converter driven oscillations have been attributed to the interaction between fast control loops or control loops and passive grid elements and instability arising from non optimal control tuning. Thus, in a scenario with high penetration of CIGs, converter driven stability will likely be a critical issue. Due to the unique features of an islanded power system the stability phenomena in such systems understandably vary in their urgency as compared to an interconnected power system. A summary of comparison between islanded power system and interconnected power for different phenomena based on the discussions in \cite{Hoke2021, ochoa-correa_pathways_2025,milano_foundations_2018,EASE2022Islands}  is provided in Table \ref{Table_comp}. 
 
 Notice that power system stability must be guaranteed and this means that all stability phenomena are relevant. Nevertheless, some stability phenomena are more critical in island power systems due to their characteristics, as will be explained in Section  \ref{Stability_review}.


\section{Review of Stability Analysis in Islanded Power Systems and GFM Converters} \label{Stability_review}

The reduction of fossil fuel dependency in isolated power systems, commonly referred to as islanded power systems (IPS), is primarily driven by the high costs, logistical complexity, and environmental unsustainability associated with supplying and delivering such fuels to insular generation facilities~\cite{Island_book,paperOzanErdinc,Meschede2022}. Consequently, the utilization of local renewable energy resources emerges as a promising pathway to mitigate this dependency, aiming to increase the penetration of RES within IPS. However, the limited size, lack of external support, and unique operational constraints make IPS more vulnerable than interconnected systems, especially as they approach high shares of inverter-based generation~\cite{Hoke2021,Meschede2022}.

\subsection{Key Stability Phenomena in Islanded Power Systems}

Compared to large interconnected power systems, IPS are inherently weak due to their higher sensitivity to disturbances such as voltage and frequency deviations. This vulnerability stems from their limited size, reduced number of generation units, and restricted electrical interconnections—conditions exacerbated under high RES penetration, where generation variability is significant. The scarcity of conventional synchronous generation units results in low system inertia, making IPS highly susceptible to large voltage and frequency excursions, ultimately compromising system reliability and security~\cite{Island_book,Meschede2022,Hoke2021,Tielens2016}.

\begin{figure}[h]
    \centering
    \includegraphics[width=0.95\textwidth]{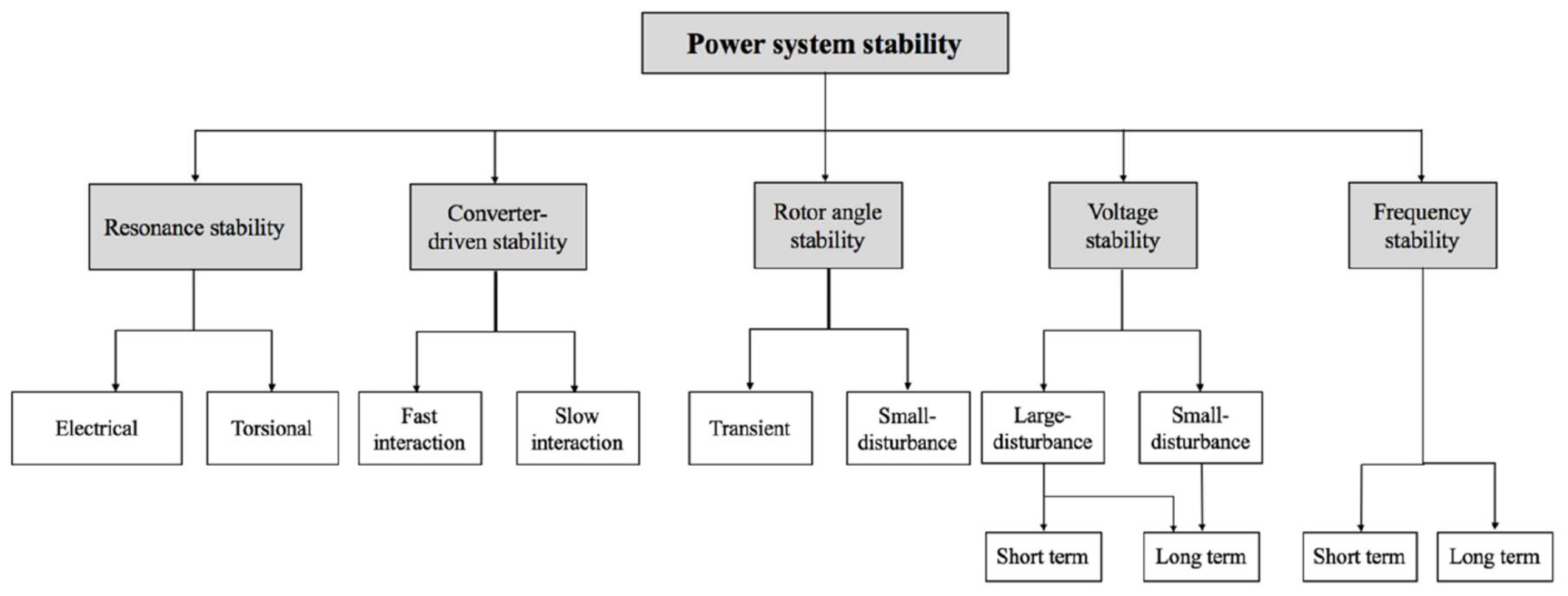}
    \caption{Classification of modern power system stability. Figure from \cite{Hatziargyriou2021}.}
    \label{fig:clas_stab}
\end{figure}

The predominant RES deployments in IPS are based on PV and wind generation technologies, each introducing specific technical and non-technical challenges to ensure stability under various disturbance scenarios. In addition, due to the power-electronics-based nature of these RES, new stability issues arise from the distinct dynamic characteristics of CIG compared to conventional synchronous generators. Consequently, novel concepts and stability classifications have been proposed and extended to incorporate converter dynamics, as illustrated in Fig.~\ref{fig:clas_stab}.

While all stability phenomena are relevant, some of them are especially critical in islanded systems. According to recent reviews and field experience~\cite{Island_book,Hatziargyriou2021,Meschede2022,Hoke2021}, probably the most critical stability phenomena in IPS are:
\begin{itemize}
    \item \textbf{Frequency stability} (traditional problem): due to low inertia, limited frequency reserve and high impact of the lack of external support (single contingencies).
    \item \textbf{Converter-driven stability} (new problem): fast/slow interactions and control-induced oscillations, especially with high shares of GFL and GFM VSC converters.
\end{itemize}

In this report, a review of previous work regarding stability issues on island power systems is presented. Some key references include Sigrist et al.~\cite{Island_book}, Collados-Rodriguez et al.~\cite{Collados2022} (Balearic Islands with LCC/VSC-HVDC), Trondheim \& Nielsen~\cite{Trondheim2023} (Faroe Islands, 100\% wind with GFM BESS),  Tielens \& Van Hertem~\cite{Tielens2016} (relevance of inertia), and Castro \& Acha \cite{Acha_2016} (VSC-HVDC system with GFM connected to an islanded power system). Practical examples for the stability issues include Madeira and Azores (Portugal), Canary and Balearic Islands (Spain), Hawaii, and others~\cite{Meschede2022,Hoke2021}.

\subsubsection{Frequency Stability}

\begin{figure}[h!]
    \centering
    \includegraphics[width=0.75\textwidth]{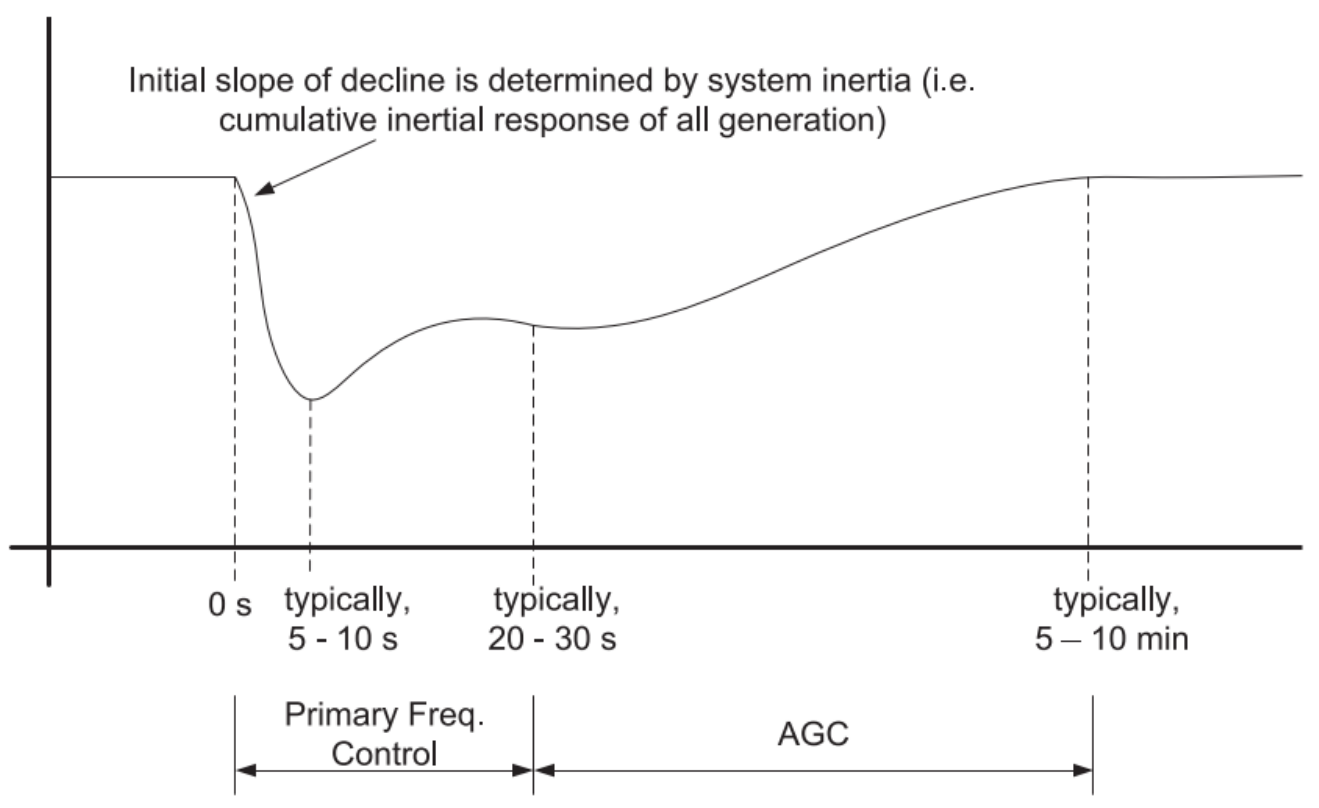}
    \caption{Stages of frequency control response in islanded systems. Figure from~\cite{Hatziargyriou2021}.}
    \label{fig:clas_freq}
\end{figure}

Frequency stability refers to the ability of a power system to maintain its operating frequency within acceptable limits following disturbances such as generator tripping. In IPS, this is a central concern due to low inertia and the high relative impact of generator outages~\cite{Island_book,Hatziargyriou2021,Meschede2022,Tielens2016}. The dynamic response to a frequency disturbance typically involves three stages: (i) the initial inertial response provided by synchronous generators, (ii) the primary frequency control through governor action and load damping, and (iii) secondary control via automatic generation control (AGC) to restore nominal frequency. Stability is achieved if remaining generators possess sufficient reserve capacity and can increase output rapidly enough to prevent frequency from falling below underfrequency protection thresholds, thereby avoiding cascading outages \cite{paperLasanthaMeegahapola,paperCIGREstabilitydefinitions}. Fig.~\ref{fig:clas_freq} shows the different periods during a frequency disturbance event. 

\begin{figure}[h!]
    \centering
    \includegraphics[width=0.65\textwidth]{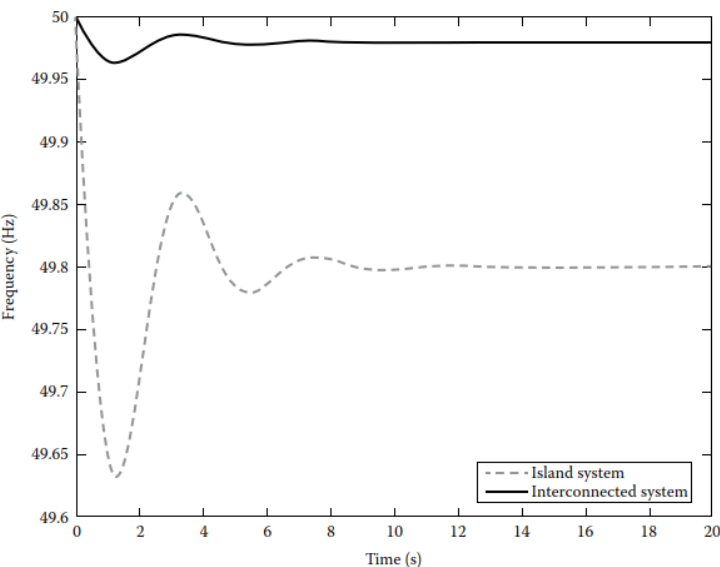}
    \caption{Comparison of frequency excursions between interconnected and islanded power systems. Figure from~\cite{Island_book}.}
    \label{fig:freq_comparisson}
\end{figure}

\begin{figure}[h!]
    \centering
    \includegraphics[width=0.75\textwidth]{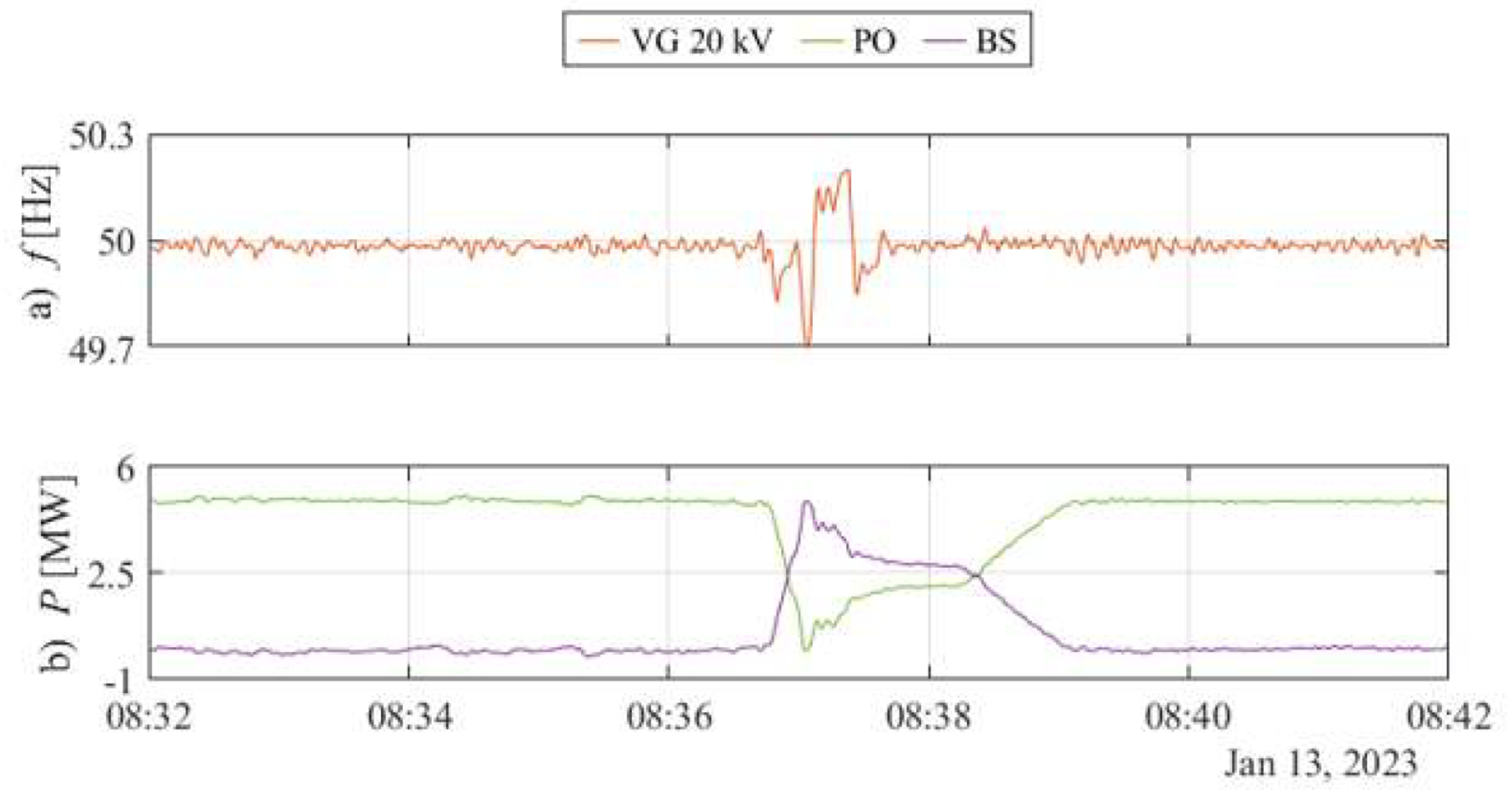}
    \caption{Measured response in Suðuroy (Faroe Islands) during a suddenly drop of 100\% wind operation with BESS and synchronous condenser. a) Frequency and, b) Wind Plant (PO) and Battery System (BS) active power during a loss of full production within 15 s. Figure from~\cite{Trondheim2023}.}
    \label{fig:faroe_freq}
\end{figure}

In IPS, the rate of frequency decline is faster, and excursions are more pronounced during severe disturbances, as illustrated in Fig.~\ref{fig:freq_comparisson}. Along this line, in IPS the disconnection of a generator could cause  a power imbalance which may be relatively big in comparison with the total generation of the system, making frequency stability a critical issue. Real-world events in the Faroe Islands~\cite{Trondheim2023} and Hawaii~\cite{Hoke2021} are shown in Figs. \ref{fig:faroe_freq} and \ref{fig:hawaii_freq}, respectively. To mitigate these risks, underfrequency and rate-of-change-of-frequency (ROCOF) protections are integrated into underfrequency load-shedding (UFLS) schemes that disconnect portions of demand to arrest frequency decay. UFLS schemes in islands play a more active role than in large systems, due to the higher risk of cascading outages due to the smaller size of islanded power systems~\cite{Dreidy2017,Meschede2022}. For example, in the Faroe Islands, the combination of a large BESS with GFM control and synchronous condenser has enabled stable operation with 100\% wind power, but the system remains highly sensitive to large disturbances~\cite{Trondheim2023}.

\begin{figure}[h!]
    \centering
    \includegraphics[width=0.75\textwidth]{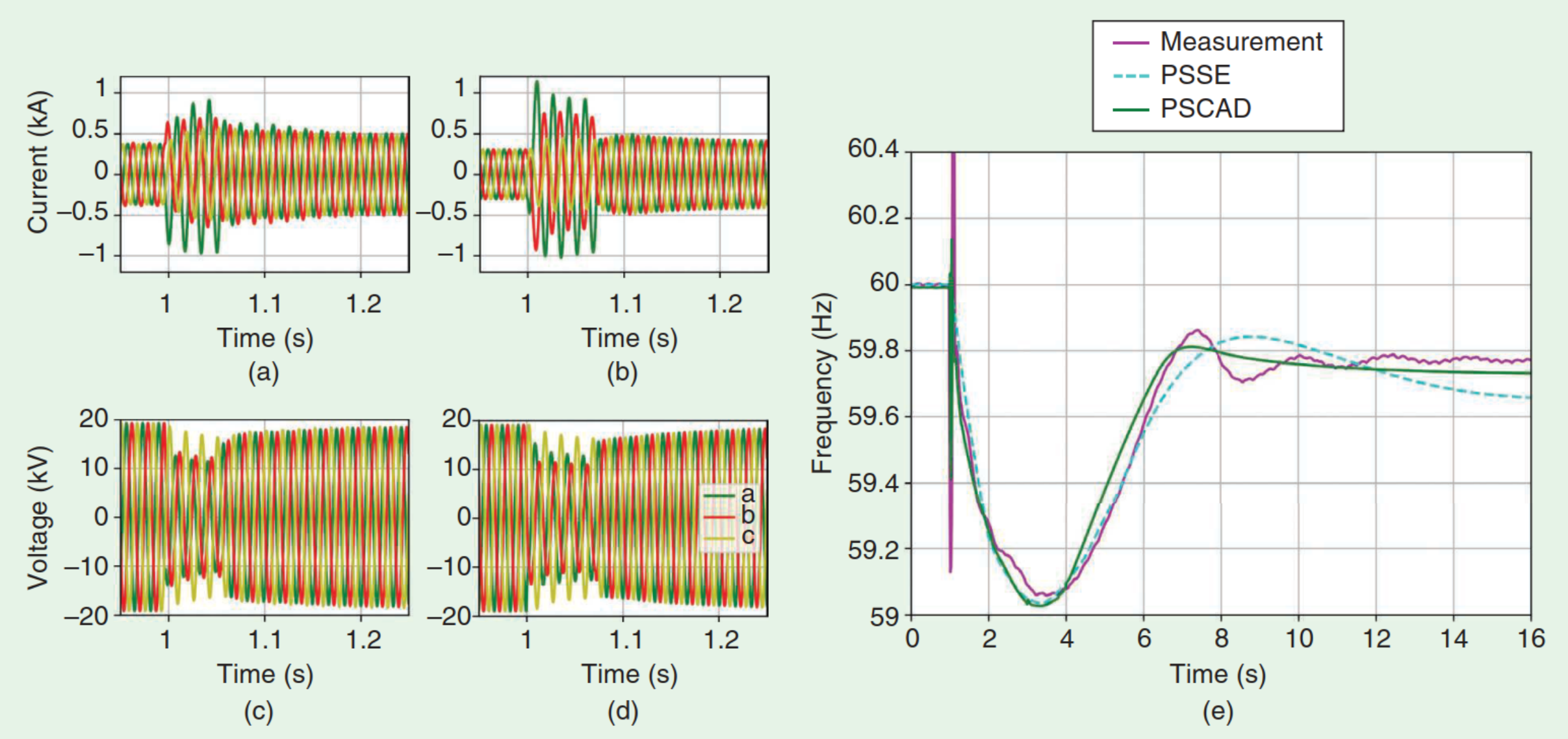}
    \caption{EMT model of Maui against field data for a 2 March 2017 fault followed by a generation trip event. (a) Measured currents in line 200-3. (b) PSCAD-simulated currents in line 200-3. (c) Measured voltages at the Kahului 23-kV bus. (d) PSCAD-simulated voltages at the Kahului 23-kv bus. (e) System frequency, measured and simulated. Figure from~\cite{Hoke2021}.}
    \label{fig:hawaii_freq}
\end{figure}

\paragraph{Underfrequency Load Shedding (UFLS)}

UFLS is a system-wide protection scheme designed to arrest frequency decline and prevent cascading outages by disconnecting load blocks when frequency or its ROCOF falls below predefined thresholds \cite{Sigrist_2011, Sigrist_2012}. This is particularly critical in islanded systems, where the loss of a single generator represents a large share of total demand, causing severe frequency deviations. UFLS schemes are divided as:
\begin{itemize}
    \item \textbf{Manual}: Rarely used; mainly for post-disturbance restoration.
    \item \textbf{Automatic}: Widely applied for rapid frequency containment.
\end{itemize}

Automatic UFLS is further divided into~\cite{Island_book,Dreidy2017,Meschede2022}:
\begin{itemize}
    \item \textbf{Conventional Schemes}:
        \begin{itemize}
            \item \textit{Static}: Simple and fast; sheds fixed load amounts but risks over/undershedding during unexpected contingencies. Typically, UFLS are implemented using different steps, for different frequency levels.
            \item \textit{Semi-adaptive}: Adds ROCOF measurement to distinguish disturbance severity, enabling earlier shedding for large events. However, ROCOF is sensitive to oscillations, requiring integrated or averaged measurements and careful tuning.
        \end{itemize}
    \item \textbf{Advanced Schemes}:
        \begin{itemize}
            \item \textit{Adaptive}: Calculates load to shed dynamically based on real-time conditions such as initial ROCOF and power imbalance.
            \item \textit{Centralized}: Eliminates staged logic and uses multiple variables (frequency, ROCOF, voltage, generation) for precise control.
        \end{itemize}
\end{itemize}

Advanced schemes are increasingly necessary as IBRs displace synchronous machines, and as GFM converters enable new forms of synthetic inertia and fast frequency response~\cite{Hoke2021,Hatziargyriou2021}.

\paragraph{Inertia and Fast Frequency Response}

The increasing penetration of CIG, such as wind and solar, introduces new challenges. Unlike synchronous machines, CIGs do not inherently provide inertial response. This displacement reduces system inertia, causing frequency excursions to occur more rapidly and increasing the likelihood of instability~\cite{Tielens2016,Dreidy2017}. However, modern wind turbines and BESS with GFM controls can provide synthetic inertia and fast frequency response (FFR), partially compensating for the decline in natural inertia~\cite{Trondheim2023,Hoke2021}.

For example, in the Faroe Islands, a 6.3 MW/7.5 MWh BESS with GFM control and an 8 MVA synchronous condenser have enabled stable operation with 100\% wind power, but only with careful tuning of frequency controls schemes~\cite{Trondheim2023}. In Hawaii, BESS and advanced inverter controls are still investigated to provide FFR and support frequency stability during large disturbances~\cite{Hoke2021,Akiba2024}.

\paragraph{UFLS and GFM Coordination}

Effective coordination between UFLS and GFM-ESS controls is essential. If UFLS is not properly coordinated with the fast response of GFM converters and BESS, there is a risk of unnecessary load shedding or failure to arrest frequency decline in time~\cite{Dreidy2017,Hoke2021}. Recent field experience in the Faroe Islands and Hawaii demonstrates the importance of integrating real-time measurements, adaptive thresholds, and communication between protection and control systems.

\subsubsection{Converter-Driven Stability}

With the increasing share of power-electronics-based generation, new stability phenomena arise from the distinct dynamic characteristics of CIG compared to conventional synchronous generators~\cite{Hatziargyriou2021,paperLasanthaMeegahapola,Collados2022}. Converter-driven stability, including both fast and slow interactions, is now recognized as a critical issue in IPS, especially under high penetration of GFM and GFL converters. 

Recent events in real-world island power systems have shown that poorly tuned converter controls, weak grid conditions, or adverse interactions between multiple IBRs can lead to oscillations, loss of synchronism, or even blackouts~\cite{Fan2023,Dong2025,Hoke2021}. For example, a 1 Hz oscillation event in Kauai (Hawaii) was traced to the power plant controller of a GFM plant, highlighting the need for careful tuning and validation of converter controls~\cite{Dong2025}. 

\begin{figure}[h]
    \centering
    \includegraphics[width=0.65\textwidth]{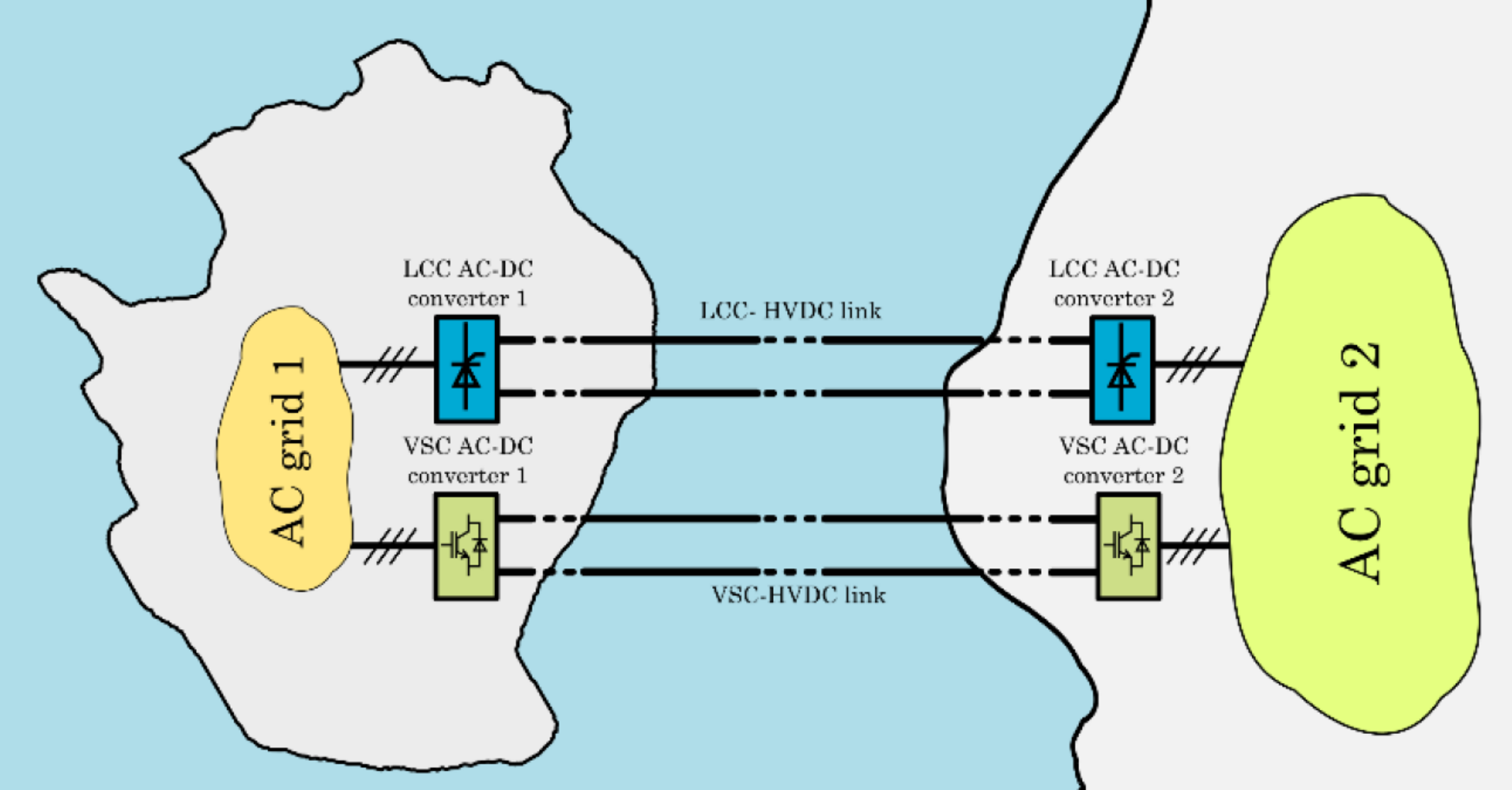}
    \caption{Example of dual-infeed HVDC system with LCC and VSC links in an islanded grid. Figure from~\cite{Collados2022}.}
    \label{fig:hvdc_island}
\end{figure}

In the Balearic Islands, the coexistence of LCC-HVDC and VSC-HVDC links has required new approaches to stability analysis and control coordination~\cite{Collados2022}. Fig. \ref{fig:hvdc_island} shows a diagram of the Balearic power system with the existing LCC-HVDC link and the planned VSC-HVDC link.  

Phase Locked Loop (PLL) synchronization becomes problematic in systems with low SCR. Under these conditions, the PLL may introduce negative admittance, destabilizing the system by generating phase-shifted currents and worsening voltage magnitude and angle. Mitigation strategies include tuning PLL parameters, reducing internal controller gains, adopting alternative control schemes, or reinforcing the grid with devices such as synchronous condensers \cite{paperCIGREstabilitydefinitions}.

Converters also face constraints related to power transfer limits in weak grids. When generated power exceeds the maximum transferable power, synchronism is lost, and the phase angle continues to increase, leading to instability. This phenomenon can be mitigated through fault-ride-through schemes that dissipate excess generation. Additionally, inverter current limits may restrict maximum power injection, especially under degraded grid conditions \cite{paperCIGREstabilitydefinitions}.

\subsubsection{Voltage Stability}

Voltage stability refers to the ability of a power system to maintain acceptable voltage levels at all buses following a disturbance. In IPS, voltage stability is strongly related to the reactive power compensation capability of the network, given the type and amount of power supply. RES typically lack the inherent voltage support capability provided by conventional synchronous generators. A common cause of voltage instability includes load loss in a specific area or the disconnection of a transmission line or other network elements, leading to voltage imbalance and triggering protective relay actions, which may result in cascading outages~\cite{Hatziargyriou2021,Collados2022}. Voltage stability in island power systems is relevant, and enough reactive-power reserve in the system is needed. However, there are also some aspects in voltage stability that are much more challenging in large-scale power systems, such as voltage control in different areas, for example. In an islanded power system, the  system is smaller, and maintaining the voltage in admissible values in a small set of buses may be enough to guarantee voltage stability of the island power system. 

\paragraph{Short-Term Voltage Stability}

Short-term voltage stability involves the dynamic behavior of fast-acting load components such as induction motors, electronically controlled loads, HVDC links, and IBRs. The time frame for this phenomenon is on the order of a few seconds (typically 1–10~s), often referred to as slow interactions, and is triggered by large disturbances such as short circuits, motor stalling, HVDC link disturbances, and delayed voltage recovery induced by faults \cite{Hatziargyriou2021}. This challenge becomes more pronounced in IPS with high penetration of RES, where the number of generation units is limited and voltage support capabilities are significantly weaker compared to interconnected mainland systems.

\paragraph{Long-Term Voltage Stability}

Long-term voltage stability involves slow-acting equipment (typically 10–20~s after a disturbance), such as protection systems, transformer on-load tap changers, generator rotor controls, and current limiters. It generally manifests as a gradual voltage decline at certain buses following the disconnection of transmission or generation equipment after fault clearance. Maximum power transfer and voltage support become further constrained when generators reach their temporary overload limits for field or armature current, prompting associated controllers to reduce reactive power output (and sometimes active power), ultimately leading to voltage instability \cite{Hatziargyriou2021}. 

In IPS with high penetration of IBRs, the impact differs because converters do not exhibit these characteristics and do not inherently provide inertia or voltage support. However, advanced control strategies such as GFL and GFM can enhance long-term voltage stability if designed to supply reactive power and voltage control. Poor coordination of these controllers, however, may exacerbate stability issues due to their dynamic behavior differing from that of traditional synchronous machines \cite{paperMevludinGlavic}.

\subsubsection{Resonance Stability}

Resonance stability refers to conditions where energy exchange occurs periodically in an oscillatory manner. In IPS, resonance stability becomes increasingly relevant due to the high penetration of converter-interfaced generation and the absence of strong grid damping. While traditional SSR was primarily associated with large interconnected networks using synchronous machines and series compensation, modern IPS can experience similar or even more severe resonance phenomena due to converter-grid interactions~\cite{Hatziargyriou2021,Collados2022}. This is because:
\begin{itemize}
    \item Some IPS may have series compensated lines, which may be subject to subsynchronous oscillations (SSO).
    \item Converter-based resources, such as doubly-fed induction generators (DFIGs) and GFM inverters, interact with network impedance and control loops, creating new resonance paths.
    \item Limited system inertia and weak grid conditions amplify the effect of negative damping introduced by converter controls, increasing the risk of subsynchronous control interaction.
\end{itemize}
The consequences include large oscillations in voltage and current, potential damage to converter hardware, and mechanical stress on turbine-generator assemblies. Mitigation strategies for IPS require advanced control design, supplemental damping controllers, and impedance-based stability analysis to ensure robust operation under varying load and generation conditions.

\subsubsection{Rotor Angle Stability}

Rotor-angle stability refers to the capability of synchronous generators to maintain synchronism when subject to disturbances. Rotor-angle stability can be divided into~\cite{Tielens2016,Hatziargyriou2021}:
\begin{itemize}
    \item Rotor-angle stability under small disturbances (electromechanical oscillations): They include local and inter-area oscillations.
    \item Rotor-angle stability under large disturbances (transient stability). 
\end{itemize}
\color{black}
In general, angle stability is a key limiting factor when large amounts of power are transmitted over long distances. Therefore, rotor-angle stability (under small and large disturbances) is more critical in large-scale power systems than in IPS~\cite{Tielens2016,Hatziargyriou2021}.

\subsubsection{Lessons Learned from Real-World Islanded Systems}

Several islands have demonstrated both the challenges and solutions for stability under high RES penetration:
\begin{itemize}
    \item \textbf{Faroe Islands}: Operation with 100\% wind power, stabilized by BESS with GFM control and synchronous condenser, but highly sensitive to large disturbances~\cite{Trondheim2023}.
    \item \textbf{Hawaii}: High shares of PV and wind, with BESS with GFM control, with advanced controls providing FFR and synthetic inertia; UFLS and GFM coordination are critical~\cite{Hoke2021,Akiba2024}.
    \item \textbf{Canary/Balearic Islands}: Potential converter-driven stability and resonance phenomena observed with the integration of MMC-HVDC and VSC-HVDC links~\cite{Collados2022,Fan2023}.
    \item \textbf{Madeira, Azores, Greek Islands}: High RES shares, use of BESS with GFM control, synchronous condensers, and adaptive UFLS schemes~\cite{Meschede2022}.
    \item \textbf{Jeju, Gotland}: VSC-HVDC interconnections and stability challenges with high wind penetration~\cite{Collados2022,Tielens2016}.
\end{itemize}

In summary, IPS face unique and acute stability challenges due to their small size, low inertia, and high penetration of inverter-based resources. Frequency stability and converter-driven stability are especially critical, requiring advanced control strategies, adaptive protection schemes, and robust coordination between GFM converters, BESS, and UFLS. Practical experience from islands such as the Faroe Islands, Hawaii, and the Canary/Balearic Islands demonstrates both the challenges and the solutions available for achieving reliable and resilient operation under high shares of renewables. Voltage and resonance stability are also important issues in IPS and may be considered in the stability analysis, but they are not as broadly critical as the previous two.

\subsection{Stability Analysis Methodologies for Islanded Power Systems} \label{sec:stability_methodologies}

The methodologies and classifications adopted in this report are fully aligned with the latest international standards and best practices, as outlined in CIGRE TB 928~\cite{cigreTB928}, Cheah-Mane et al.~\cite{cheahmane2022}, and IEEE/PES~\cite{Hatziargyriou2021}. This ensures that the stability analysis of converter-dominated and IPS is rigorous, consistent, and applicable to real-world scenarios. Following \cite{cigreTB928} and \cite{cheahmane2022}, stability analysis methods are classified as:
\begin{itemize}
    \item \textbf{Non-Linear Time-Domain (TD) Simulations:} EMT, RMS, real time simulations, HIL, co-simulation, frequency sweep based on frequency scan.
    \item \textbf{Small-Signal (Modal) Analysis:} Eigenvalue-based, participation factors, modal sensitivity.
    \item \textbf{Frequency-Domain (FD) Methods:} Transfer-function, impedance-based, passivity-based, positive-net-damping.
\end{itemize}

\subsubsection{Non-Linear Time-Domain Simulations}

Non-Linear time-domain simulations provide the most accurate representation of system dynamics because they capture nonlinearities, control saturation, and switching effects that are not visible in linearized models. These methods are essential for validating stability under large disturbances and for confirming the results obtained from small-signal analysis. According to CIGRE TB 928~\cite{cigreTB928}, TD approaches are typically used to assess transient (large-disturbance) stability and to validate robustness margins derived from linear analyses.

\paragraph{Electromagnetic Transient (EMT) Simulations}
EMT simulations solve the full set of differential equations of the power system using very small time steps (typically in the microsecond range). This allows the capture of phenomena from fundamental frequency to high-frequency introduced by PWM switching. In EMT models, the power grid is represented by its differential equations and converters' controllers are included in detail. The general form of the EMT model is:
\begin{equation}
\mathbf{C}\frac{d\mathbf{v}(t)}{dt} + \mathbf{G}\mathbf{v}(t) = \mathbf{i}(t)
\end{equation}
where $\mathbf{C}$ is the capacitance matrix, $\mathbf{G}$ the conductance matrix, $\mathbf{v}(t)$ the node voltages, and $\mathbf{i}(t)$ the injected currents. EMT tools such as PSCAD/EMTDC$\circledR$~\cite{pscad2020,emtdc2019}, Matlab-Simulink-Simpscape~\cite{mathworks_2026}, among others, are widely used for detailed converter modeling. In islanded systems with high IBR penetration, EMT is indispensable to reveal fast control interactions, PLL behaviours, and wideband resonances documented by recent CIGRE work~\cite{cigreTB928}.

It is important to highlight that EMT models include models with different level of detail, such as average or switching models. The level of detail of EMT models depends also of the electromagnetic phenomenon of interest. For example, an average EMT model could be enough to represent subsynchronous oscillations, but more detailed or switching EMT model is necessary to analyze harmonic stability. 

\paragraph{Root-Mean-Square (RMS) Simulations}

RMS (phasor) time-domain simulations approximate voltages and currents by their fundamental-frequency phasors and typically assume near-sinusoidal steady state. They are efficient for system-wide studies in the electromechanical range (roughly $\sim$0.1--10~Hz) and are commonly applied and used for power system studies, including transient stability phenomenon, electromechanical oscillations or short-term voltage stability. In converter-dominated systems, RMS applicability is bounded: under weak-grid or strong converter-control interactions, RMS simulation may miss some relevant dynamics and EMT simulations should be used instead~\cite{kuri2022,cheahmane2022}. In RMS models, the power grid is represented by algebraic equations and some controllers of power converters are included (e. g., outer controllers, power plant controllers, etc...), but some other controllers are typically neglected (inner current loops, PLLs).

A general RMS network representation can be expressed in compact phasor form as
\begin{equation}
    \mathbf{I} = \mathbf{Y}(\mathbf{x})\,\mathbf{V},
\end{equation}
where $\mathbf{V}$ and $\mathbf{I}$ denote nodal voltage and current phasors at the fundamental frequency, and $\mathbf{Y}(\mathbf{x})$ is the network admittance matrix, possibly dependent on slow states $\mathbf{x}$ (e.g., tap positions, load models, or converter control set-points). Time-domain waveforms are approximated as
\begin{equation}
    v(t) \approx \Re\{\tilde{V} e^{j\omega_0 t}\}, \qquad
    i(t) \approx \Re\{\tilde{I} e^{j\omega_0 t}\},
\end{equation}
highlighting that RMS simulations operate solely on the fundamental-frequency phasors. 

In RMS simulation, the power system is represented by a set of differential-algebraic equations:
\begin{eqnarray}
    \dot{\mathbf{x}} &=& \mathbf{f}(\mathbf{x},\mathbf{u}), \\
    0 &=& \mathbf{g}(\mathbf{x},\mathbf{u}).
\end{eqnarray}
where $\mathbf{x}$ are the state variables and $\mathbf{u}$ are the algebraic variables.

\paragraph{Frequency Response (FR) model}
For the assessment of frequency stability, the average value of the frequency of interest of the system (or the frequency of the center of inertia, COI) and the network can be neglected \cite{SigristBalearic}. The so-called Frequency Response (FR) model of the power system is widely used to analyze frequency stability, in which all the generators and loads are aggregated and relevant dynamics of the generators (inertial response and turbine and speed regulators) and UFLS schemes are included \cite{Egido_2009, Sigrist_2011}. 

\paragraph{Quasi-Static (QSS) Simulations}

Long-term voltage stability involves slow dynamics evolving over minutes. Steady-state analysis is often used for long-term voltage stability, but power-flow algorithms may present convergence problems when the operating point is close to voltage collapse. For this reason, specific steady-state methods can be used to assess long-term voltage stability \cite{Echavarren_part1_2011, Echavarren_part2_2011}. In some cases, dynamic simulation is required to capture slow dynamic behaviour (e.g., voltage stability and tap-changer dynamics). RMS models can be used, but the simulation horizon of several minutes makes the computational burden significant.

In this context, Quasi-Static (QSS) models provide an efficient alternative. The QSS model is obtained by assuming that fast electrical variables reach equilibrium instantaneously, while slow variables evolve dynamically. Formally, the full long-term dynamic model can be written as a singularly perturbed system \cite{kundur1994, chiang2011direct,vancutsem1998voltage}:
\begin{equation}
    \begin{aligned}
\dot{x} &= f(x, y, z), \\
\epsilon \dot{y} &= g(x, y, z), \\
0 &= h(x, y, z),
\end{aligned}
\end{equation}
where $x$ represents slow states (e.g., tap-changer positions, load recovery), $y$ represents fast electrical states (e.g., generator internal states), $z$ are algebraic network variables, and $\epsilon$ is a small parameter. The QSS approximation is obtained by taking the limit $\epsilon \to 0$, which forces the fast dynamics to satisfy the algebraic constraint $g(x,y,z)=0$ at all times. The resulting QSS model is therefore:
\begin{equation}
\begin{aligned}
    \dot{x} &= f(x, y, z), \\
0 &= g(x, y, z), \\
0 &= h(x, y, z),
\end{aligned}
\end{equation}
which neglects fast transients while preserving the slow evolution of the system. As described in \cite{chiang2011direct, vasconcelosAdvancedControlSolutions2015}, QSS models are particularly suitable for long-term voltage stability studies when steady-state analysis is insufficient to reproduce slow dynamic behaviour, but full RMS simulations are computationally restricted.

\paragraph{Real-time simulation}

In real-time simulation, EMT models are used. It can be considered as fast EMT simulations (in almost real time), and the same models could be used for HIL simulations. 

\paragraph{Hardware-in-the-Loop (HIL) Simulations}

HIL simulations integrate real controller hardware into the simulation environment. The plant model runs in real-time on a digital simulator, while the actual controller hardware executes its control algorithms. This approach validates the performance of control systems before deployment and ensures that stability margins are maintained under realistic conditions \cite{paperGermano,cigreTB928}.

\paragraph{Co-Simulation of Multi-Vendor Systems}

In modern power systems, especially islanded microgrids, components from different vendors often operate together. Co-simulation frameworks allow the integration of different simulation tools (e.g., EMT models for converters and RMS models for the grid) to analyze interactions. This is critical for systems where proprietary models are unavailable, as it enables stability assessment without full disclosure of internal control details \cite{paperGermano,cigreTB928}.

\paragraph{Frequency Scan and Frequency Response from EMT Models}

Frequency scanning techniques inject sinusoidal perturbations at various frequencies into the EMT model to derive impedance or admittance profiles. These profiles are then used to build small-signal models or validate impedance-based stability criteria. The frequency response is obtained by:
\begin{equation}
H(j\omega) = \frac{\text{Output Response}}{\text{Input Perturbation}}
\end{equation}

This method bridges the gap between time-domain and frequency-domain analysis, providing accurate data for stability studies \cite{paperGermano,cigreTB928,paperMolinas2017toolimpedance, garciareyes2026siadtool}.

\paragraph{Implications for Islanded Power Systems}

IPS that rely heavily on power electronic converters for generation, storage, and load interfacing introduce fast dynamics and nonlinear interactions into the network. This dependency makes the system particularly sensitive to disturbances such as faults or sudden load changes, which can trigger instability. To accurately capture these phenomena, time-domain simulations are indispensable. EMT and HIL simulations provide a detailed representation of converter behavior and validate control robustness under varying conditions, while frequency scanning techniques enable impedance-based analysis for resonance detection and stability assessment.

\medskip
\paragraph{Short- vs Long-Term Studies.} Following IEEE/PES~\cite{Hatziargyriou2021}, short-term studies (seconds) in IPS typically require EMT/RMS with controller detail. Long-term studies (minutes) use QSS/RMS coupled with slower dynamics (OLTCs, AGC, load recovery). When converter-driven oscillations emerge, EMT or frequency-domain methods are generally required~\cite{cigreTB928}.

\subsubsection{Small-Signal Analysis (Modal Analysis)}

Small-signal stability analysis is a fundamental approach for evaluating the dynamic performance of power systems under small perturbations around a steady-state operating point. Modern power systems, especially IPS, are increasingly dominated by power electronic converters, which introduce complex dynamics across a wide frequency range. Operating in isolation from the main grid, IPS face unique stability challenges. Unlike traditional synchronous generator-based systems, these converter-dominated networks exhibit low inertia and limited fault current capability, making stability assessment more challenging. Small-signal analysis provides a systematic way to identify oscillatory modes, assess damping, and design controllers that ensure reliable operation under varying conditions without resorting to exhaustive time-domain simulations. Similar to large interconnected systems, stability in IPS is categorized into large-disturbance stability (e.g., faults) and small-signal stability, which deals with minor perturbations such as load changes or controller interactions~\cite{paperGermano,paperShuyaWang,bookMJGibbard,kundur1994,cheahmane2022}.

The relevance of small-signal analysis in IPS lies in its ability to address:
\begin{itemize}
\item High penetration of power electronics, which introduces fast dynamics and potential resonance and passivity issues.
\item Limited fault current and protection complexity, making stability margins tighter.
\item Interaction between multiple DG units and storage, often from different vendors, creating black-box modeling challenges.
\item Wide frequency stability concerns, from sub-synchronous to harmonic ranges, due to converter switching and filter dynamics.
\end{itemize}

By applying small-signal techniques, operators and designers can predict system behavior under small disturbances, optimize control strategies, and maintain stability in these low-inertia, converter-dominated environments—an essential step for the reliable integration of renewables and achieving decarbonization goals. In this section, the different small-signal methodologies for analyzing stability in IPS are discussed.

The mathematical foundation of small-signal analysis begins with the representation of the nonlinear system using differential-algebraic equations \cite{kundur1994,cheahmane2022}:
\begin{equation}
\begin{aligned}
\dot{x} &= f(x,u), \\
y &= g(x,u),
\end{aligned}
\end{equation}
where $x$ is the state vector, $u$ the input vector, and $y$ the output vector. These equations are inherently nonlinear due to factors such as magnetic saturation, controller limits, and coupling between electrical and mechanical dynamics. To apply small-signal analysis, the system is linearized around an operating point, resulting in:
\begin{equation}
\begin{aligned}
\Delta \dot{x} &= A \Delta x + B \Delta u, \\
\Delta y &= C \Delta x + D \Delta u,
\end{aligned}
\end{equation}
where $A$ is the state matrix, $B$ the input matrix, $C$ the output matrix, and $D$ the direct transmission matrix. These matrices, also referred to as Jacobian matrices, are evaluated at the operating point. This linearized model forms the basis for state-space analysis and eigenvalue computation \cite{kundur1994,cheahmane2022}.

\paragraph{State-Space and Eigenvalue-Based Analysis}
State-space analysis is one of the most widely used methodologies for small-signal stability studies. Eigenvalues and participation factors (PFs) are conventional tools used to analyze the impact of oscillatory modes in power system stability. Here, the linearized system is expressed as \cite{paperGermano,cheahmane2022}:
\begin{equation}
\begin{aligned}
\dot{x} &= A x + B u, \\
y &= C x + D u,
\end{aligned}
\end{equation}
where:
\begin{itemize}
\item $x \in \mathbb{R}^n$ is the state vector (e.g., rotor angles, converter control states).
\item $u \in \mathbb{R}^m$ is the input vector (e.g., voltage references, power commands).
\item $y \in \mathbb{R}^p$ is the output vector (e.g., terminal voltages, currents).
\item $A \in \mathbb{R}^{n \times n}$ is the state matrix, which governs system dynamics.
\end{itemize}

The eigenvalues of matrix \(A\) determine the stability of the linearized system. 
If all eigenvalues have negative real parts, the system is locally asymptotically stable around the operating point. However, global stability cannot be guaranteed by eigenvalues alone; it requires that the original nonlinear system be globally stable. This typically involves proving the existence of a Lyapunov function that is valid over the entire state space \cite{khalil2002nonlinear}. 

For oscillatory modes, eigenvalues appear as complex conjugate pairs:
\begin{equation}
\lambda = \sigma \pm j\omega
\end{equation}
where $\sigma$ is the real part, representing damping, and $\omega$ is the imaginary part, representing oscillation frequency. The damping ratio $\xi$ and frequency $f$ of each mode are computed as:
\begin{equation}
\xi = -\frac{\sigma}{\sqrt{\sigma^2 + \omega^2}}, \quad f = \frac{\omega}{2\pi}
\end{equation}

Unstable modes correspond to $\sigma > 0$, while poorly damped modes have $\sigma \approx 0$. If $\sigma < 0$, the mode is stable. Here, the smaller $\sigma$ is, the more
stable the mode is \cite{paperQifanChen}.

\paragraph{Participation Factors}

To identify which states contribute most to each mode, participation factors (PFs) are calculated using the left and right eigenvectors of $A$ following the original formulation \cite{IPA_1982,verghese1982,pagola1989}:
\begin{equation}
p_{ik} = w_{ki} \, v_{ik},
\end{equation}

where $v_{k}$ and $w_{k}$ are, respectively, the right and left eigenvectors associated with mode $k$, and $p_{ik}$ denotes the participation of state $i$ in mode $k$. Proper normalization (e.g., $w_k^\top v_k = 1$) is assumed to ensure consistent interpretation. PFs provide insight into the interaction between system components. For example, in IPS, high PFs associated with converter control states indicate that oscillatory modes are driven by control dynamics rather than network elements \cite{paperShuyaWang}. This can provide a full dynamics detail and metrics for assess the stability.

\paragraph{Limitations}
Eigenvalue analysis offers a clear picture of system stability and is highly effective for designing controllers such as Automatic Voltage Regulators (AVRs), Power System Stabilizers (PSSs), and power oscillation damping controllers (POD) in converters. However, this method requires detailed models of all components, which may not be available in multi-vendor environments where converters are provided as black-box models \cite{garciareyesluisa_fitss2025}. Additionally, the analysis is valid only near the linearization point and does not capture nonlinear phenomena during large disturbances.

\paragraph{On SSA-EMT vs SSA-RMS models}

When constructing linearized state-space models for small-signal stability analysis (SSA models, for short), two families are common in converter-rich systems \cite{Cheah_2026}:
\begin{itemize}
\item \textbf{SSA-EMT}: they are linear models derived from EMT-level representations (including wideband converter and filter dynamics, possibly via harmonic state-space or linear time-periodic approximations). These models preserve frequency couplings and are suitable to analyze converter-driven and wideband phenomena. Notice that SSA-EMT models can also have different level of detail: first-harmonic SSA-EMT models are accurate enough to analyze SSOs or some other slow control interactions, while more detailed SSA-EMT models are needed to analyze harmonic stability (oscillations in the kHz range).
\item \textbf{SSA-RMS:} they are linear models derived from RMS/phasor representations (assuming balanced sinusoidal operation and focusing on electromechanical-range dynamics). These are effective for low-frequency modes and system-wide interactions but cannot represent sub-cycle harmonics. SSA-RMS models are useful to analyse electromechanical oscillations in power systems.
\end{itemize}
Therefore, SS-EMT is recommended where controllers, PLLs, and filters shape modes beyond the electromechanical band, while SS-RMS is adequate to assess electromechanical oscillations or other slow oscillations \cite{cheahmane2022,kuri2022,Hatziargyriou2021}.

\subsubsection{Frequency-Domain Methods}

Frequency-domain methods analyze system dynamics using algebraic representations derived from Laplace transforms. These approaches are particularly useful for converter-dominated systems and IPS because they allow stability assessment without requiring full disclosure of proprietary control models. The general idea is to represent the system as a set of transfer functions or impedance models and evaluate stability using classical control criteria such as Nyquist or Bode plots \cite{paperGermano,cheahmane2022,paperMolinas2017}.

\paragraph{Transfer-Function Based Analysis}

The system can be represented as a multiple-input multiple-output (MIMO) control system, where the open-loop transfer function is:
\begin{equation}
L(s) = G(s)H(s)
\end{equation}
and the closed-loop transfer function is:
\begin{equation}
T(s) = \frac{G(s)}{1 + L(s)}
\end{equation}
For a MIMO system, $L(s)$ is a matrix, and stability is assessed by analyzing the eigenvalues $\lambda_k$ of $L(j\omega)$ across the frequency range. The General Nyquist Criterion (GNC) is applied to each eigenvalue to determine stability margins:
\begin{equation}
\text{Stability} \iff \text{no clockwise encirclements of } (-1,0) \text{ by } \lambda_k(L(j\omega)).
\end{equation}

This method provides insight into the frequency-dependent behavior of the system and is widely used for converter control design. However, it requires accurate transfer-function models, making the definition of $G(s)$ and $H(s)$ very important for an accurate analysis \cite{skogestad2005,harnefors2007}.

\paragraph{Impedance-Based Analysis}

Impedance-based methods are among the most practical for converter-dominated systems. The system is divided into two subsystems: the source (e.g., grid or upstream converter) and the load (e.g., downstream converter or microgrid). Each subsystem is represented by its impedance or admittance in the frequency domain:
\begin{equation}
\begin{bmatrix}
i_x \\ i_y
\end{bmatrix}
=
\begin{bmatrix}
Y_{xx}(s) & Y_{xy}(s) \\
Y_{yx}(s) & Y_{yy}(s)
\end{bmatrix}
\begin{bmatrix}
v_x \\ v_y
\end{bmatrix}
\end{equation}
The inverse of the admittance matrix defines the impedance matrix $Z(s)$. Stability is evaluated using the minor loop gain:
\begin{equation}
L(s) = Z_{\text{source}}(s) \cdot Y_{\text{load}}(s)
\end{equation}
Nyquist plots of $L(s)$ indicate stability if there are no clockwise encirclements of the point $(-1,0)$. For IPS, impedance-based analysis is highly relevant because it works with terminal measurements and does not require internal control details. It also captures high-frequency dynamics introduced by PWM and filter interactions \cite{paperMolinas2017,paperMolinas2017toolimpedance,sun2011,harnefors2007}.

\paragraph{Impedance Ratio Analysis}

The impedance ratio method simplifies the evaluation by considering the ratio of source and load impedances:
\begin{equation}
\text{Impedance Ratio} = \frac{Z_{\text{source}}(s)}{Z_{\text{load}}(s)}
\end{equation}
Stability is assessed using Bode plots of the impedance ratio. If the phase margin and gain margin meet predefined criteria, the system is considered stable. This method is straightforward but provides limited insight into the origin of instability \cite{cheahmane2022,paperJianSun}.

\paragraph{Passivity-Based Analysis}

Passivity-based methods ensure stability by verifying that each subsystem is passive, meaning it does not generate energy. A system is passive if:
\begin{equation}
\Re\{Z(s)\} \geq 0 \quad \forall \omega
\end{equation}
For MIMO systems, passivity is checked using the passivity index:
\begin{equation}
\lambda_{\min}\left( \frac{F(j\omega) + F^\top(j\omega)}{2} \right) \geq 0,
\end{equation}

where $F(s)$ is the transfer matrix of the subsystem. If both subsystems are passive, the interconnected system is stable. This method is conservative but useful for screening critical frequency ranges \cite{beza2019,beza2020,sainz2017}.

\paragraph{Limitations}
Although frequency-domain approaches are powerful for converter-dominated systems, they present several limitations:
\begin{itemize}
\item They require accurate frequency-domain models or measurements, which may not always be available for black-box devices.
\item Operating-point dependency can lead to incorrect conclusions if the system conditions change significantly.
\item Computational complexity increases for large-scale MIMO systems, especially when applying Nyquist or singular value analysis.
\item Passivity-based criteria are conservative and do not guarantee optimal performance, only stability margins.
\item These methods do not directly provide time-domain responses or nonlinear behavior under large disturbances.
\end{itemize}
For IPS, these limitations imply that frequency-domain methods should be complemented with time-domain simulations to validate stability under realistic conditions.

\subsubsection{Limitations and Recommendations for Islanded Environment}

Consistent with the IEEE/PES extended definitions~\cite{Hatziargyriou2021} and the CIGRE multi-frequency perspective~\cite{cigreTB928}, Table~\ref{tab:methodMap} summarizes which methodology is suited to which stability type in IPS, indicating time scale and typical models.

\paragraph{Practical Guidance for Method Selection in IPS}

\begin{itemize}
\item Use \textbf{EMT/TD} whenever PLLs, current controllers, or filter resonances are suspected to drive instability; complement with \textbf{FD impedance} (minor-loop gain) for root-cause and margin quantification \cite{cigreTB928,sun2011}.
\item Prefer \textbf{SS-EMT} for modal insight in wideband converter interactions; prefer \textbf{SS-RMS} for electromechanical-range oscillations and short-term voltage stability \cite{cheahmane2022,kuri2022}.
\item Apply \textbf{QSS/RMS} to analyse long-term voltage/frequency phenomena (minutes) with slow devices (OLTCs, AGC) \cite{Hatziargyriou2021}.
\item In multi-vendor IPS, combine \textbf{co-simulation} and \textbf{terminal measurements} (frequency scans) to build FD models and perform \textbf{black-box} stability screening \cite{cigreTB928}.
\end{itemize}

\paragraph{Recommended Workflow and Best Practices}

\begin{figure}[H]
    \centering
\includegraphics[width=0.82\textwidth,angle=90]{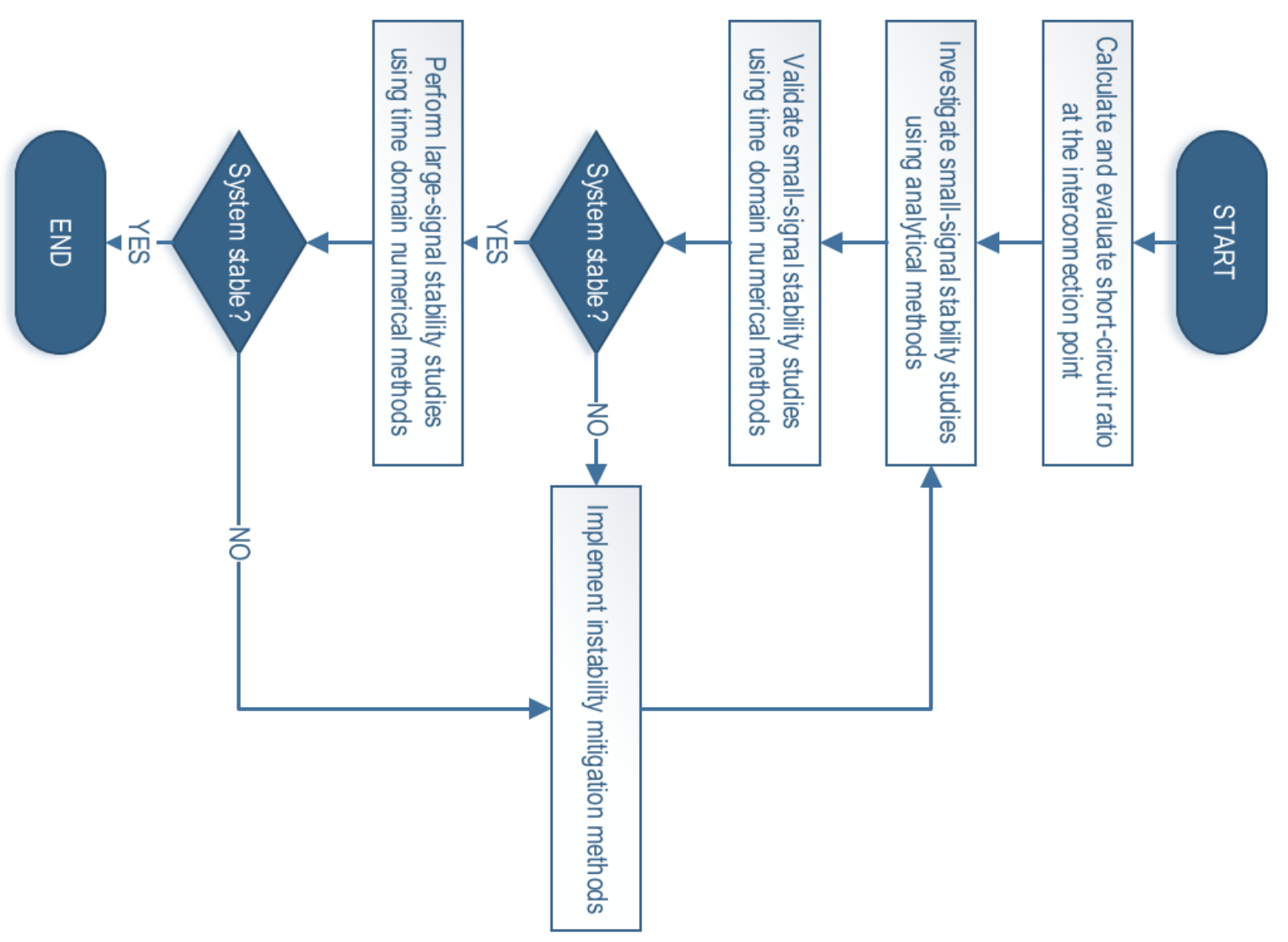}
    \caption{Recommended stability analysis workflow for converter-based power systems. Figure from CIGRE TB 928 \cite{cigreTB928}.}
    \label{fig:workflow}
\end{figure}

Following the recommended workflow in CIGRE TB 928 (see Fig.~\ref{fig:workflow}), stability analysis and mitigation should proceed as:
\begin{enumerate}
    \item Perform small-signal stability analysis (impedance, eigenvalue, passivity).
    \item Validate results with time-domain simulations.
    \item Apply mitigation: control tuning, passive/active filters, SCR reinforcement, etc.
\end{enumerate}
\newcommand{\wA}{0.20} 
\newcommand{\wB}{0.16} 
\newcommand{\wC}{0.24} 
\newcommand{\wD}{0.24} 
\begin{table}[H]
\centering
\caption{Mapping of stability types to recommended methodologies in islanded, converter-dominated systems (based on the review of section \ref{Stability_review}).}
\label{tab:methodMap}
\renewcommand{\arraystretch}{1.25}
\setlength{\tabcolsep}{6pt}
\normalsize
\begin{tabular}{
  p{\wA\linewidth}
  p{\wB\linewidth}
  p{\wC\linewidth}
  p{\wD\linewidth}
}
\toprule
\textbf{Stability type} 
& \textbf{Time scale} 
& \textbf{Primary methods} 
& \textbf{Notes (IPS context)} \\
\midrule

Frequency stability 
& seconds (short-term), minutes (long-term)
& FR model
& Short-term: details of droop, virtual inertia, PLL, UFLS and FFR. Long-term: dynamics involve AGC and load recovery \cite{Hatziargyriou2021}. \\

Converter-driven (wideband) 
& slow interactions: 0.1–2.5 Hz; SSOs: 2.5–50 Hz; fast interactions:  $>50$ Hz
& SS-EMT, FD impedance-based, passivity analysis, TD EMT
& Dominant in IBR-rich IPS. Captures PLL-induced negative damping, control interactions, weak-grid effects, and resonance phenomena. Requires MIMO Nyquist or passivity screening \cite{cigreTB928,sun2011,harnefors2007}. \\

Voltage stability 
& seconds (short-term), minutes (long-term)
& Short-term: TD RMS; Long-term: steady-state analysis or QSS
& Short-term driven by induction motors, HVDC/IBR dynamics. Long-term dominated by LTCs, load recovery and reactive power limits. QSS suitable when RMS becomes computationally heavy. \\

Harmonic / Interharmonic
& $>100$ Hz
& FD impedance analysis, SS-EMT, TD EMT
& EMT frequency scans used to obtain $Z(s)$. Apply MIMO Nyquist or passivity-based criteria \cite{paperMolinas2017toolimpedance,cigreTB928}. \\

Rotor angle (synchronous) 
& 0.1–10 s
& Large disturbance: TD RMS; small disturbance: SSA-RMS; TD EMT
& Relevant in hybrid IPS with synchronous machines. Electromechanical modes captured by RMS; EMT required for unbalanced or faulted conditions \cite{Hatziargyriou2021}. \\

\bottomrule
\end{tabular}
\end{table}

\subsection{Review on GFM Converter Technology} \label{GFM_control}

\subsubsection{Motivation for GFM Control}

The growing integration of renewable energies increases converter-based generation in the power system. Power systems increasingly encounter weak grid and islanded operating scenarios, especially in remote locations such as offshore wind farms and large PV plants. Under these conditions, due to a lack of sufficient synchronous machine support, the system experiences low inertia and reduced voltage and frequency stability. Consequently, robust control strategies are required to ensure reliable operation and maintain grid stability.

\begin{figure}[!htbp]
  \centering
  \includegraphics[width=1\linewidth]{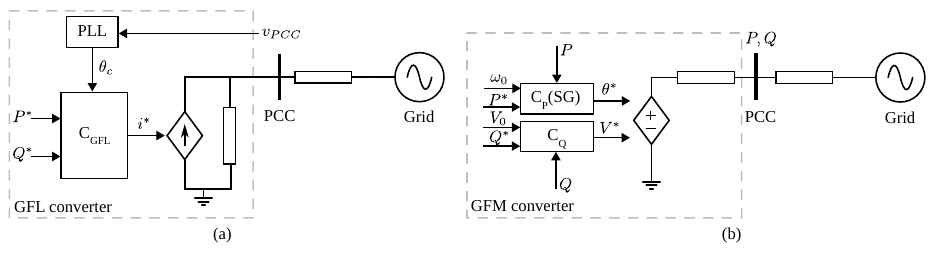} 
  \caption{GFL and GFM converter control structure.}
  \label{fig:GFM-vs-GFL-control}
\end{figure}

Most power-electronic voltage source converters (VSC) in current power systems operate in grid-following (GFL) mode, where the converter synchronize to the grid using a Phase-Locked Loop (PLL) and control the dq frame currents based on reference points, as shown in \autoref{fig:GFM-vs-GFL-control}(a). As a result, GFL converters act as controlled current sources and rely on a stable external voltage at the grid-connection point for phase reference. In weak grid or islanded conditions, the grid voltage can become unstable, leading to phase tracking errors and even loss of synchronism \cite{verbeComparativeStudyGFMgrid2021}. Furthermore, GFL converters lack inherent frequency or voltage regulation capabilities and cannot provide black-start functionality that is essential for islanded systems.

The limitations of the GFL control motivate the development of grid-forming technologies (GFM). The typical structure of the GFM control is illustrated in \autoref{fig:GFM-vs-GFL-control}(b). In contrast to GFL control, phase information is obtained by emulating synchronous generators; typical approaches include power-based synchronization (PSC) and virtual synchronous machine (VSM) control. As a result, GFM converters actively establish and regulate voltage and frequency at their connection points by setting the phase (and thus frequency) and voltage, behaving like a controlled voltage source 
\cite{liRevisitingGridFormingGridFollowing2022}. GFM converters can operate independently or in parallel with other units. They regulate voltage and frequency, providing synthetic inertia, and enable black-start and system restoration. In weak grid or islanded scenarios, GFM converters are essential for maintaining stable operation.

The following sections review major control strategies, functional capabilities, and challenges of GFM control. 

\subsubsection{GFM Control Strategies}

The swing equation of a synchronous generator links imbalance in active power to the frequency dynamics, 
\begin{align}
2H\,\frac{d\,\Delta\omega}{dt} \;=\; P_{m}-P_{e}-D_{pu}\,\Delta\omega ,
\label{eq:swing-equation}
\end{align}
where $H$ is the inertia constant, $D_{pu}$ is the per-unit damping, $\Delta\omega=(\omega-\omega_0)/\omega_0$ is the frequency deviation, $P_m$ and $P_e$ are mechanical input and electrical output power. 
Most GFM controllers generate the internal referece angle by emulating the swing equation based on active-power imbalance and frequency. They can be grouped as follows:

\begin{itemize}
    \item $P-f$ droop droop control and power-synchronization control (PSC) set the frequency based on active power errors, implementing the damping term in the swing equation. 
    \item Virtual synchronous generator (VSM) control introduces inertia and damping for improved frequency dynamics. 
    \item Matching control uses the DC-link energy (voltage) as the power deviation, coupling AC frequency to the DC-link energy balance so that the DC-link voltage is regulated. 
    \item Virtual-oscillator control (VOC) does not implement the swing equation explicitly; however, the oscillator dynamics reproduce $P-\omega$ droop-like behaviors and provide fast and robust synchronization. 
    \item Advanced controllers combine or optimize strategies to improve performance across a wide range of grid conditions.
\end{itemize}

\paragraph{$P-f$ Droop Control}

\begin{figure}[!htbp]
  \centering
  \includegraphics[width=0.4\linewidth]{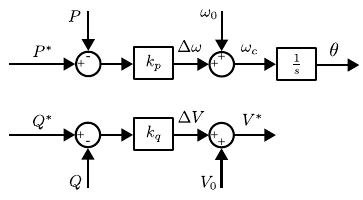} 
  \caption{Droop control of active and reactive power.}
  \label{fig:droop-control}
\end{figure}

\begin{figure}[!htbp]
  \centering
  \includegraphics[width=0.35\linewidth]{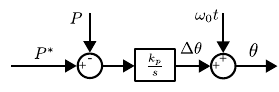} 
  \caption{Power-based synchronization control.}
  \label{fig:PSC-control}
\end{figure}

$P-f$ droop control has been widely used in microgrids for power sharing and voltage control \cite{liDesignAnalysisRealtime2004}. A typical control structure is shown in \autoref{fig:droop-control}. The deviation of converter's frequency from the nominal frequency, $\Delta \omega=(\omega_c-\omega_{0})$, is proportional to the error in active power: 
\begin{align}
-k_{p}(P-P^{*})  = \omega_{c} - \omega_{0}=\Delta \omega, 
\label{eq:p-droop}
\end{align}
where $k_{p}$ is the droop coefficent for active power. The power-based synchronization \cite{zhangPowerSynchronizationControlGridConnected2010}, shown in \autoref{fig:PSC-control}, is equivalent to the droop control with the integration moved to the back. $P-f$ droop control is in accordance with the $P-\omega$ droop of SGs by omitting the inertia term in \eqref{eq:swing-equation}. $P-f$ droop realizes robust synchronization without a PLL and supports parallel operation. However, $P-f$ droop control only emulates the steady-state $P-\omega$ characteristics of SGs, therefore, it provides no inertial support. In transmission systems, the droop coefficient is often tightly bounded by the TSO (e.g., $2\%$-$12\%$ according to ENTSO-E \cite{ENTSOE2013-RfG-IGD}), limiting the freedom for parameter tuning. As a result, there can be a non-trivial trade-off between dynamic performance and grid-code compliance.

Similar to $P-f$ droop, voltage regulation and reactive power sharing are realized with reactive power droop control (\autoref{fig:droop-control}):
\begin{align}
-k_{q}(Q-Q^{*})  = V^{*} - V_{0},
\label{eq:q-droop}
\end{align}
where and $k_{p}$ is the droop coefficent for reactive power and $V_0$ is the nominal terminal voltage. 

\paragraph{Virtual Synchronous Machine (VSM)}

\begin{figure}[!htbp]
  \centering
  \includegraphics[width=0.95\linewidth]{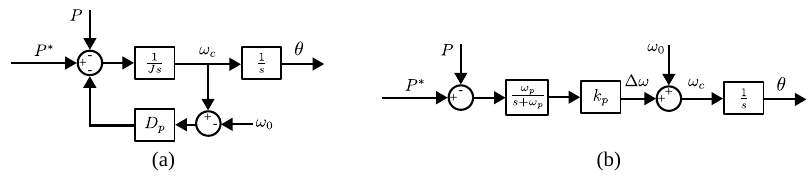} 
  \caption{Different realizations of VSM. (a) Classic. (b) $P-f$ droop with low-pass filtered active power error.}
  \label{fig:VSM-control}
\end{figure}

Virtual synchronous machine (VSM) is realized by introducing inertia to droop control \cite{darcoVirtualSynchronousMachines2013}, as shown in \autoref{fig:VSM-control}(a). The frequency dynamics are: 
\begin{align}
J\dot{\omega_{c}} = (P^{*}-P)-D_{P}(\omega_{c}-\omega_{0}),
\label{eq:VSM}
\end{align}
which emulates the swing equation with synthetic inertia, $J$, and damping, $D_P$. 

$P-f$ droop control with the active power low-pass filtered, as shown in \autoref{fig:VSM-control}(b), achieves frequency dynamics equivalent to that of VSM \cite{darcoEquivalenceVirtualSynchronous2014}: 
\begin{align}
\frac{1}{\omega_{p}k_{p}}\dot{\omega}_{c} =(P^{*}-P)-\frac{1}{k_{p}}(\omega_{c}-\omega_{0}), 
\label{eq:P-LPF}
\end{align}
where $\omega_p$ is the cutoff frequency of the low-pass filter. The frequency dynamics of P-LPF \eqref{eq:P-LPF} is equivalent to that of the VSM \eqref{eq:VSM} with a parameter mapping: $J=\frac{1}{\omega_p k_p},\, D_p=\frac{1}{k_p}$.

\begin{figure}[!htbp]
  \centering
  \includegraphics[width=0.5\linewidth]{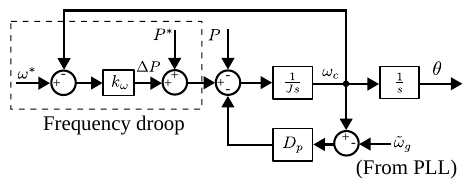} 
  \caption{VSM-PLL control with external frequency droop.}
  \label{fig:VSM-PLL-control}
\end{figure}

In practical implementations, the VSM scheme often uses a PLL for grid frequency estimation (VSM-PLL) \cite{darcoVirtualSynchronousMachine2015}, as shown in \autoref{fig:VSM-PLL-control}. The constant grid frequency reference $\omega_0$ in the damping is replaced by the estimated grid frequency $\tilde{\omega}_g$ obtained via a PLL. This configuration ensures accurate damping power based on the measured frequency deviation. Additionally, an external frequency droop loop is introduced to regulate the steady-state frequency. However, relying on a PLL introduces potential instability in weak grids compared to PLL-free implementations that use a fixed reference.

\begin{figure}[!htbp]
  \centering
  \includegraphics[width=0.4\textwidth]{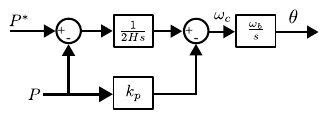} 
  \caption{PLL-free integral-proportional (IP) control.}
  \label{fig:IP-control}
\end{figure}

The control structure of VSM can be further optimized. A PLL-free integral-proportional (IP) control architecture (\autoref{fig:IP-control}) was proposed in \cite{qoriaPLLFreeGridFormingControl2020a}. Unlike the classic VSM where inertia and damping parameters are coupled, the IP controller decouples these dynamics and therefore facilitates parameter tuning and exhibits adequate small-signal and transient performance. This approach simplifies the design and modeling process and provides inertial support without requiring a dedicated PLL.

\begin{figure}[!htbp]
  \centering
  \includegraphics[width=0.6\linewidth]{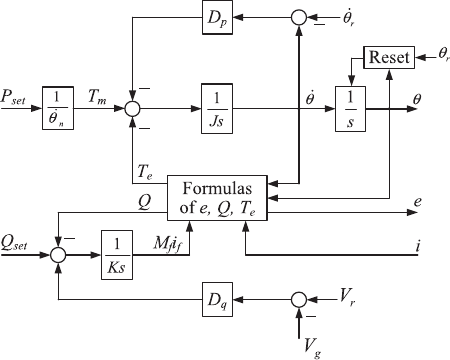} 
  \caption{Synchronverter control. Figure from \cite{zhongSelfSynchronizedSynchronvertersInverters2014}).}
  \label{fig:synchronverter-control}
\end{figure}

The synchronverter further extends the VSM concept by emulating additional SG quantities (e.g., electromagnetic torque) and damping paths \cite{zhongSelfSynchronizedSynchronvertersInverters2014}, as shown in \autoref{fig:synchronverter-control}. It retains the swing equation in \eqref{eq:swing-equation} while providing a more comprehensive emulation of SGs.

Compared with droop control, the VSM-type controllers provide inertial support and contribute to frequency regulation. 
However, the added synthetic inertia and damping can make the control more susceptible to disturbances. Careful tuning of inertia and damping parameters is essential for stable operation \cite{panTransientStabilityVoltageSource2020}.  

%
%

\paragraph{Virtual-oscillator Control (VOC)}

\begin{figure}[!htbp]
  \centering
  \includegraphics[width=0.6\linewidth]{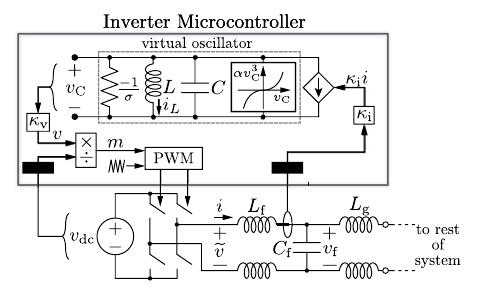} 
  \caption{VOC control. Figure from \cite{johnsonSynthesizingVirtualOscillators2016}).}
  \label{fig:VOC-control}
\end{figure}

Virtual-oscillator control (VOC) is a time-domain nonlinear control strategy that realizes grid synchronization by mimicking digitally the behavior of an oscillator \cite{johnsonSynthesizingVirtualOscillators2016}. As shown in \autoref{fig:VOC-control},
the current measurement is scaled and used as the excitation of the nonlinear virtual oscillator (VO).  The capacitor voltage of the VO is then scaled to generate the converter's voltage reference. Because VOC operates directly in the time domain, it does not rely on explicit voltage or frequency measurements, enabling rapid response and robust synchronization. The original VOC does not have embedded power setpoints; a dispatchable VOC (dVOC) with configurable power setpoints is proposed to enable decentralized power sharing \cite{seoDispatchableVirtualOscillator2018}. 

Although VOC does not explicitly emulate the swing equation, it exhibits droop-like steady-state behavior. 
classical VOC/dVOC does not provide inertia support, and the nonlinear dynamics complicate parameter selection. The original single-loop voltage control is also challenging for the implementation of fault ride-through and current limiting control.

\paragraph{Advanced Control Strategies}

\begin{figure}[!htbp]
  \centering
  \includegraphics[width=0.5\linewidth]{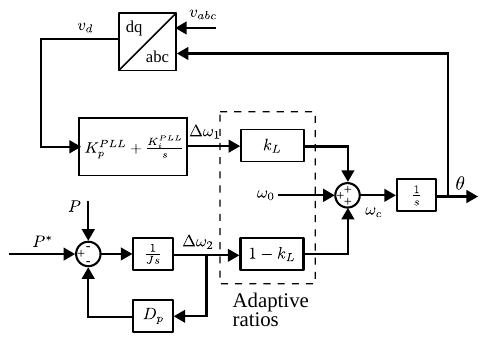} 
  \caption{Hybrid control that combines VSM and PLL.}
  \label{fig:hybrid-GFM-GFL-control}
\end{figure}

Recent developments of GFM control aim to improve robustness and dynamic performance across different grid conditions. 
One direction is to develop a hybrid and universal control that utilizes both GFM and GFL mechanisms. For example, a hybrid control is proposed in \cite{harneforsUniversalControllerGridConnected2021} that combines PSC and PLL for synchronization. This hybrid control is further developed in \cite{liuAdaptiveHybridGridForming2024} and the ratio of the VSM and PLL are adaptively selected based on grid strength (\autoref{fig:hybrid-GFM-GFL-control}) so that the stability is improved in various grid conditions. 

Multivariable control synthesis is applied to the GFM converters by treating all the control loops, such as frequency, DC voltage, and current, as a MIMO control structure. This enables joint parameter optimization using methods such as $H_\infty$ \cite{chenGeneralizedMultivariableGridForming2022}. Model predictive control (MPC) can also be employed to achieve fast dynamic response, reduced harmonic distortion, fault ride-through, and fast current limiting \cite{zhangModelPredictiveDualControlLoop2025, carnieluttiFixedSwitchingFrequency2025}.

\section{Technology Enablers for the Integration of RES in Islanded Power System}

A technology enabler is a technology or a capability that can provide a solution to an existing challenge. In the context of an islanded power system, technology enablers are the key technologies, controls, and system-level capabilities that makes the operation of the system stable and reliable. Some of the possible technology enablers for islanded power system are provided below:
\begin{itemize}
    \item AC interconnections.
    \item HVDC interconnections.
    \item Synchronous condensers (SYNCON).
    \item GFM control of power converters.
    \item Energy storage systems (ESS) through power converters (ESS, ultracapacitors, flywheels, pumped hydro power plants, hydrogen). 
    \item FACTS (STATCOMs, E-STATCOMs).
    \item Must-run units
\end{itemize}

\subsection{AC Interconnections}

In the context of an islanded power system, AC interconnection can be both between the islands and between the island and the mainland. AC interconnection to the mainland can be an effective measure to ensure that the island power system does not become an energy island while interconnecting islands can increase the RES participation with in the island system \cite{BalearicAC}. \color{black} In the paper \cite{SigristBalearic}, the authors have studied the value of interconnection in remote islands considering using realistic case studies. The authors conclude that use of interconnection links enable cheaper power flow between islands, but also significantly contribute to the fulfillment of reserve constraints which leads to a cheaper and more sustainable island operation. \color{black} Study done on Portugese archipelago of Azores showed that interconnection of the power systems of the two islands, Pico and Faial,  increased RES penetration by 50 \% \cite{AzoresAC}. The \color{black} work \color{black} in \cite{Renedo2024BalearicHVDC} specifies AC interconnection between islands in the Balearic archipelago as a technology enabler for increasing transmission capacity between the island and strengthening the overall system. In paper \cite{GreekAC}, the authors conclude that the interconnection of Greek islands to the mainland offers both economic and environmental benefits to the whole Greek power sector.

\subsection{HVDC Interconnections}

HVDC technology, with its bulk power transmission capability over long distances, provides an excellent option for interconnecting islands which are far away from the mainland. This report by Red Eléctrica presents the Romulo project \cite{Romulo}, an LCC-HVDC connection between the Spanish mainland and Mallorca in the Balearic archipelago, and how it increases both the quality and reliability of the electricity supply in the Balearic Islands. To further improve the reliability of the Balearic power system and increase the transmission capacity while decarbonising the overall system, Red Eléctrica plans to build a VSC-HVDC line between the mainland and the Balearic archipelago \cite{Renedo2024BalearicHVDC}. Another example of application of HVDC technology in islanded power system is the Jeju island in South Korea where two LCC-HVDC links are responsible for frequency regulation in the island power system \cite{Jejuinertia}.

\subsection{Synchronous Condensers (SYNCON)}
\color{black}
Synchronous condensers can be considered as \emph{old devices to solve new problems} \cite{Rouco_2024} and they are a key technology enabler in islanded power systems. SYNCONs are used for:
\begin{itemize}
    \item Increase of the short-circuit power, to increase the grid strength and short-circuit currents.
    \item Contribute to system inertia.
    \item Voltage control.
\end{itemize}

For example, SYNCONs have been planned in the Canary and Balearic Islands (Spain) \cite{redlectrica_planif_2021, Renedo2024BalearicHVDC}.

Notice that SYNCONs are energy storage systems. However, they are included in a different subsections, for the sake of clarity.

\color{black}

\subsection{Energy Storage Systems (ESS)}

Energy storage systems (ESS) have been touted as a crucial enablers to allow for higher renewable energy penetration in the islands by providing \color{black} technical capabilities in the context of large amounts of  renewable energy sources \cite{Sigrist_2013, ochoa-correa_pathways_2025}. \color{black} In this context, the ESS technologies discussed are battery energy storage system (BESS), pumped-hydro storage (PHS), flywheels, ultracapacitor and hydrogen storage. 

\color{black} The use of ESS as technology enabler for specific applications would depend on the owner of the facility (TSO or other stakeholder) and regulation in each country. For example, according to Article 63 of European Union (EU) Directive 2019/944 \cite{EU2019_944}, ESS facilities owned by TSOs must be defined as fully-integrated network components, which means that they can provide certain technical capabilities related to power system reliability and security, but they cannot be used for balancing or congestion management. On the contrary, ESS owned by other stakeholders could provide technical capabilities related to balancing, for example, such as primary frequency control.  \color{black}

(\textbf{BESS}) have been identified as one of the solutions because of their versatility and fast dynamic response  and their ease of installation \cite{Hoke2021}. \color{black} The BESS is connected to the grid through a power converter, \color{black} \color{black} BESS can be used to supply/consume energy in case of contingencies, to maintain, security of the system, for peak shaving, for frequency support, or to provide grid strength by means of GFM control, among other applications. As discussed previously, the application of the BESS must comply with the regulation of each country, depending on the owner of the facility. The BESSs are treated as fully-integrated network components \color{black}

\color{black} For example, a BESS with GFM control was installed in the Faroe Islands, enabling stable operation in scenarios with 100\% wind power generation~\cite{Trondheim2023}. \color{black} A study conducted in an Island in Taiwan showed the resilience of the BESS system during transient events in a renewable dominated microgrid \cite{BESSisland}. In the Greek island of Astypalaia, a hybrid system consisting of BESS with wind turbine was shown to deliver a quick response to grid \color{black} disturbance \color{black} while in a similar context the Maui island in Hawaii used BESS for peak shaving and load curve smoothing in a distribution grid with high PV penetration \cite{ochoa-correa_pathways_2025}. \color{black} In the Balearic Archipelago (Spain), two BESSs will be installed in Ibiza (90 MW/67.5 MW-h) and Menorca (50 MW/37.5 MW-h) islands. The two BESS are defined as fully-integrated network components \cite{EU2019_944} and they are planned to inject zero active power in normal operation and they will inject active power only in case of tripping of interconnection links between the Iberian Peninsula and the Balearic Archipelago, or in case of the trip of the AC interconnections between the islands of the Balearic power system \cite{Renedo2024BalearicHVDC}. \color{black}

\textbf{Ultracapacitor(UC)} is another storage technology that has been proposed for specific use cases in islanded system. \color{black} The UC is connected to the grid through a power converter. \color{black} Compared to batteries, they have a higher power density and thus have a very fast response to loads. Thus, their use case has been mostly identified in \color{black} conjunction \color{black} with other storage systems such as BESS and fuel cells \cite{Island_book}. \color{black}Applications of UCs in isolated power systems are frequency support and inertial response \color{black}. In terms of application in islanded power system, the impact of a UC on the frequency stability of Guadeloupe Islands has been analyzed for two scenarios of different penetration of solar PV and wind \cite{guadeloupe}.

Similarly, \textbf{Flywheels} \color{black} consist of a rotational mass which is connected to the grid though power converters. Flyweels are used to provide inertial response, by proper control strategies, and it is useful in renewable dominated low inertia islanded power systems \cite{ochoa-correa_pathways_2025}. \color{black} An implementation in El Hierro island showed that combining flywheels with wind and PHS reduced frequency fluctuations and improved power supply quality \cite{sarasua_hybrid_2022}. Flywheels have been shown to be a valuable technology when used in combination with other storage facilities like BESS, providing both short and long term energy storage \cite{ochoa-correa_pathways_2025}. In the study conducted in Jeju island \cite{Jejuinertia} it was shown that flywheel-connected synchronous condenser can ensure frequency stability in a system which is almost saturated with penetration of RES.

(\textbf{Pumped hydro power plants (PHS)}) is another energy storage technology which has been shown to be effective for higher adoption of RES in islands while maintaining the system stablity and reliability. One such example is the island of El Hierro in the Canary islands, where a wind-PHS hybrid system allowed for an operation with even 100\% RES on specific days \cite{ochoa-correa_pathways_2025}. This was made possible by the adoption of sophisticated frequency control which is paramount for a low inertia grid \cite{elhierrocontrol}. Similarly, on Crete island in Greece, PHS system stored excess wind energy by converting it to hydraulic energy, maintaining a reliable energy reserve during periods of low renewable generation while integrating PHS with large-scale wind energy production \cite{PHScrete} and on the island of Lesbos, it was shown that PHS can reduce renewable energy curtailment and replace energy imports from the mainland \cite{PHSlesbos}.

\textbf{Hydrogen} storage is especially mentioned for its ability to provide long term energy storage and has been shown to complement other storage technologies such as BESS, offering a more economical and scalable long term storage \cite{ochoa-correa_pathways_2025}. An example is the Yong Shu Island in the South China Sea, where hydrogen storage integrated with BESS significantly improved the system efficiency and reduced carbon emission \cite{ye_towards_2017}. Similarly, in a study conducted in Aegean island it was demonstrated how fuel cells and hydrogen storage can effectively support large-scale wind energy integration \cite{hydrogenaegean}. 

Different ESS technologies have their own characteristics and suitability. The BESS and PHS have a high energy density making them suitable for long duration storage while UCs and flywheels have a high power density making them suitable for applications where fast response is required but for shorter duration. They also have different efficiencies and lifespans. Thus, previous work done on this area not only focus on the technology point of view but also on optimizing the energy storage based on an optimal sizing as well as using different ESS technologies in combination to achieve the best possible performance \cite{ochoa-correa_pathways_2025,Reviewnoninterconn,Island_book}.

\subsection{GFM as a Technology Enabler for Islanded Power Systems}

GFM control of VSCs \color{black} is a key technology enabler in power systems dominated by non-synchronous generation. GFM power converters behave as a voltage source behind a series impedance and allows to create the grid. They can provide grid strength, inertial response, while maintaing the synchronisation with the rest of the system. GFM power converters could be used also for black-start applications. \color{black} It is because of those capabilities the interest around GFM controlled VSC is high and they are expected to help greatly in the energy transition. While a lot of work has been done in academic environments regarding GFM control of VSCs, including different synchronization strategies and current limitation, it is still a nascent technology in terms of real world application. \color{black} In the context of energy transition, \color{black} Islanded power \color{black} systems \color{black} on the other hand require measures to strengthen their grid, and GFM technology could play a vital role in it while also providing real world testing ground before they are deployed in larger interconnected network. Below are some of the application areas of GFM technology in islanded power system. 

 Battery energy storage systems(\textbf{BESS}) have been promoted as one of the solutions because of their versatility and fast dynamic response  and their ease of installation compared to some other alternatives like pumped hydro storage (PHS) \cite{Hoke2021}. A report by Energy System Integration Group (ESIG) group \cite{ESIG2022GFM} states that GFM implementation in utility scale batteries could provide a low cost solution for increasing grid strength and providing synthetic inertia. Utility scale GFM batteries have already been installed in places like Australia and Scotland to support the grid as their grid goes through energy transition. Paper \cite{cai_evaluation_2024} proposes as framework on the evaluation of GFM capabilities of BESS for different services and scenarios such as voltage regulation, frequency regulation and blackstart. In paper \cite{Hoke2021}, the authors have referred to a pilot experiment, where the effectiveness of GFM BESS for blackstart in isolated power system was shown. In islanded power systems affected by low levels of interconnection and low inertia, it is a more urgent issue to provide necessary support to the grid and GFM batteries could provide an effective solution. Recently, some system operators for islanded power systems have started including BESS system with GFM capability in their future plan. The Hawai'i Public Utilities Commission (PUC) in their integrated grid plan (IGP) explicitly requires GFM BESS to provide GFM functionalities such as virtual inerti, fast-frequency response (FFR), voltage regulation, ride-through and black-start \cite{IGP2024Update}.

\color{black} \textbf{VSC-HVDC} interconnections could also have GFM control \color{black} to support the island power grid. While most of the installed VSC-HVDC links are operated with GFL technology, more recently studies have studied the use case of GFM technology in VSC-HVDC converters. In \cite{derviskadic_opportunities_2023}, the authors present opportunities and challenges for grid-forming implementation in VSC-HVDC system. In recent years, various European TSOs have included the GFM requirement from power-pack modules (PPM) which also includes HVDC systems while the European project InterOPERA has provided in depth details on Grid-forming functional requirements for HVDC converter stations and DC connected power park modules \cite{InterOPERA2024D22}. While these requirements are for a VSC-HVDC links in continental grid, they could be adopted for islanded power system albeit with some modifications taking into account the peculiarity of islanded power systems. In case of real world application, \color{black} the Spanish TSO \color{black} has planned to build a VSC-HVDC link connecting the Spanish mainland and the Balearic Islands to decarbonize the Balearic islands. The island side VSC-HVDC link is planned to be operated in GFM control mode to provide voltage and inertia support to the weak island grid while increasing the power transmission capacity to the Balearc archipelago \cite{Renedo2024BalearicHVDC}. \color{black} Furthermore, the work in \cite{Collados2022} showed \color{black} that GFM controlled VSC can support the island power system during transients including in instances when the island power system is operated with 100\% converter interfaced resources.

\color{black}
\textbf{FACTS} could also be used in islanded power systems . STATCOMs are a proven transmission grid solution for dynamic reactive power compensation and voltage regulation. They normally operate in grid-following control mode to inject or absorb reactive power as required. But, STATCOMs with GFL control may not be a suitable technology for weak grids because of the inherent limitations of GFL control in such conditions. Enhanced STATCOMs (E-STATCOM) are an interesting application, where certain amount of energy is stored in the DC side (with UCAPs). In this way, the E-STATCOM could have GFM control and provide gird strength, which makes it a very interesting application in islanded power systems with large amounts of non-synchronous generation. Some examples of GFM control in STATCOMs and E-STATCOMs can be found in the literature. In \cite{shen_grid-forming_2025} the capability of a GFM STATCOM in suppressing the commutation failure in LCC-HVDC in weak grid condition is studied. In the paper \cite{halonen_grid_2024}, the evolution of GFM technology in STATCOM is introduced with reference to two implementations in real world scenario. Another study \cite{rodriguez-amenedo_rms_2025} presents a study with GFM control in E-STATCOM in an isolated power system, \color{black}showing that the E-STATCOM improves power system stability, while complying with grid codes. \color{black} 

\textbf{Must-Run units} are synchronous generation that are connected to the system to guarantee power system stability and they are commnoly used in current islanded power systems. This means that, even the operating costs of the system would be lower with other generator dispatched, they need to keep connected. In some cases, these must-run units are based on thermal generation. Precisely, technology enablers described above are aimed to replace/reduce the need of must-run units and allow energy transition.
\color{black}

\section{Stability Challenges Across Real-World Islanded Power Systems}

\subsection{Spanish Non-Peninsular System}

The Spanish non-peninsular \color{black} system includes the power systems \color{black} of the Canary Islands and \color{black} the Balearic Islands. \color{black} The  \textbf{Canary Islands} is an archipelago in the Atlantic ocean consisting of a group of seven islands in the Atlantic ocean, 1400 km off mainland Spain. Thus, the power system of Canary Islands is isolated and there is very limited interconnection between the islands in the archipelago due to geographical reasons. Historically, like most of the islanded power systems, the energy demand in the archipelago has been fulfilled by fossil-fuel imports. However, recently the local government have made a push towards higher renewable energy integration with an ambitious plan to go 100\% renewable by 2040 \cite{Endesa2020CanaryIslands100}. According to the most recent data by Spanish transmission system operator (TSO) Red Eléctrica, renewables accounted for 21.2 \% of total generation in the islands. Thus, the future power system of the Canary Islands could pose new challenges due to a combination of being an isolated weakly interconnected network with high penetration of renewable energy. \color{black} Synchronous condensers have been planned in the Canary island to enable energy transition, in Gran Canaria (Arucas 25 MVA), Tenerife (Drago 25 MVA) and Lanzarote (Jares 25 MVA) \cite{redlectrica_planif_2021}. Besides, a pumped-hydro power plant has been planned in Gran Canaria island (Chira-Soria PHS). 
\color{black}

Similarly, another archipelago in Spain, the \textbf{Balearic Archipelago} consisting of four islands in the Mediterranean sea (Mallorca, Menorca, IBiza and Formentera), presents a different structure of islanded power system but some similar challenges. Historically, its power system has relied on fossil fuel imports as the major source of electricity generation. The Balearic Islands are interconnected with each other with AC cables and to the mainland grid through a \color{black} 2x200 MW \color{black} bipolar LCC-HVDC line since 2012. \color{black} Large amounts of solar PV generation are expected in the Balearic Islands. In this context, Balearic power system faces the same challenges as any other low-inertia power system: frequency control, low inertia, low short circuit power and voltage control. The situation is even more challenging due to the existing LCC-HVDC link, which needs a certain grid stiffness for stable operation. Technology enablers planned for energy transition include \cite{Renedo2024BalearicHVDC}: a new 2x200 MW VSC-HVDC link with GFM control, five SYNCON (5x100 MVA) and two ESSs (90 MW/67.5 MW-h and 50 MW/37.5 MW-h) defined as fully-integrated network components \cite{EU2019_944}. \color{black}

\subsection{Hawaii Archipelago}

The Hawaiian archipelago, comprising eight main islands, is a paradigmatic example of an isolated, multi-grid power system facing the challenges of energy transition, high renewable penetration, and stability under real-world operational constraints~\cite{Lee2020,Island_book,Meschede2022}. Each major island operates its own self-contained grid, with no interconnection to neighboring islands or the mainland United States. Historically, Hawaii has relied on imported petroleum products for over 85\% of its primary energy, resulting in some of the highest electricity prices in the US—often double the national average~\cite{Lee2020,Akiba2024}. This economic vulnerability, coupled with environmental concerns and the imperative of climate change mitigation, has driven Hawaii to become a global leader in renewable energy policy and grid innovation.

Hawaii's energy transition is underpinned by a robust, multi-level policy framework. The Hawaii Clean Energy Initiative (HCEI), launched in 2008, set the state on a path to 100\% renewable electricity by 2045—the first such mandate in the US~\cite{Lee2020,Akiba2024}. This ambitious target is supported by a suite of regulatory measures, including Renewable Portfolio Standards (RPS), energy efficiency mandates, and performance-based utility regulation. The transition is coordinated through a collaborative governance model involving the Hawaii Public Utilities Commission (PUC), the Hawaii State Energy Office, the Hawaii Natural Energy Institute (HNEI), utilities (HECO, MECO, HELCO, KIUC), and a broad coalition of stakeholders~\cite{Lee2020}.

Each island grid is characterized by a unique mix of generation resources, load profiles, and operational challenges. Oahu, the largest grid, has a peak demand of over 1,200 MW, while Maui, Hawaii (Big Island), and Kauai operate smaller, more isolated systems. Historically, generation was dominated by oil-fired steam and diesel plants, with some coal and biomass. Over the past decade, rapid deployment of utility-scale wind, distributed and utility-scale solar PV, and BESS has transformed the generation mix~\cite{HawaiiSolarIntegration,Akiba2024}. By 2023, several islands regularly operate with instantaneous renewable penetration exceeding 60–80\%, and Kauai has achieved periods of 100\% renewable supply with BESS and GFM inverters~\cite{Akiba2024,Meschede2022}.

The transition to high shares of IBRs has fundamentally altered the frequency dynamics of Hawaii's grids. The displacement of synchronous machines by PV and wind reduces system inertia, increasing the ROCOF and the risk of underfrequency events following contingencies~\cite{HawaiiSolarIntegration,Island_book,Akiba2024}. Studies and operational experience show that, under high renewable scenarios, frequency excursions can be more severe and occur more rapidly than in traditional systems. For example, on Oahu, scenarios with >50\% non-synchronous generation have demonstrated frequency nadirs and ROCOF values that challenge existing protection schemes and generator ride-through capabilities~\cite{HawaiiSolarIntegration}.

To address these challenges, Hawaii has pioneered the deployment of BESS with GFM and FFR capabilities. On Maui and Kauai, BESS are used to provide synthetic inertia, primary frequency response, and GFM services, enabling stable operation even during large disturbances~\cite{Akiba2024,Meschede2022}. Figure~\ref{fig:hawaii_freq} illustrates the modeled frequency response of the Maui grid under high renewable penetration, showing the impact of BESS and advanced inverter controls. Maui's thermal units can adjust their output for most periods of high variability of solar and wind resources, but, occasionally, 
the BESS kicks in to keep the frequency range within a $\pm$0.1-Hz band around the nominal 60-Hz frequency.

\begin{figure}[h]
    \centering
    \includegraphics[width=0.8\textwidth]{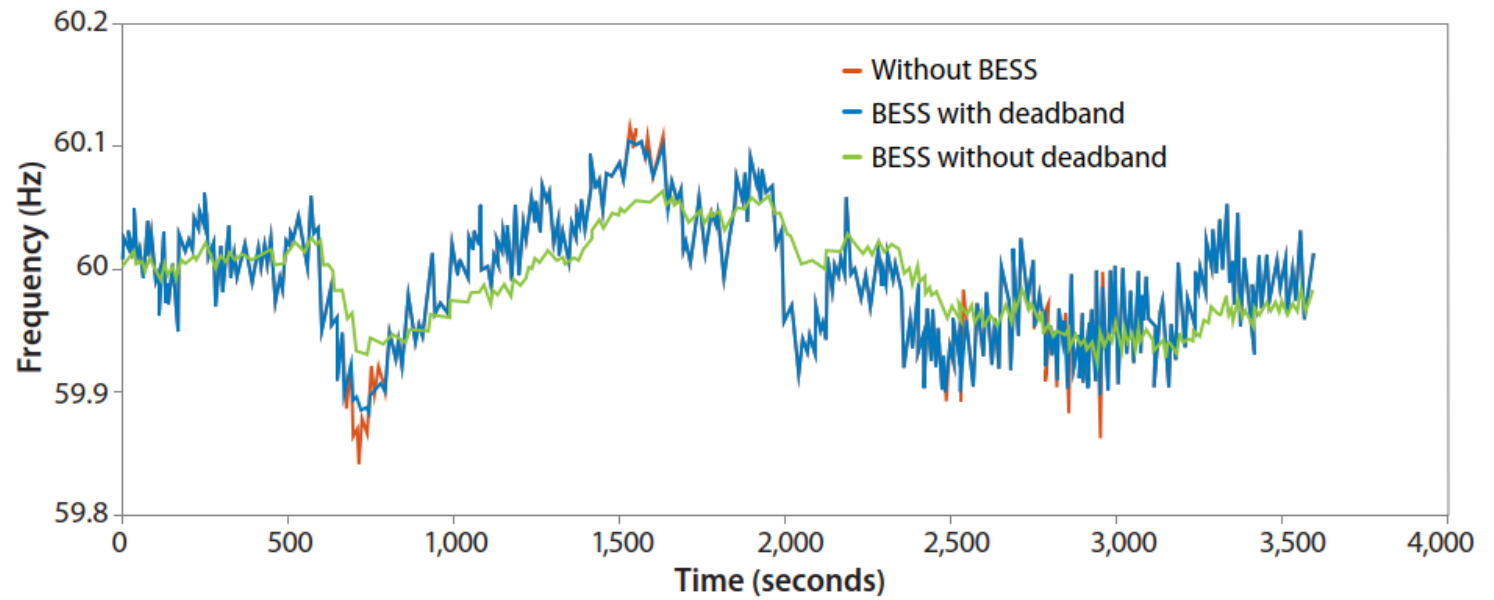}
    \caption{Modeled frequency response in Maui under high renewable penetration, with and without BESS/FFR. Figure from~\cite{HawaiiSolarIntegration}.}
    \label{fig:hawaii_freq}
\end{figure}

 The Hawaii Solar Integration Study found that distributed PV, while less variable than central PV, presents unique challenges due to the inability of the utility to curtail output, leading to periods of overgeneration and voltage excursions~\cite{HawaiiSolarIntegration}.

 Real-world incidents in Hawaii and other high-renewable islands have documented oscillatory events (e.g., 0.1–1 Hz) linked to poorly tuned inverter controls, weak grid conditions, and interactions between multiple IBRs~\cite{SurveyGridIncidents,Akiba2024}.

Hawaii's pathway to 100\% renewable electricity is characterized by a phased approach: (1) initial integration of wind and PV alongside fossil generation, (2) deployment of flexible storage (BESS, pumped hydro, hydrogen), (3) grid modernization and smart inverter adoption, and (4) sector coupling (transport, water, buildings)~\cite{Lee2020,Technologies2025,Meschede2022}. Modeling studies and real-world pilots (e.g., Kauai, Maui) have shown that fully renewable operation is technically feasible and can be economically competitive with fossil-based systems~\cite{Technologies2025,Meschede2022}. Key enablers include:

\begin{itemize}
    \item Large-scale BESS with grid-forming and FFR capabilities.
    \item Advanced inverter controls (volt-var, ride-through, synthetic inertia).
    \item Demand response, flexible loads, and sector coupling.
    \item Robust regulatory and policy frameworks.
    \item Community engagement and multi-stakeholder governance.
\end{itemize}

\subsection{Faroe Island}

Faroe Islands are located in the North Atlantic Ocean and have a population of approximately 55,000. Their power system is fully isolated and is managed by SEV. The power system of Faroe Islands contains several electrically separated areas: the main grid which connects the central islands, and the smaller grid of Suðuroy, and 5 other small grids. The main grid supplies roughly 90\% of the total demand, while the Suðuroy grid supplies about 10\%. Other small grids account for only 0.2\% of the demand and therefore are often neglected in studies. This structure allows Suðuroy to function as a testbed to validate a system with high-penetration of renewables before deployment to the main system. Historically, the energy supply of the archipelago relied heavily on imported fossil fuels.  To minimize both the economic and environmental dependency on fossil imports, SEV has set an ambitious target of achieving 100\% renewable electricity generation by 2030 \cite{katsaprakakisFaroeIslands1002019}. Recently, demand ranges between 22 MW and 60 MW \cite{trondheim100SustainableElectricity2021}. Operational data further indicate a generation mix with an approximately 50/50 split between thermal plants and renewables (primarily hydropower and wind). 

As the Faroe Islands transition to higher renewable penetration, the primary stability challenges arise from the isolated grid and the characteristics of IBRs. With the reduction in synchronous generation, the grid becomes more sensitive to reactive power fluctuations and requires robust voltage support. The high variability of the North Atlantic wind and the seasonal fluctuation of wind and rain significantly challenge the balancing of the system, particularly during summer when low wind speed and precipitation coincide. 

To manage these challenges while pursuing the 100\% renewable target, multiple has been proposed. Most studies utilize the Suðuroy grid to validate these strategies. The Suðuroy system integrates synchronous condensers to provide essential physical inertia and fault current contribution \cite{trondheim2021frequency, ahmedDynamicGridStability2023}. BESS units have been utilized for Frequency Restoration Control (FRC) \cite{reusFrequencyRestorationControl2024}. PHS has been proven to stabilize the grid frequency and enable a higher wind integration \cite{raufIntegratingRenewablesInsights2025}. 
Advanced control strategies such as MPC are also studied in the simulation work of Suðuroy grid that account for forcast uncertainties, reduce operational costs, and enhance system robustness \cite{alferinkRobustModelPredictive2024}.

\subsection{Madeira}

Madeira archipelago includes 4 islands: Madeira, Porto Santo, Desertas and Selvagens. Desertas and Selvagens are protected nature reserves with no permanent population. Because only Madeira and Porto Santo host utility electricity networks, most power system studies focus on these two islands. The two islands are two electrically isolated systems operated by Empresa de Electricidade da Madeira (EEM) with no interconnection to the mainland of Portugal. In 2022, the peak demand was about 146 MW on Madeira (total installed capacity about 377 MW) and about 7.9 MW on Porto Santo (total installed capacity about 20.6 MW). The generation mix is transitioning toward higher renewable penetration, aiming at 55\% share of renewable generation by 2030; EEM reported renewables at around 31\% in 2022 \cite{eem_ce4eui_isolated_services_2023}. To achieve this 55\% target, the planned renewable expansion will focus on solar and wind. In particular, installed solar PV is expected to expand from about 15 MW to 75 MW, while wind capacity is expected to grow from 45 MW to 70 MW \cite{eem_ce4eui_isolated_services_2023}.

To support higher PV and wind penetration and reduce thermal synchronous generation, Madeira relies on a set of stability resources: BESS, reversible pumped hydro (Calheta III), and synchronous condensers. A 15~MW/10~MWh BESS has already been operational in Madeira and is providing essential stabilizing functionalities, improving frequency and voltage service quality by and delivering fast response and dynamic voltage support \cite{vasconcelosAdvancedControlSolutions2015,miguelEnergyStorageRenewable2017, paivaEffectsBatteryEnergy2024, ntomalisDynamicModelingSimulation2020}. 
Storage at longer timescales is provided by the Calheta~III reversible PHS. Calheta~III is a variable-speed hydroplant that covers roughly 10\% of Madeira's electricity demand \cite{fotopoulouReviewEnergyStorage2024}, capable of providing fast frequency support due to the the converter interface. Finally, synchronous condensers are recognized by EEM as one of the important enablers for renewable integration. In general, this can be achieved by retrofitting existing generators at thermal plants as synchronous condensers \cite{beiresNeedSynchronousInertia2018, eem_ce4eui_isolated_services_2023}. The effectiveness of these stabilizing approaches is illustrated by a disturbance event reported by EEM. On 23 June 2023, the trip of engines at C.T. Caniçal caused a 16-MW generation loss at an instantaneous load of 73 MW. The recorded response shows the BESS providing fast power injection, while hydropower ramped up to compensate for the generation loss, rebalancing the generation and demand, and supporting frequency recovery \cite{eem_ce4eui_isolated_services_2023}.

Beyond these utility-scale solutions, flexibility from transport electrification is also being explored through smart charging and Vehicle-to-Grid (V2G) pilot projects. On Porto Santo, the Smart Fossil Free Island program has tested coordinated EV charging and V2G discharging during peak hours, alongside reusing second-life electric vehicle batteries as stationary storage to support renewable integration  \cite{renault_porto_santo_smart_island_2018}. Similarly, Madeira Island serves as a demonstrator for the Horizon INSULAE project, which deploys smart charging and V2G to provide grid services. This project involves four 10-kW bi-directional chargers and an innovative 50-kW Silicon Carbide (SiC) fast charger with V2G, two 50-kW quick chargers with smart charging (V1G) for comparison. These chargers are integrated with a 100-kW BESS and a unified charging management system to provide frequency support and voltage regulation \cite{ martins_madeira_replication_action_plan_2024}.


\subsection{Case Study on GFM Implementation in Islanded Power System}
\begin{figure}[h]
    \centering
    \includegraphics[width=0.7\linewidth]{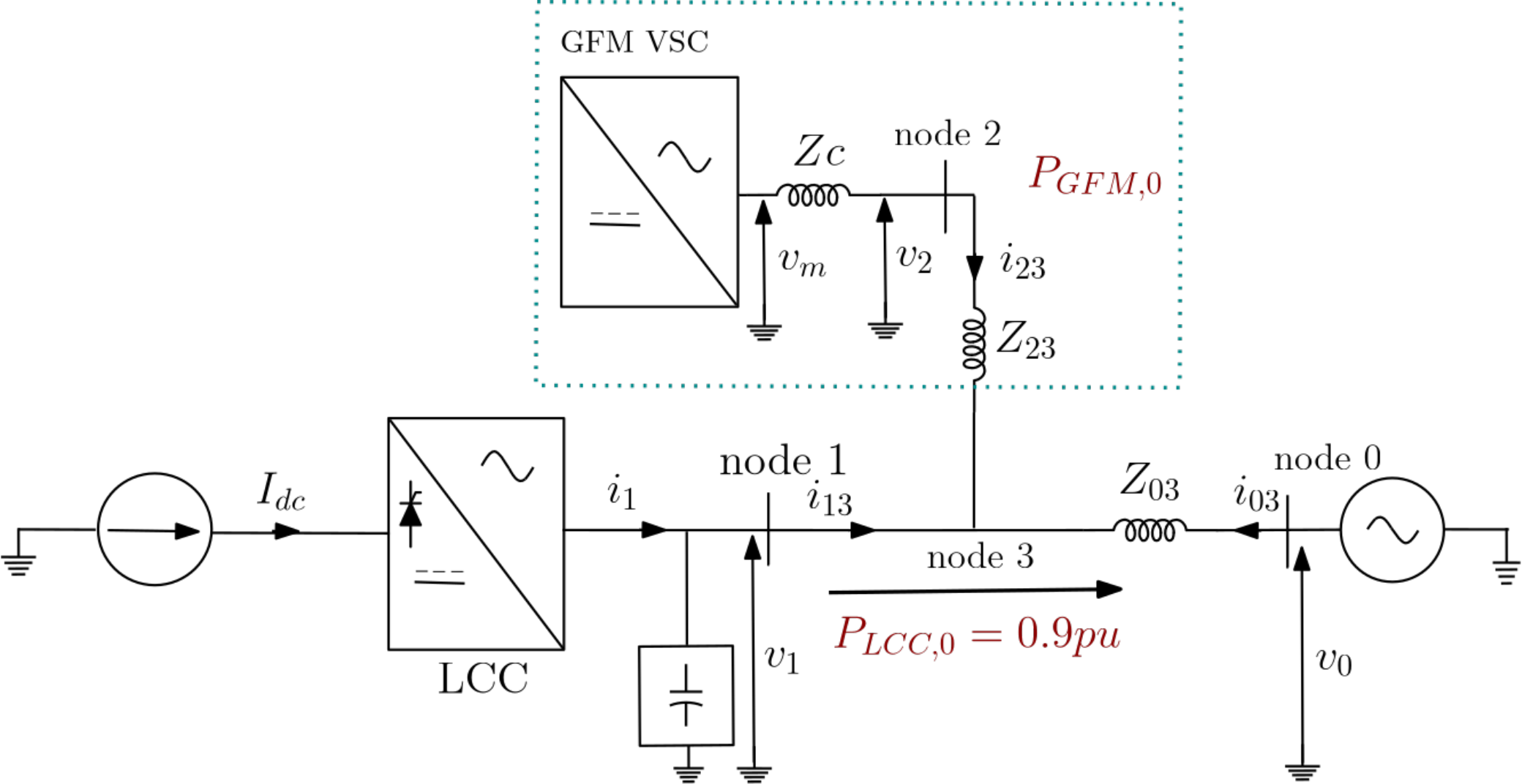}
    \caption{Test case}
    \label{fig:benchmark}
\end{figure}

In this case study, the stability enhancement capability of GFM converter in weak grid conditions, which is characteristic of an islanded power system, is studied. The test system system shown in Fig. \ref{fig:benchmark}. This test case consists of a simplified LCC-HVDC link, a GFM-VSC and a thevenin equivalent representing the island grid. This test case is based on the simplified model of the Balearic island power system. In the study, the stabilizing effect of the GFM-VSC on the LCC-HVDC link will be studied from a small-signal stability perspective. The objective of this example is not to cover all aspects related to islanded power systems, just: an LCC-HVDC connected to a weak grid, the stabilizing effect of GFM power converters and the work just focuses on small-signal stability and interactions. 

\begin{figure}[H]
    \centering
    \includegraphics[width=0.7\linewidth]{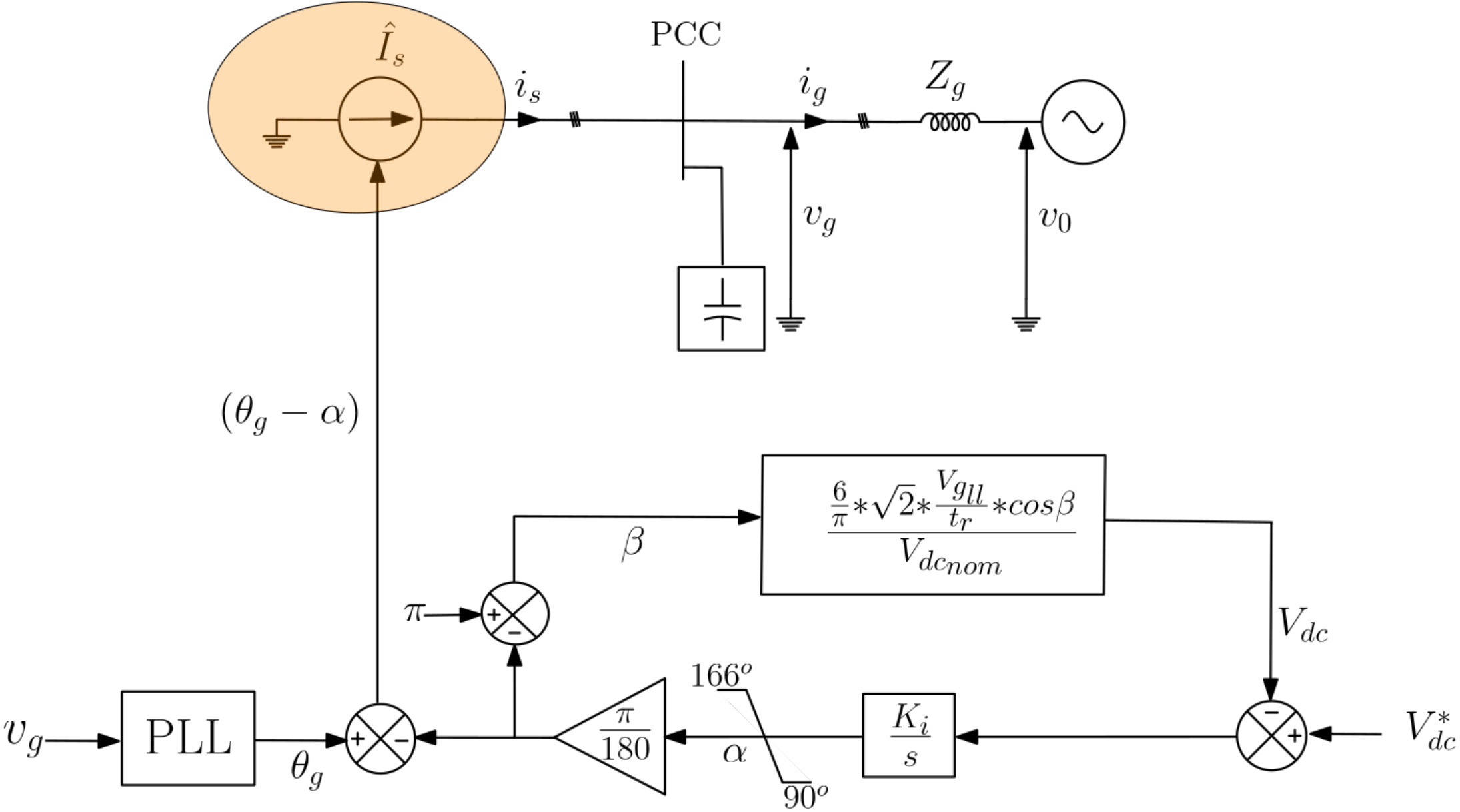}
    \caption{simplified model-nonlinear}
    \label{fig:simp}
\end{figure}

\begin{figure}[h]
     \centering
     \includegraphics[width=0.85\linewidth]{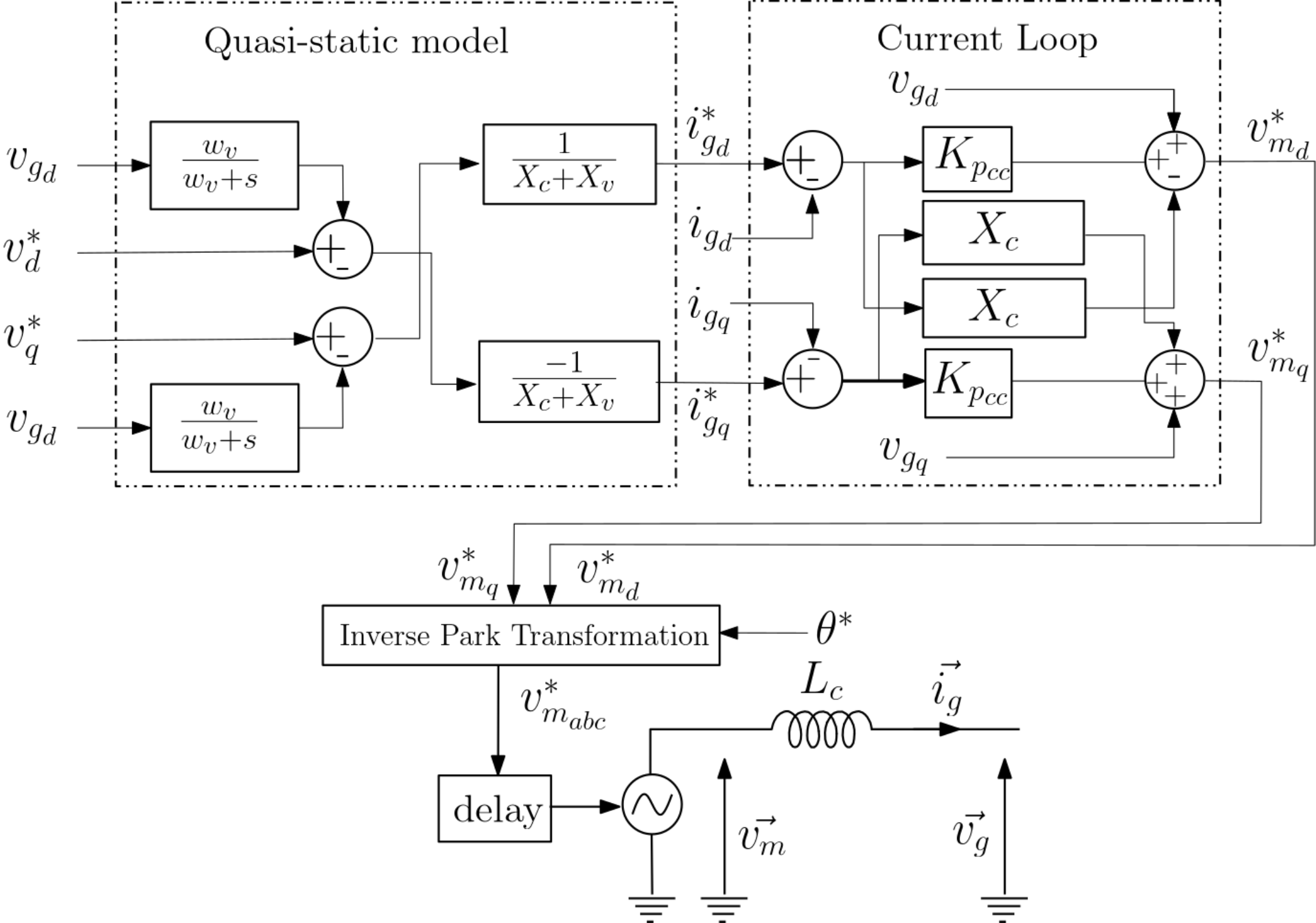}
     \caption{ccGFM configuration}
     \label{fig:CC-GFM}
 \end{figure}

The model of LCC-HVDC and GFM-VSC is provided in Fig. \ref{fig:simp} and Fig \ref{fig:CC-GFM} respectively. In Fig \ref{fig:simp} an averaged model, hereby called the simplified model- nonlinear, of LCC-HVDC considering only the inverter side is shown. 

\begin{figure}[h]
     \centering
     \includegraphics[width=0.6\linewidth]{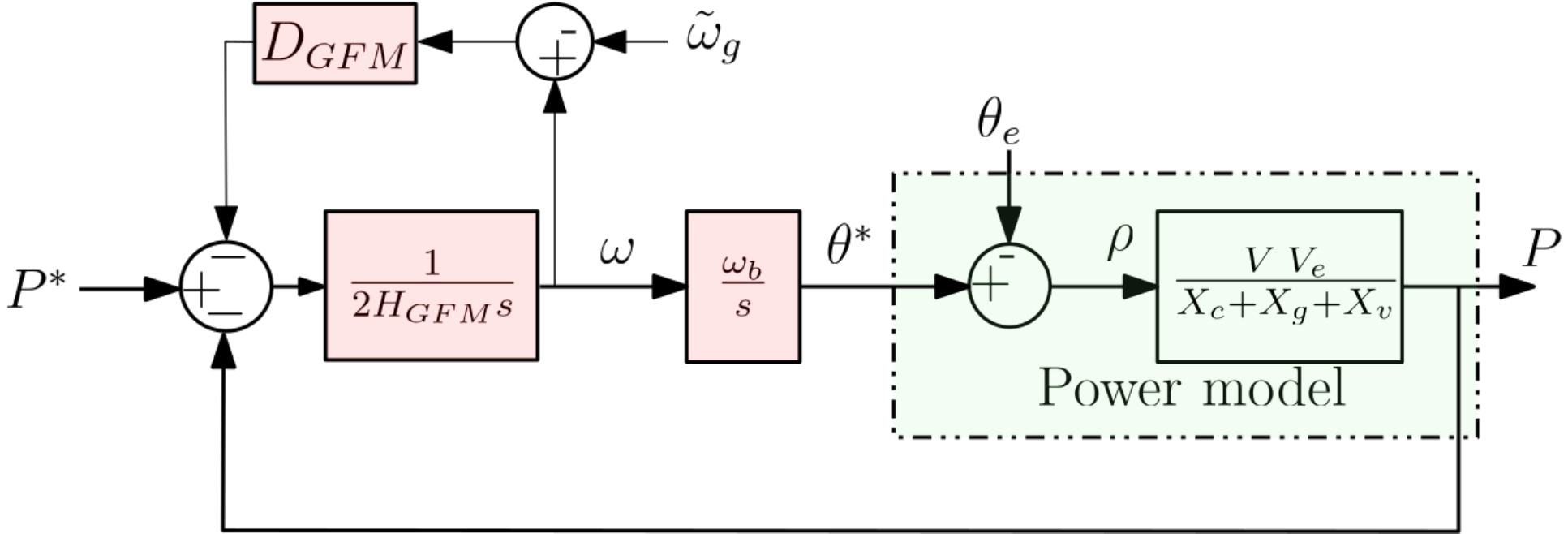}
     \caption{VSM control with PLL}
     \label{fig:VSM}
 \end{figure}

The inverter is represented by a three-phase controlled current source which is modeled to provide the same modulated AC current in fundamental frequency as the actual inverter. A  constant DC voltage control is employed in the inverter. The controller is implemented as a pure integrator. The state-space model of LCC-HVDC for small-signal study is based on this averaged model. The small-signal state-space model of GFM controlled VSC is based on the model presented in \cite{lamrani_stabilizing_2024}. The configuration adopted is the one with current control called the current controlled GFM or ccGFM Ref Fig.\ref{fig:CC-GFM}. and the virtual synchronous machine (VSM) with PLL (Ref. Fig.\ref{fig:VSM}) is used for power synchronization Fig.\ref{fig:CC-GFM}.

The details on system and control parameters of the LCC-HVDC and the GFM-VSC used in this study are provided in the Table \ref{tab:LCC_combined} and Table \ref{tab:combinedGFM} respectively.

\begin{table}[h]
\caption{LCC-HVDC, AC grid, and control parameters}
\begin{center}
\scalebox{0.85}{
\begin{tabular}{lll}
\toprule
\textbf{Model Component} & \textbf{Parameter} & \textbf{Value} \\
\midrule

\multicolumn{3}{l}{\textbf{System Parameters}} \\[3pt]

LCC-HVDC & $V_{dc_{nom}}$ & 500 kV \\
         & $I_{dc_{nom}}$ & 2000 A \\
         & $P_{n,LCC}$ & 1000 MW \\

Converter Transformer (ideal) & $V_{ll,p}$ & 400 kV \\
                              & $V_{ll,s}$ & 220 kV \\

Capacitor Bank & $Q_c$ & 600 Mvar \\

AC Grid & $U_n$ & 400 kV \\
        & $Z_g$ & Variable \\[6pt]

\multicolumn{3}{l}{\textbf{Control Parameters}} \\[3pt]

DC Voltage Control & Response Time & 300 ms \\
                   & $K_i$ & 1000 deg/pu/s \\

PLL & $\omega_n$ & 50 rad/s \\
    & $\zeta$ & 0.707 \\

\bottomrule
\end{tabular}
}
\label{tab:LCC_combined}
\end{center}
\vspace{0.2cm}
\end{table}

\begin{table}[h]
\caption{GFM-VSC system and control parameters}
\begin{center}
\scalebox{0.85}{
\begin{tabular}{lll}
\toprule
\textbf{Model Component} & \textbf{Parameter} & \textbf{Value} \\
\midrule

\multicolumn{3}{l}{\textbf{System Parameters}} \\[3pt]

GFM & $S_{n_{GFM}}$ & 1000 MVA \\
    & $R_c$ / $L_c$ & 0.005 pu / 0.15 pu \\

OHL & $R_{23}$ / $L_{23}$ &0.0072 pu / 0.144 pu \\[6pt]

\multicolumn{3}{l}{\textbf{Control Parameters}} \\[3pt]

VSM Control & $D_{GFM}$ / $H_{GFM}$ & 118.15 pu / 5 s \\
            &  & $(\zeta = 1)$ \\

Virtual Impedance & $X_v$ & 0.3 pu \\

Voltage Filter & $\tau_{f_{CC}}$ & 40 ms \\

Current Controller & $\omega_{CC}$ & 1200 rad/s \\

\bottomrule
\end{tabular}
}
\label{tab:combinedGFM}
\end{center}
\end{table}

\begin{figure}[h!]
     \centering
     \includegraphics[width=0.83\linewidth]{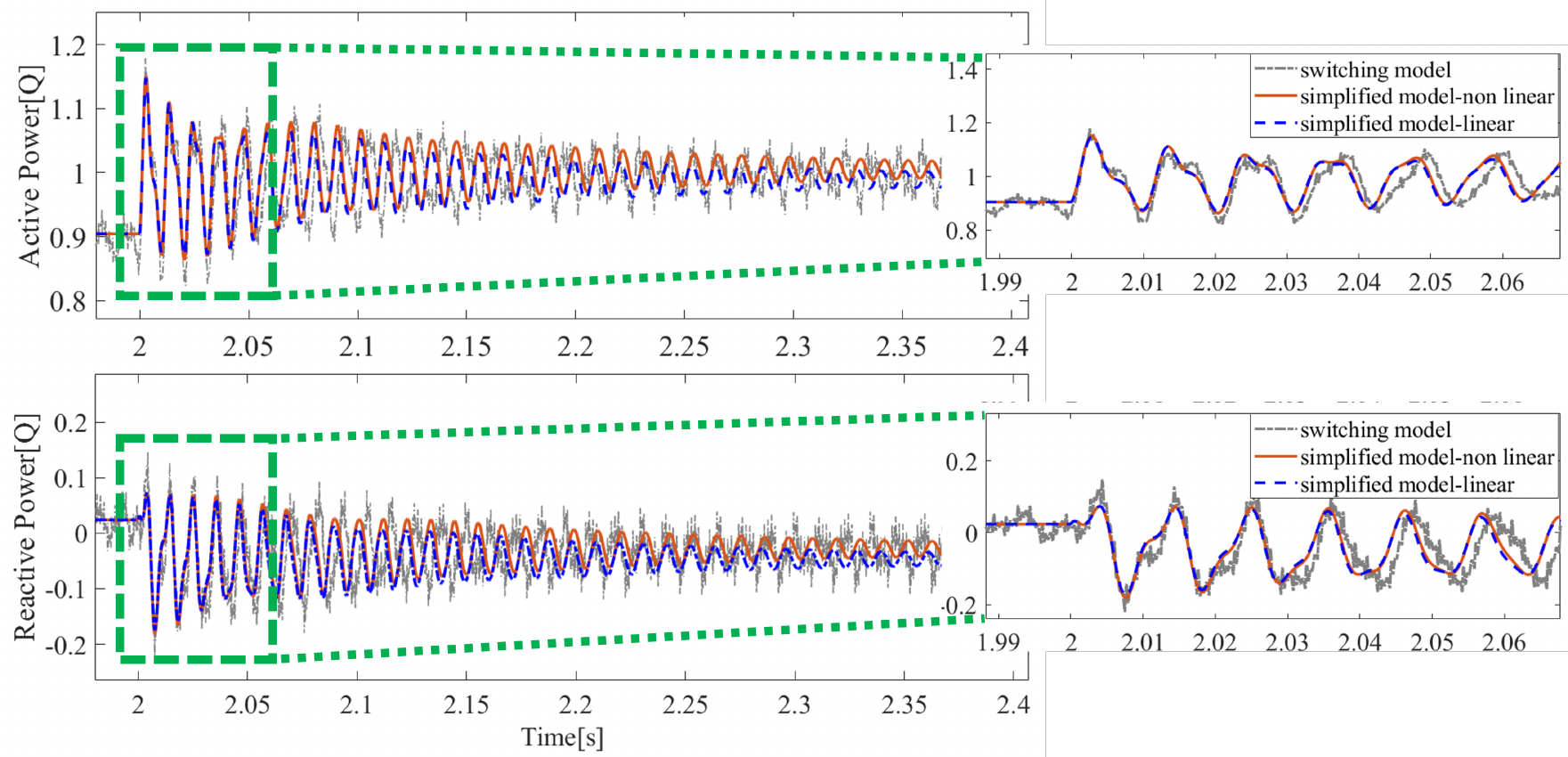}
     \caption{Model Validation: PQ response.}
     \label{fig:LCC_thev_validation}
 \end{figure}

First, the LCC-HVDC link connected to an infinite grid, without GFM-VSC, is considered. The representation of this system is the one in  Fig.\ref{fig:benchmark} without the part bounded by the dotted rectangle.
A small-signal state-space model of this system is developed and validated. The results show a good agreement as shown in Fig. \ref{fig:LCC_thev_validation}.



The sensitivity to grid strength is studied by increasing the grid impedance, $Z_g$. In a thevenin equivalent setup, the grid impedance in per-unit and SCR have an inverse relation. Thus, increasing the grid impedance is equivalent to decreasing the SCR. 

\begin{figure}[h!]
    \centering

\begin{subfigure}[b]{0.9\columnwidth}
    \centering
    \includegraphics[width=\linewidth]{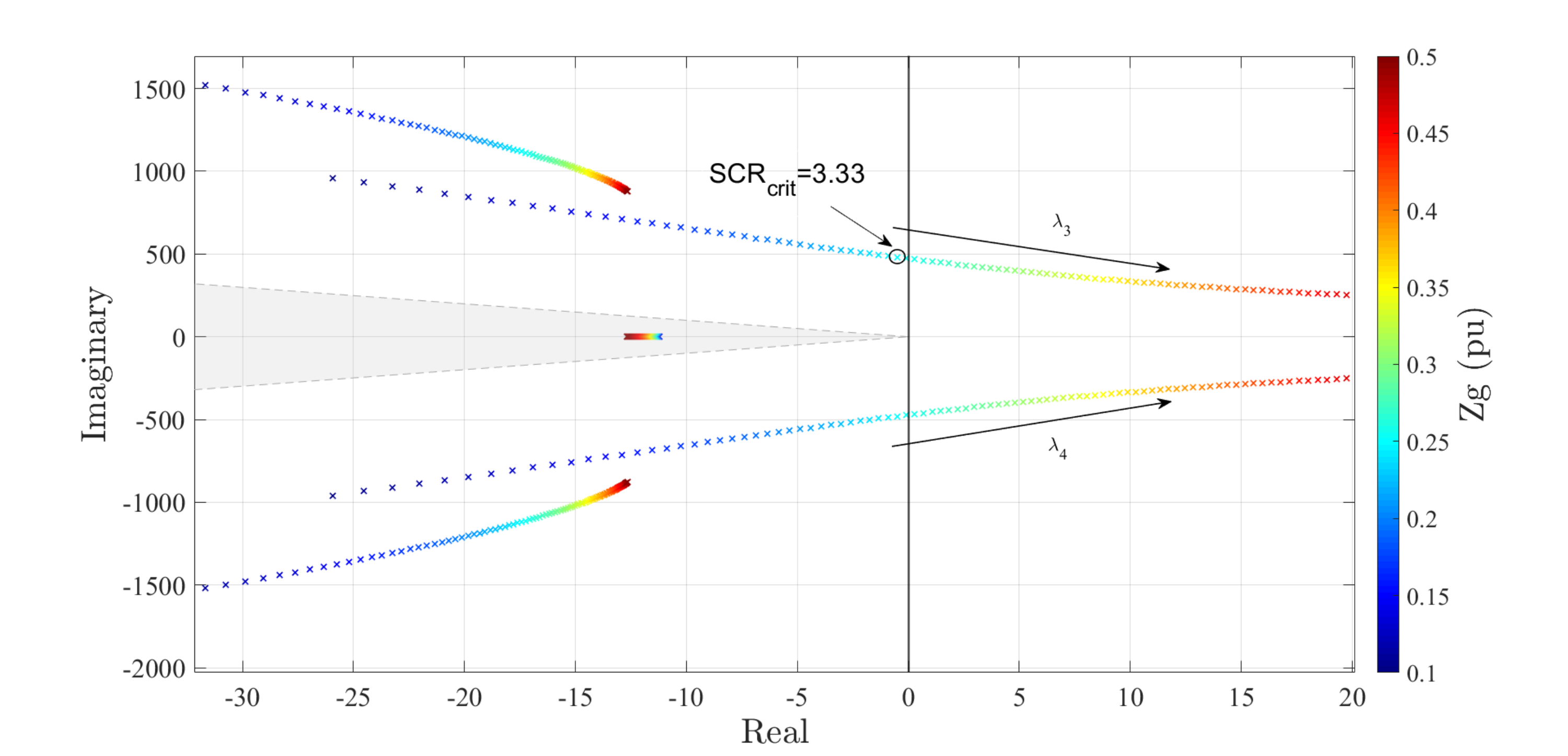}
    \caption{Sensitivity to grid strength}
    \label{fig:Sensitivity_Lg}
\end{subfigure}

\vspace{0.3cm}

\begin{subfigure}[b]{0.75\columnwidth}
    \centering
    \includegraphics[width=\linewidth]{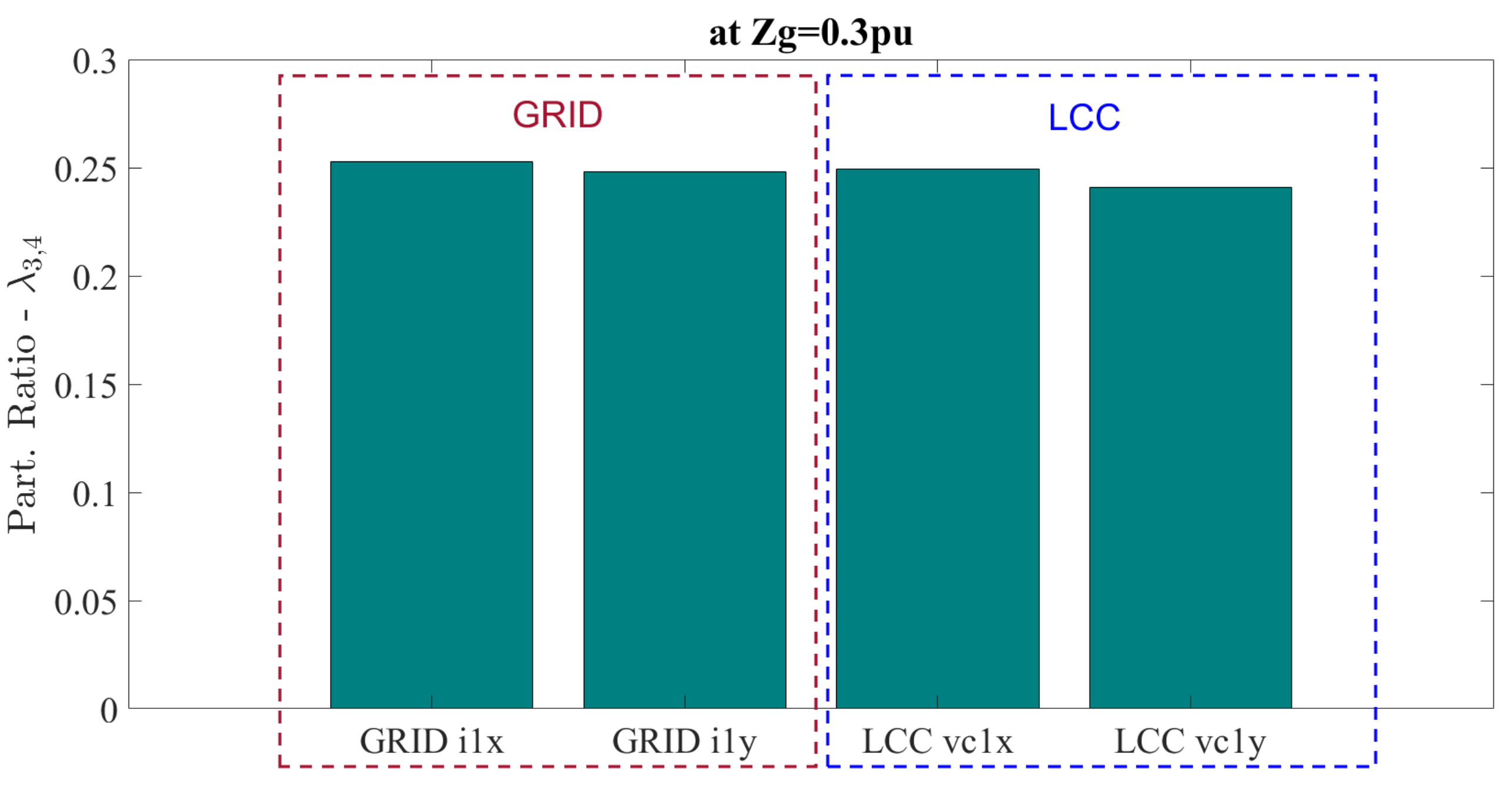}
    \caption{PF Analysis}
    \label{fig:PF_LCC}
\end{subfigure}

    \caption{Small-signal stability study of LCC-HVDC connected to a thevenin equivalent}
    \label{fig:LCC_ss}
\end{figure}

From the sensitivity analysis, the stability limit is found to be at SCR of 3.33 as shown in figure \ref{fig:Sensitivity_Lg}. The eigenvalue pair $\lambda_{3,4}$ is the most critical mode. The participation factor (PF) analysis in figure \ref{fig:PF_LCC} showed the participation of both the LCC and grid states in the dominant mode.

\begin{figure}[h!]
    \centering

    \begin{subfigure}[b]{0.85\columnwidth}
        \centering
        \includegraphics[width=\linewidth]{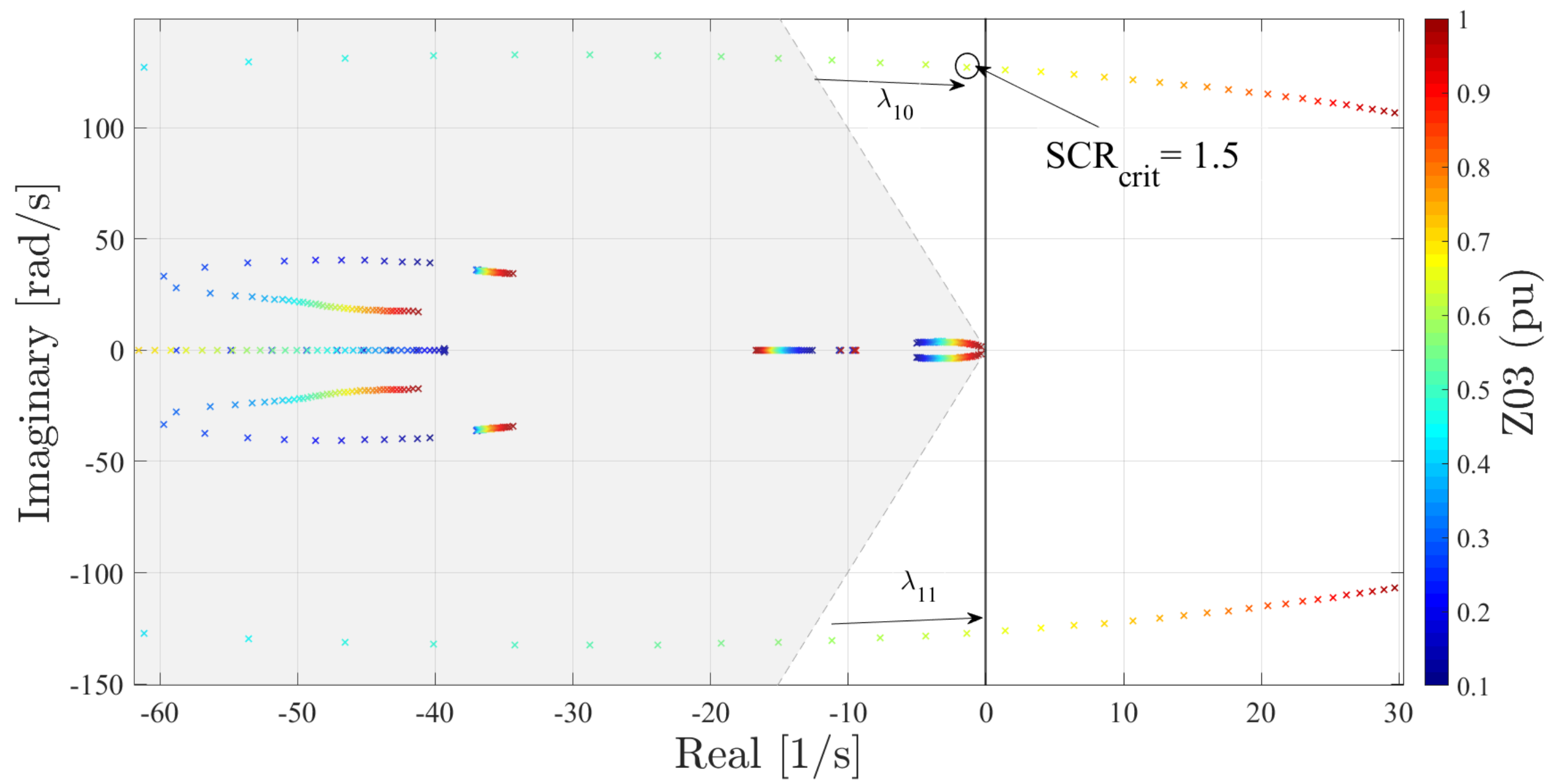}  
         \caption{Sensitivity to grid strength}
        \label{fig:Sensitivity_Z03}
    \end{subfigure}%
    \vspace{0.2cm}
    \begin{subfigure}[b]{0.95\columnwidth}
        \centering
        \includegraphics[width=\linewidth]{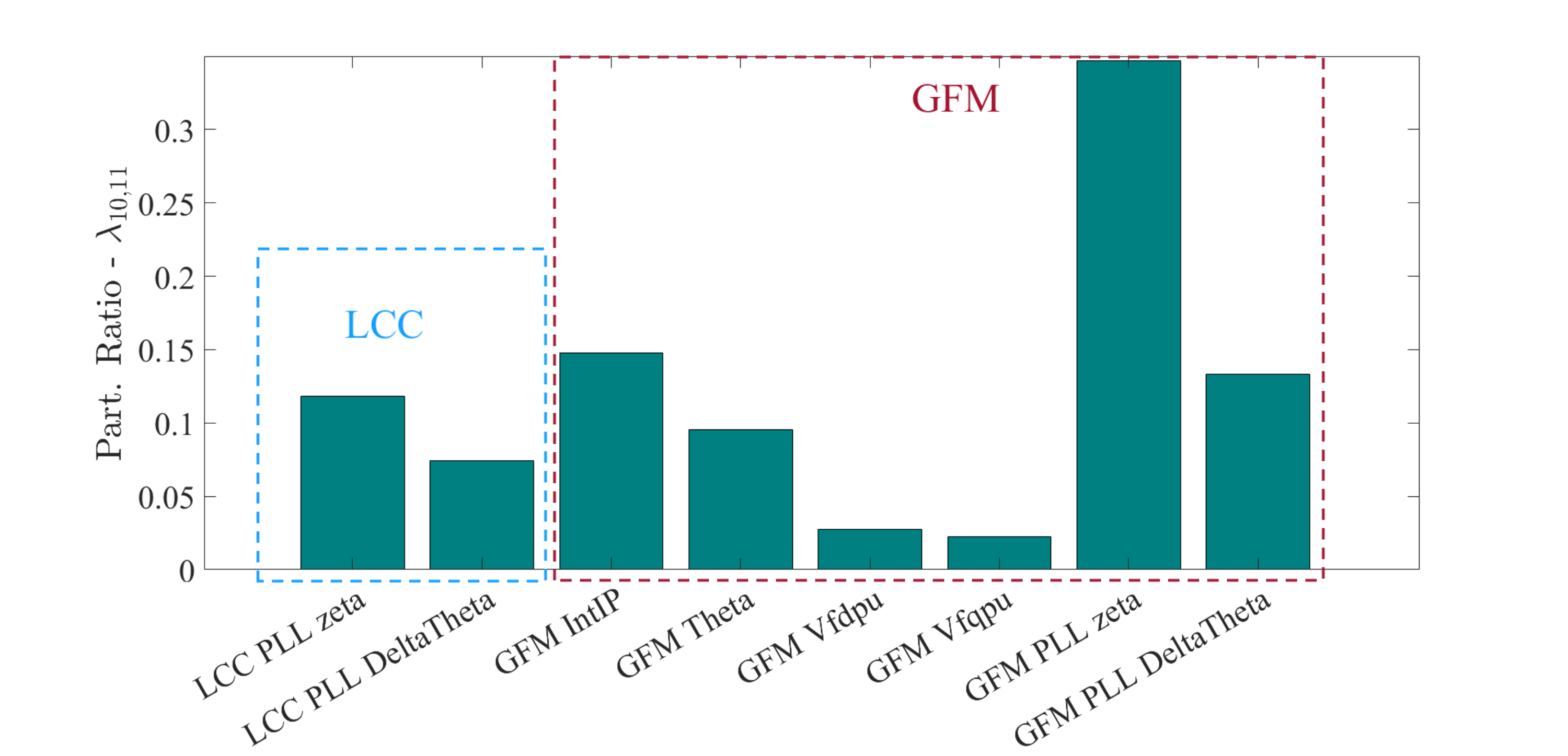} 
        \caption{PF Analysis with GFM-VSC }
        \label{fig:PF_LCC_GFM}
    \end{subfigure}

    \caption{Small-signal stability improvement with GFM-VSC}
    \label{fig:LCC_GFM_ss}
\end{figure}

To study the stabilizing impact of GFM-VSC we again go back to the full setup of the test case in Fig \ref{fig:benchmark}. The power flow is set up so that there is no active power contribution from the GFM-VSC. Thus, the GFM converter \color{black} could represent a BESS or an E-STATCOM, for example. \color{black} Following the same methodology as before, a small-signal state-space model of this system is developed and validated. The sensitivity analysis results to grid strength are shown in Fig. \ref{fig:LCC_GFM_ss}, where the stability limit improved from SCR=3.33 to SCR=1.5 with the addition of GFM VSC.

Additionally, the analysis of participation factor of the critical mode ($\lambda_{10,11}$) showed a much higher participation of the states of GFM VSC. This provides an important conclusion: The contribution
of LCC states in the critical mode decreases significantly with the addition of GFM VSC while the absence of grid's states show the decoupling of LCC-HVDC from weak grid and the dynamics governed by the GFM-VSC's control.




\section{Conclusion and Future Work}

Islanded power systems face unique stability challenges due to their particular characteristics that are intensifying under the energy transition. As the share of inverter-based resources increases and synchronous machines decline, fundamental grid attributes—such as inertia, voltage stiffness, and fault-current support—are diminished. This amplifies vulnerability across all major stability domains: frequency stability becomes critical; voltage excursions become larger in weak grid conditions; converter-driven interactions emerge as dominant modes; and the system’s ability to withstand large disturbances may be reduced.

The review presented in this report shows that conventional stability mechanisms cannot be relied upon in future high CIG penetrated islanded grids. Technology enablers are essential to ensure higher RES penetration in the islanded power system while maintaining the stability and reliablity of the system.  Some of the most relevant technology enablers in the context of islanded power systems with large amounts of renewable energy sources include: 
\begin{itemize}
    \item AC and HVDC interconnections.
    \item Synchronous condensers (SYNCON).
    \item GFM control of power converters.
    \item Energy storage systems (ESS) through power converters (ESS, ultracapacitors, flywheels, pumped hydro power plants, hydrogen). 
    \item FACTS (STATCOMs, E-STATCOMs).
    \item Must-run units
\end{itemize}

The analysis of control strategies and functional capabilities demonstrates that GFM converters are able to provide system supporting and strengthening capabilities to ensure reliable operation of islanded grids as it goes through the energy transition.

However, the transition toward GFM‑centric islanded systems is not without challenges. Operational headroom requirements, current‑limit constraints, protection coordination, interoperability among multiple IBRs, and the absence of standardized grid codes still create uncertainty and limit deployment. Addressing these issues demands coordinated efforts from system operators, manufacturers, and regulators so that GFM technology can effectively tackle the operational challenges of high‑RES islanded systems. Ultimately, while GFM is a key enabler for RES‑dominated IPS, its success will rely on continued advances in modeling, validation, hardware capability, system‑level planning, and regulatory evolution. Pilot projects, standardized performance definitions, and coordinated control frameworks—together with close collaboration among all stakeholders—will be essential to unlock reliable, fully decarbonized island power systems.

In the future, the scope of this work can be extended to include case studies to \color{black} illustrate technical capabilities of GFM power converters, as well as small/large disturbance studies and black-start capability in an islanded grid. \color{black} The test system can also be expanded to include multi converter setup with GFM and GFL controls to study the interactions in the system.

\addcontentsline{toc}{section}{References}
 \printbibliography

@article{lamrani_stabilizing_2024,
	title = {On the stabilizing contribution of different grid‐forming controls to power systems},
	volume = {18},
	issn = {1751-8687, 1751-8695},
	url = {https://ietresearch.onlinelibrary.wiley.com/doi/10.1049/gtd2.13269},
	doi = {10.1049/gtd2.13269},
	number = {23},
	journal = {IET Generation, Transmission \& Distribution},
	author = {Lamrani, Yahya and Colas, Frédéric and Van Cutsem, Thierry and Cardozo, Carmen and Prevost, Thibault and Guillaud, Xavier},
	month = dec,
	year = {2024},
	pages = {3863--3877},
}

@article{cross_benchmarking_2017,
	title = {Benchmarking island power systems: {Results}, challenges, and solutions for long term sustainability},
	volume = {80},
	issn = {13640321},
	url = {https://linkinghub.elsevier.com/retrieve/pii/S1364032117307724},
	doi = {10.1016/j.rser.2017.05.126},
	journal = {Renewable and Sustainable Energy Reviews},
	author = {Cross, Sam and Padfield, David and Ant-Wuorinen, Risto and King, Phillip and Syri, Sanna},
	month = dec,
	year = {2017},
	pages = {1269--1291},
}

@article{ochoa-correa_pathways_2025,
	title = {Pathways to 100\% {Renewable} {Energy} in {Island} {Systems}: {A} {Systematic} {Review} of {Challenges}, {Solutions} {Strategies}, and {Success} {Cases}},
	volume = {13},
	issn = {2227-7080},
	url = {https://www.mdpi.com/2227-7080/13/5/180},
	doi = {10.3390/technologies13050180},
	number = {5},
	journal = {Technologies},
	author = {Ochoa-Correa, Danny and Arévalo, Paul and Martinez, Sergio},
	month = may,
	year = {2025},
	pages = {180},
}

@book{Island_book,
author = {Sigrist, Lukas and Lobato, E. and Echavarren, Francisco and Egido, Ignacio and Rouco, Luis},
year = {2016},
month = {12},
pages = {1-272},
title = {Island Power Systems},
isbn = {9781315368740},
publisher = {CRC Press},
doi = {10.1201/9781315368740}
}

@inproceedings{Renedo2024BalearicHVDC,
  author       = {Javier Renedo and Silvia Sanz Verdugo and Antonio Cord{\'o}n and Bel{\'e}n Segura and David Casta{\~n}eda and Rosal{\'i}­a Rivas and Patricia Labra},
  title        = {New VSC-HVDC Interconnection Between the Iberian Peninsula and Balearic Archipelago to Enable Energy Transition},
  booktitle    = {CIGR{\'E} Session 2024 – Paris, France – B4 Working Group: DC Systems and Power Electronics},
  year         = {2024},
  pages        = {--},  
  publisher    = {CIGRÉ},
  doi          = {10.5809/CIGRE.B4.10623.2024},  
}

@inproceedings{shen_grid-forming_2025,
	address = {Yantai, China},
	title = {Grid-{Forming} {STATCOM} for {Commutation} {Failure} {Mitigation} in {LCC}-{HVDC} {System}},
	isbn = {979-8-3315-2403-6},
	url = {https://ieeexplore.ieee.org/document/11148988/},
	doi = {10.1109/ICIEA65512.2025.11148988},
	booktitle = {2025 {IEEE} 20th {Conference} on {Industrial} {Electronics} and {Applications} ({ICIEA})},
	publisher = {IEEE},
	author = {Shen, Xiaolin and Du, Shangan and Liu, Yiqi and Wu, Yucheng and Yu, Xiaohan and Gu, Fangning},
	month = aug,
	year = {2025},
	pages = {1--5},
}

@inproceedings{milano_foundations_2018,
	address = {Dublin, Ireland},
	title = {Foundations and {Challenges} of {Low}-{Inertia} {Systems} ({Invited} {Paper})},
	isbn = {978-1-910963-10-4},
	url = {https://ieeexplore.ieee.org/document/8450880/},
	doi = {10.23919/PSCC.2018.8450880},
	booktitle = {2018 {Power} {Systems} {Computation} {Conference} ({PSCC})},
	publisher = {IEEE},
	author = {Milano, Federico and Dörfler, Florian and Hug, Gabriela and Hill, David J. and Verbič, Gregor},
	month = jun,
	year = {2018},
	pages = {1--25},
}

@article{rodriguez-amenedo_rms_2025,
	title = {{RMS} {Modeling} and {Control} of a {Grid}-{Forming} {E}-{STATCOM} for {Power} {System} {Stability} in {Isolated} {Grids}},
	volume = {15},
	issn = {2076-3417},
	url = {https://www.mdpi.com/2076-3417/15/6/3014},
	doi = {10.3390/app15063014},
	number = {6},
	journal = {Applied Sciences},
	author = {Rodriguez-Amenedo, José Luis and Montilla-DJesus, Miguel E. and Arnaltes, Santiago and Arredondo, Francisco},
	month = mar,
	year = {2025},
	pages = {3014},
}

@inproceedings{halonen_grid_2024,
	address = {Seattle, WA, USA},
	title = {Grid {Forming} – an {Evolution} in {STATCOM} {Technology} for {Today}’s {Power} {Grids}},
	isbn = {979-8-3503-8183-2},
	url = {https://ieeexplore.ieee.org/document/10688873/},
	doi = {10.1109/PESGM51994.2024.10688873},
	booktitle = {2024 {IEEE} {Power} \& {Energy} {Society} {General} {Meeting} ({PESGM})},
	publisher = {IEEE},
	author = {Halonen, Mikael and Eriksson, Emil and Urrutia, Leonardo Alvarez and Boström, Anders and Majji, Vasista},
	month = jul,
	year = {2024},
	pages = {1--5},
}

@inproceedings{derviskadic_opportunities_2023,
	address = {Rome, Italy},
	title = {Opportunities and {Challenges} in the {Mediterranean} {Region} with {Grid} {Forming} {Control} for {HVDC} {Systems}},
	isbn = {979-8-3503-2489-1},
	url = {https://ieeexplore.ieee.org/document/10179078/},
	doi = {10.1109/AEITHVDC58550.2023.10179078},
	booktitle = {2023 {AEIT} {HVDC} {International} {Conference} ({AEIT} {HVDC})},
	publisher = {IEEE},
	author = {Derviskadic, Arman and Persico, Alberto and Isabegovic, Emir and Thorslund, Anton},
	month = may,
	year = {2023},
	pages = {1--6},
}

@inproceedings{cai_evaluation_2024,
	address = {Chengdu, China},
	title = {Evaluation for {Grid}-{Forming} {Capability} of {Battery} {Energy} {Storage} {Systems}},
	copyright = {https://doi.org/10.15223/policy-029},
	isbn = {979-8-3315-3086-0},
	url = {https://ieeexplore.ieee.org/document/10949993/},
	doi = {10.1109/ICPEE64457.2024.10949993},
	booktitle = {2024 8th {International} {Conference} on {Power} and {Energy} {Engineering} ({ICPEE})},
	publisher = {IEEE},
	author = {Cai, Yongxiang and Wang, Yang and Zhang, Song and Shi, Shaoxin and Peng, Qiao},
	month = dec,
	year = {2024},
	pages = {222--226},
}

@article{SigristBalearic,
author = {Lobato, E. and Sigrist, Lukas and Rouco, Luis},
year = {2017},
month = {10},
pages = {192-200},
title = {Value of electric interconnection links in remote island power systems: The Spanish Canary and Balearic archipelago cases},
volume = {91},
journal = {International Journal of Electrical Power \& Energy Systems},
doi = {10.1016/j.ijepes.2017.03.014}
}

@ARTICLE{guadeloupe,
  author={Delille, Gauthier and Francois, Bruno and Malarange, Gilles},
  journal={IEEE Transactions on Sustainable Energy}, 
  title={Dynamic Frequency Control Support by Energy Storage to Reduce the Impact of Wind and Solar Generation on Isolated Power System's Inertia}, 
  year={2012},
  volume={3},
  number={4},
  pages={931-939},
  doi={10.1109/TSTE.2012.2205025}}

@article{Reviewnoninterconn,
author = {Fotopoulou, Maria and Pediaditis, Panagiotis and Skopetou, Niki and Rakopoulos, Dimitrios and Christopoulos, Sotirios and Kartalidis, Avraam},
year = {2024},
month = {02},
pages = {1572},
title = {A Review of the Energy Storage Systems of Non-Interconnected European Islands},
volume = {16},
journal = {Sustainability},
doi = {10.3390/su16041572}
}

@article{ye_towards_2017,
	title = {Towards a 90\% renewable energy future: {A} case study of an island in the {South} {China} {Sea}},
	volume = {142},
	issn = {01968904},
	shorttitle = {Towards a 90\% renewable energy future},
	url = {https://linkinghub.elsevier.com/retrieve/pii/S0196890417302522},
	doi = {10.1016/j.enconman.2017.03.038},
	language = {en},
	urldate = {2026-01-08},
	journal = {Energy Conversion and Management},
	author = {Ye, Bin and Zhang, Kai and Jiang, JingJing and Miao, Lixin and Li, Ji},
	month = jun,
	year = {2017},
	pages = {28--41},
}

@inproceedings{hydrogenaegean,
author = {Kaldellis, John and Kavadias, Kosmas and Zafirakis, Dimitrios and Andreadis, Dimitrios and Alyssandrakis, A},
year = {2013},
month = {10},
pages = {1-14},
title = {The Role of Hydrogen-Based Energy Storage in the Support of Large-Scale Wind Energy Integration in Island Grids},
volume = {34},
journal = {International Journal of Sustainable Energy},
doi = {10.1080/14786451.2013.846342}
}

@article{sarasua_hybrid_2022,
	title = {Hybrid frequency control strategies based on hydro‐power, wind, and energy storage systems: {Application} to 100\% renewable scenarios},
	volume = {16},
	issn = {1752-1416, 1752-1424},
	shorttitle = {Hybrid frequency control strategies based on hydro‐power, wind, and energy storage systems},
	url = {https://ietresearch.onlinelibrary.wiley.com/doi/10.1049/rpg2.12326},
	doi = {10.1049/rpg2.12326},
	language = {en},
	number = {6},
	urldate = {2026-01-08},
	journal = {IET Renewable Power Generation},
	author = {Sarasua, José Ignacio and Martínez‐Lucas, Guillermo and García‐Pereira, Hilel and Navarro‐Soriano, Gustavo and Molina‐García, Ángel and Fernández‐Guillamón, Ana},
	month = apr,
	year = {2022},
	pages = {1107--1120}
}

@article{elhierrocontrol,
author = {Sarasúa, José and Martinez de Lucas, Guillermo and Lafoz, M.},
year = {2019},
month = {06},
pages = {817},
title = {Analysis of alternative frequency control schemes for increasing renewable energy penetration in El Hierro Island power system},
volume = {113},
journal = {Electric Power Systems Research},
doi = {10.1016/j.ijepes.2019.06.008}
}

@article{PHSlesbos,
author = {Kapsali, M. and Anagnostopoulos, John},
year = {2017},
month = {07},
pages = {},
title = {Investigating the role of local Pumped-hydro energy storage in interconnected island grids with high wind power generation},
volume = {114},
journal = {Renewable Energy},
doi = {10.1016/j.renene.2017.07.014}
}

@article{PHScrete,
author = {Nikolaou, T. and Stavrakakis, George and Tsamoudalis, Konstantinos},
year = {2020},
month = {05},
pages = {2705},
title = {Modeling and Optimal Dimensioning of a Pumped Hydro Energy Storage System for the Exploitation of the Rejected Wind Energy in the Non-Interconnected Electrical Power System of the Crete Island, Greece},
volume = {13},
journal = {Energies},
doi = {10.3390/en13112705}
}

@ARTICLE{BESSisland,
  author={Ku, Te-Tien and Li, Chung-Sheng},
  journal={IEEE Transactions on Industry Applications}, 
  title={Implementation of Battery Energy Storage System for an Island Microgrid With High PV Penetration}, 
  year={2021},
  volume={57},
  number={4},
  pages={3416-3424},
  doi={10.1109/TIA.2021.3075655}}

@article{BalearicAC,
author = {Blanco, David and Berna, Cesar and Rivera, Yago and Muñoz-Cobo, José},
year = {2025},
month = {12},
pages = {118866},
title = {Decarbonizing energy end-uses in semi-isolated islands via renewable and nuclear integration: The Balearic Islands by 2040},
volume = {139},
journal = {Journal of Energy Storage},
doi = {10.1016/j.est.2025.118866}
}

@article{AzoresAC,
author = {Alves, M. and Segurado, Raquel and Costa, Mário},
year = {2019},
month = {06},
pages = {},
title = {Increasing the penetration of renewable energy sources in isolated islands through the interconnection of their power systems. The case of Pico and Faial islands, Azores},
volume = {182},
journal = {Energy},
doi = {10.1016/j.energy.2019.06.081}
}

@article{GreekAC,
author = {Georgiou, Paraskevas and Mavrotas, G. and Diakoulaki, Danae},
year = {2011},
month = {08},
pages = {2607-2620},
title = {The effect of islands' interconnection to the mainland system on the development of renewable energy sources in the Greek power sector},
volume = {15},
journal = {Renewable and Sustainable Energy Reviews},
doi = {10.1016/j.rser.2011.03.007}
}

@article{Jejuinertia,
author = {Jung, Jaeyeop and Park, Jeonghoo and Lee, Hwanik and Lee, Byongjun},
year = {2023},
month = {10},
pages = {1374-1383},
title = {Evaluation of inertia resource for securing nadir frequency in HVDC interconnected system with high penetration of RES},
volume = {9},
journal = {Energy Reports},
doi = {10.1016/j.egyr.2023.05.192}
}

@article{oscillationsmicrogrid,
author = {Saridaki, Georgia and Paspatis, Alexandros and Kotsampopoulos, Panos},
year = {2023},
month = {10},
pages = {109610},
title = {An investigation of factors affecting Fast-Interaction Converter-driven stability in Microgrids},
volume = {223},
journal = {Electric Power Systems Research},
doi = {10.1016/j.epsr.2023.109610}
}

@techreport{Romulo,
  author      = {{Red Eléctrica de España}},
  title       = {Rómulo project, the Spanish peninsula-Balearic Islands electricity interconnection},
  institution = {Red Eléctrica de España},
  year        = {2022},
  url         = {https://www.ree.es/sites/default/files/downloadable/romuloprojecteng.pdf},
}

@misc{EU2019_944,
  author       = {{European Parliament and Council of the European Union}},
  title        = {Directive (EU) 2019/944 of the European Parliament and of the Council of 5 June 2019 on common rules for the internal market for electricity},
  year         = {2019},
  note         = {Official Journal of the European Union, L 158, pp. 125--199, 14 June 2019},
}

@techreport{ESIG2022GFM,
  author      = {{Energy Systems Integration Group (ESIG)}},
  title       = {Grid-Forming Technology in Energy Systems Integration},
  institution = {ESIG},
  year        = {2022},
  url         = {https://www.esig.energy/wp-content/uploads/2024/10/ESIG-GFM-report-2022.pdf},
}

@techreport{EASE2022Islands,
  author      = {{European Association for Storage of Energy (EASE)}},
  title       = {Study on Power System Challenges of Islands and Isolated Systems with High Shares of Variable Renewables},
  institution = {EASE},
  address     = {Brussels, Belgium},
  year        = {2020},
  url         = {https://energystorageeurope.eu/wp-content/uploads/2020/04/EASE-Study-on-Power-System-Challenges-of-Islands-and-Isolated-Systems.pdf},
}

@techreport{InterOPERA2024D22,
  author      = {{InterOPERA Consortium}},
  title       = {Grid-forming functional requirements for HVDC converter stations and DCconnected power park modules (PPMs) in multi-terminal multivendor HVDC systems},
  institution = {InterOPERA, Horizon Europe Project},
  year        = {2024},
  url         = {https://interopera.eu/wp-content/uploads/files/deliverables/InterOPERA-D2.2-GFM-functional-requirements-202401.pdf},
}

@techreport{IGP2024Update,
  author      = {{Hawaiian Electric}},
  title       = {Integrated Grid Plan: Action Plan Annual Update},
  institution = {Hawaiian Electric},
  address     = {Honolulu, Hawaii, USA},
  year        = {2024},
  month       = july,
  url         = {https://hawaiipowered.com/igpreport/IGP_Action_Plan_Annual_Update_2024-06-24.pdf},
}

@online{Endesa2020CanaryIslands100,
  author       = {Endesa},
  title        = {A Plan for the Canary Islands to be 100\% Renewable by 2040},
  year         = {2019},
  url          = {https://www.endesa.com/en/press/press-room/news/energy-transition/a-plan-for-the-canary-islands-to-be-100-renewable-by-2040},
  note         = {Accessed: 2025-11-23}
}

@article{paperOzanErdinc,
  author    = {Ozan Erdinc and Nikolaos G. Paterakis and Jo{\~a}o P. S. Catal{\~a}o},
  title     = {Overview of insular power systems under increasing penetration of renewable energy sources: Opportunities and challenges},
  journal   = {Renewable and Sustainable Energy Reviews},
  volume    = {52},
  pages     = {333--346},
  year      = {2015},
  doi       = {10.1016/j.rser.2015.07.104},
  url       = {https://doi.org/10.1016/j.rser.2015.07.104}
}

@article{Hatziargyriou2021,
  author = {Hatziargyriou, N. and Milanovic, J. and Rahmann, C. and et al.},
  title = {Definition and classification of power system stability -- revisited \& extended},
  journal = {IEEE Transactions on Power Systems},
  volume = {36},
  number = {4},
  pages = {3271--3281},
  year = {2021}
}

@article{Meschede2022,
  author = {Meschede, H. and Bertheau, P. and Khalili, S. and Breyer, C.},
  title = {A review of 100\% renewable energy scenarios on islands},
  journal = {WIREs Energy and Environment},
  volume = {11},
  number = {6},
  pages = {e450},
  year = {2022}
}

@article{Hoke2021,
  author = {Hoke, A. and Gevorgian, V. and Shah, S. and Koralewicz, P. and Kenyon, R. and Kroposki, B.},
  title = {Island Power Systems With High Levels of Inverter-Based Resources: Stability and Reliability Challenges},
  journal = {IEEE Electrification Magazine},
  volume = {9},
  number = {1},
  pages = {74--91},
  year = {2021}
}

@article{Trondheim2023,
  author = {Trondheim, H. M. and Nielsen, T.},
  title = {Experiences with 100\% wind power generation in an isolated power system -- Suðuroy, Faroe Islands},
  journal = {CIGRE Science \& Engineering},
  volume = {30},
  pages = {1--13},
  year = {2023}
}

@article{Collados2022,
  author = {Collados-Rodriguez, C. and Cheah-Mane, M. and Cifuentes-Garcia, F. and et al.},
  title = {Integration of an MMC-HVDC Link to the Existing LCC-HVDC Link in Balearic Islands Based on Grid-Following and Grid-Forming Operation},
  journal = {IEEE Transactions on Power Delivery},
  volume = {37},
  number = {6},
  pages = {5278--5288},
  year = {2022}
}

@article{paperLasanthaMeegahapola,
  author    = {Lasantha Meegahapola and Alfeu Sguarezi and Jack Stanley Bryant and Mingchen Gu and Eliomar R. Conde D. and Rafael B. A. Cunha},
  title     = {Power System Stability with Power-Electronic Converter Interfaced Renewable Power Generation: Present Issues and Future Trends},
  journal   = {Energies},
  volume    = {13},
  number    = {13},
  pages     = {3441},
  year      = {2020},
  doi       = {10.3390/en13133441},
  url       = {https://doi.org/10.3390/en13133441}
}

@article{Fan2023,
  author = {Fan, L. and Miao, Z. and Piper, D. and et al.},
  title = {Analysis of 0.1-Hz Var Oscillations in Solar Photovoltaic Power Plants},
  journal = {IEEE Transactions on Sustainable Energy},
  volume = {14},
  number = {1},
  pages = {734--739},
  year = {2023}
}

@article{Dong2025,
  author = {Dong, S. and Tan, J. and Ding, L. and et al.},
  title = {Analysis of 1 Hz Oscillations Induced by Grid-Forming Plant in Kaua‘i Power System},
  journal = {Proc. IEEE Power \& Energy Society General Meeting},
  year = {2025}
}

@incollection{paperMevludinGlavic,
  author    = {Mevludin Glavic and Scott Greene},
  title     = {Voltage Stability in Future Power Systems},
  booktitle = {Encyclopedia of Electrical and Electronic Power Engineering},
  publisher = {Elsevier},
  pages     = {209--223},
  year      = {2023},
  doi       = {10.1016/b978-0-12-821204-2.00141-0},
  url       = {https://doi.org/10.1016/b978-0-12-821204-2.00141-0}
}

@article{Dreidy2017,
  author = {Dreidy, M. and Mokhlis, H. and Mekhilef, S.},
  title = {Inertia response and frequency control techniques for renewable energy sources: A review},
  journal = {Renewable and Sustainable Energy Reviews},
  volume = {69},
  pages = {144--155},
  year = {2017}
}

@article{Akiba2024,
  author = {Akiba, L. H.},
  title = {Hawaii Best Practices and Lessons Learned for Grid Modernization and Renewable Energy Impacts on Island Grid Systems},
  booktitle = {ERRA Conference},
  year = {2024}
}

@article{Tielens2016,
  author = {Tielens, P. and Van Hertem, D.},
  title = {The relevance of inertia in power systems},
  journal = {Renewable and Sustainable Energy Reviews},
  volume = {55},
  pages = {999--1009},
  year = {2016}
}

@ARTICLE{paperCIGREstabilitydefinitions,
  author={Hatziargyriou, Nikos and Milanovic, Jovica and Rahmann, Claudia and Ajjarapu, Venkataramana and Canizares, Claudio and Erlich, Istvan and Hill, David and Hiskens, Ian and Kamwa, Innocent and Pal, Bikash and Pourbeik, Pouyan and Sanchez-Gasca, Juan and Stankovic, Aleksandar and Van Cutsem, Thierry and Vittal, Vijay and Vournas, Costas},
  journal={IEEE Transactions on Power Systems}, 
  title={Definition and Classification of Power System Stability – Revisited \& Extended}, 
  year={2021},
  volume={36},
  number={4},
  pages={3271-3281},
  doi={10.1109/TPWRS.2020.3041774}}

@misc{cigreTB928,
  author = {Kocewiak, {\L}ukasz and Buchhagen, Christian and Blasco-Gimenez, Rafael and Kwon, J. B. and Larsson, Magnus and Sun, Yin and Wang, Xiongfei and Mills, David and Beerten, Jef and others},
  title = {Multi-frequency Stability of Converter-based Modern Power Systems},
  howpublished = {CIGRE Technical Brochure TB 928},
  year = {2024},
  note = {WG C4.49}
}

@article{cheahmane2022,
  author = {Cheah-Mañe, Marc and Egea-Alvarez, Agustí and Prieto-Araujo, Eduardo and Mehrjerdi, Hasan and Gomis-Bellmunt, Oriol and Xu, Lie},
  title = {Modeling and Analysis Approaches for Small-signal Stability Assessment of Power-electronic-dominated Systems},
  journal = {WIREs Energy and Environment},
  volume = {12},
  number = {1},
  pages = {e453},
  year = {2023},
  doi = {10.1002/wene.453}
}

@article{paperGermano,
  author = {Mugambi, Germano Rugendo and Darii, Nicolae and Khazraj, Hesam and Saborío-Romano, Oscar and Raducu, Alin George and Sharma, Ranjan and Cutululis, Nicolaos A.},
  title = {Methodologies for Offshore Wind Power Plant Dynamic Stability Analysis},
  journal = {Renewable and Sustainable Energy Reviews},
  volume = {216},
  pages = {115635},
  year = {2025},
  doi = {10.1016/j.rser.2025.115635}
}

@inproceedings{kuri2022,
  author = {Kuri, Ananya and Zurowski, Rainer and Mehlmann, Gert and Luther, Matthias},
  title = {Converter Modelling Aspects at the Boundary between EMT and RMS Domain},
  booktitle = {Proc. 2022 International Conference on Smart Energy Systems and Technologies (SEST)},
  year = {2022},
  doi = {10.1109/SEST53650.2022.9898501}
}

@article{IPA_1982,
  author = {Pérez-Arriaga, I. J. and Verghese, G. C. and Schweppe, F. C.},
  title = {Selective Modal Analysis with Applications to Electric Power Systems, Part I: Heuristic Introduction},
  journal = {IEEE Transactions on Power Apparatus and Systems},
  volume = {PAS-101},
  number = {9},
  pages = {3117 -- 3125},
  year = {1982}
}

@article{verghese1982,
  author = {Verghese, G. C. and Pérez-Arriaga, I. J. and Schweppe, F. C.},
  title = {Selective Modal Analysis with Applications to Electric Power Systems, Part II: The Dynamic Stability Problem},
  journal = {IEEE Transactions on Power Apparatus and Systems},
  volume = {PAS-101},
  number = {9},
  pages = {3126--3134},
  year = {1982}
}

@article{pagola1989,
  author = {Pagola, F. L. and Pérez-Arriaga, I. J. and Verghese, G. C.},
  title = {On Sensitivities, Residues and Participations: Applications to Oscillatory Stability Analysis and Control},
  journal = {IEEE Transactions on Power Systems},
  volume = {4},
  number = {1},
  pages = {278--285},
  year = {1989},
  doi = {10.1109/59.32489}
}

@article{sun2011,
  author = {Sun, Jian},
  title = {Impedance-Based Stability Criterion for Grid-Connected Inverters},
  journal = {IEEE Transactions on Power Electronics},
  volume = {26},
  number = {11},
  pages = {3075--3078},
  year = {2011},
  doi = {10.1109/TPEL.2011.2136439}
}

@article{harnefors2007,
  author = {Harnefors, Lennart and Bongiorno, Massimo and Lundberg, Stefan},
  title = {Input-Admittance Calculation and Shaping for Controlled Voltage-Source Converters},
  journal = {IEEE Transactions on Industrial Electronics},
  volume = {54},
  number = {6},
  pages = {3323--3334},
  year = {2007},
  doi = {10.1109/TIE.2007.899959}
}

@book{skogestad2005,
  author = {Skogestad, Sigurd and Postlethwaite, Ian},
  title = {Multivariable Feedback Control: Analysis and Design},
  edition = {2},
  year = {2005},
  publisher = {Wiley},
  address = {Chichester}
}

@misc{paperMolinas2017toolimpedance,
  author = {Molinas, Marta},
  title = {Impedance-Based Stability Analysis of Power Electronics Systems: Concept and Applications},
  howpublished = {Tutorial/Slides},
  year = {2017},
  note = {NTNU}
}

@misc{garciareyes2026siadtool,
      title={SIaD-Tool: A Comprehensive Frequency-Domain Tool for Small-Signal Stability and Interaction Assessment in Modern Power Systems}, 
      author={Luis A. Garcia-Reyes and Oriol Gomis-Bellmunt and Eduardo Prieto-Araujo and Vinícius A. Lacerda and Marc Cheah-Mañe},
      year={2026},
      eprint={2601.05519},
      archivePrefix={arXiv},
      primaryClass={eess.SY},
      url={https://arxiv.org/abs/2601.05519}, 
}

@INPROCEEDINGS{paperShuyaWang,
  author={Shuya Wang and Jianhui Su and Xiangzhen Yang and Yan Du and Yong Tu and Huadian Xu},
  booktitle={2016 IEEE 8th International Power Electronics and Motion Control Conference (IPEMC-ECCE Asia)}, 
  title={A review on the small signal stability of microgrid}, 
  year={2016},
  volume={},
  number={},
  pages={1793-1798},
  doi={10.1109/IPEMC.2016.7512566}}

@INPROCEEDINGS{garciareyesluisa_fitss2025,
  author={Garcia-Reyes, Luis A. and Prieto-Araujo, Eduardo and Lacerda, Vinicius A. and Arévalo-Soler, Josep and Gomis-Bellmunt, Oriol and Martin-Almenta, Macarena and Nuño-Martínez, Edgar and Renedo, Javier},
  booktitle={2025 IEEE PES Innovative Smart Grid Technologies Conference Europe (ISGT Europe)}, 
  title={Data-Driven State-Space Modeling for Small-Signal Stability Analysis of Black-Box Power Converters}, 
  year={2025},
  volume={},
  number={},
  pages={1-6},
  doi={10.1109/ISGTEurope64741.2025.11305456}}

@book{bookMJGibbard,
  title     = {Small-signal stability, control and dynamic performance of power systems},
  author    = {Gibbard, M. J. and Pourbeik, P. and Vowles, D. J.},
  publisher = {University of Adelaide Press},
  year      = {2015},
  address   = {Adelaide},
  isbn      = {978-1-925261-02-8},
  doi       = {10.20851/small-signal}
}

@book{kundur1994,
  author    = {P. Kundur},
  title     = {Power System Stability and Control},
  publisher = {McGraw-Hill},
  year      = {1994}
}

@book{khalil2002nonlinear,
  title     = {Nonlinear Systems},
  author    = {Khalil, Hassan K.},
  edition   = {3rd},
  publisher = {Prentice Hall},
  year      = {2002},
  isbn      = {9780130673893}
}

@ARTICLE{paperQifanChen,
  author={Chen, Qifan and Bu, Siqi and Chung, Chi Yung},
  journal={Journal of Modern Power Systems and Clean Energy}, 
  title={Small-Signal Stability Criteria in Power Electronics-Dominated Power Systems: A Comparative Review}, 
  year={2024},
  volume={12},
  number={4},
  pages={1003-1018},
  doi={10.35833/MPCE.2023.000526}}

@ARTICLE{paperMolinas2017,
  author={Amin, Mohammad and Molinas, Marta},
  journal={IEEE Transactions on Industry Applications}, 
  title={Small-Signal Stability Assessment of Power Electronics Based Power Systems: A Discussion of Impedance- and Eigenvalue-Based Methods}, 
  year={2017},
  volume={53},
  number={5},
  pages={5014-5030},
  doi={10.1109/TIA.2017.2712692}}

@ARTICLE{paperJianSun,
  author={Sun, Jian},
  journal={IEEE Transactions on Power Electronics}, 
  title={Small-Signal Methods for AC Distributed Power Systems–A Review}, 
  year={2009},
  volume={24},
  number={11},
  pages={2545-2554},
  doi={10.1109/TPEL.2009.2029859}}

@article{beza2019,
  author    = {M. Beza and M. Bongiorno},
  title     = {Identification of resonance interactions in offshore wind farms connected to the main grid by MMC-based HVDC system},
  journal   = {International Journal of Electrical Power \& Energy Systems},
  volume    = {111},
  pages     = {101--113},
  year      = {2019},
  doi       = {10.1016/j.ijepes.2019.04.004}
}

@article{beza2020,
  author    = {M. Beza and M. Bongiorno},
  title     = {Impact of converter control strategy on low- and high-frequency resonance interactions in power-electronic dominated systems},
  journal   = {International Journal of Electrical Power \& Energy Systems},
  volume    = {120},
  pages     = {105978},
  year      = {2020},
  doi       = {10.1016/j.ijepes.2020.105978}
}

@article{sainz2017,
  author = {Sainz, Luis and others},
  title = {Criterion for the Electrical Resonance Stability of Offshore Wind Power Plants Connected through HVDC links},
  journal = {IEEE Transactions on Power Systems},
  volume = {32},
  number = {6},
  pages = {4579--4589},
  year = {2017}
}

@misc{pscad2020,
  title = {PSCAD User's Guide},
  author = {{Manitoba HVDC Research Centre}},
  year = {2020},
  howpublished = {Software Manual}
}

@misc{emtdc2019,
  title = {EMTDC Theory Manual},
  author = {{Manitoba HVDC Research Centre}},
  year = {2019},
  howpublished = {Software Manual}
}

@article{panTransientStabilityVoltageSource2020,
  title = {Transient {{Stability}} of {{Voltage-Source Converters With Grid-Forming Control}}: {{A Design-Oriented Study}}},
  shorttitle = {Transient {{Stability}} of {{Voltage-Source Converters With Grid-Forming Control}}},
  author = {Pan, Donghua and Wang, Xiongfei and Liu, Fangcheng and Shi, Rongliang},
  date = {2020-06},
  journaltitle = {IEEE Journal of Emerging and Selected Topics in Power Electronics},
  volume = {8},
  number = {2},
  pages = {1019--1033},
  issn = {2168-6785},
  doi = {10.1109/JESTPE.2019.2946310},
  eventtitle = {{{IEEE Journal}} of {{Emerging}} and {{Selected Topics}} in {{Power Electronics}}}
}

@inproceedings{verbeComparativeStudyGFMgrid2021,
  title = {Comparative {{Study}} of {{GFM-grid}} and {{GFL-grid}} in {{Islanded Operation}}},
  booktitle = {2021 {{IEEE PES Innovative Smart Grid Technologies}} - {{Asia}} ({{ISGT Asia}})},
  author = {Verbe, Sunjoh Christian and Shigenobu, Ryuto and Ito, Masakazu},
  date = {2021-12},
  pages = {1--5},
  issn = {2378-8542},
  doi = {10.1109/ISGTAsia49270.2021.9715619},
  url = {https://ieeexplore.ieee.org/document/9715619/},
  urldate = {2025-10-15},
  eventtitle = {2021 {{IEEE PES Innovative Smart Grid Technologies}} - {{Asia}} ({{ISGT Asia}})}
}

@inproceedings{darcoVirtualSynchronousMachines2013,
  title = {Virtual {{Synchronous Machines}} – {{Classification}} of {{Implementations}} and {{Analysis}} of {{Equivalence}} to {{Droop Controllers}} for {{Microgrids}}},
  booktitle = {2013 {{IEEE Grenoble Conference PowerTech}}, {{POWERTECH}} 2013},
  author = {D'Arco, Salvatore and Suul, Jon},
  date = {2013-06-16},
  doi = {10.1109/PTC.2013.6652456}
}

@article{liRevisitingGridFormingGridFollowing2022,
  title = {Revisiting {{Grid-Forming}} and {{Grid-Following Inverters}}: {{A Duality Theory}}},
  shorttitle = {Revisiting {{Grid-Forming}} and {{Grid-Following Inverters}}},
  author = {Li, Yitong and Gu, Yunjie and Green, Timothy C.},
  date = {2022-11},
  journaltitle = {IEEE Transactions on Power Systems},
  volume = {37},
  number = {6},
  pages = {4541--4554},
  issn = {1558-0679},
  doi = {10.1109/TPWRS.2022.3151851},
  eventtitle = {{{IEEE Transactions}} on {{Power Systems}}}
}

@article{liDesignAnalysisRealtime2004,
  title = {Design, Analysis, and Real-Time Testing of a Controller for Multibus Microgrid System},
  author = {Li, Yunwei and Vilathgamuwa, D.M. and Loh, Poh Chiang},
  date = {2004-09},
  journaltitle = {IEEE Transactions on Power Electronics},
  volume = {19},
  number = {5},
  pages = {1195--1204},
  issn = {1941-0107},
  doi = {10.1109/TPEL.2004.833456},
  url = {https://ieeexplore.ieee.org/document/1331480/},
  urldate = {2025-10-15}
}

@article{zhangPowerSynchronizationControlGridConnected2010,
  title = {Power-{{Synchronization Control}} of {{Grid-Connected Voltage-Source Converters}}},
  author = {Zhang, Lidong and Harnefors, Lennart and Nee, Hans-Peter},
  date = {2010-05},
  journaltitle = {IEEE Transactions on Power Systems},
  volume = {25},
  number = {2},
  pages = {809--820},
  issn = {1558-0679},
  doi = {10.1109/TPWRS.2009.2032231},
  url = {https://ieeexplore.ieee.org/document/5308285/figures#figures},
  urldate = {2024-06-18},
  eventtitle = {{{IEEE Transactions}} on {{Power Systems}}}
}

@article{zhongSelfSynchronizedSynchronvertersInverters2014,
  title = {Self-{{Synchronized Synchronverters}}: {{Inverters Without}} a {{Dedicated Synchronization Unit}}},
  shorttitle = {Self-{{Synchronized Synchronverters}}},
  author = {Zhong, Qing-Chang and Nguyen, Phi-Long and Ma, Zhenyu and Sheng, Wanxing},
  date = {2014-02},
  journaltitle = {IEEE Transactions on Power Electronics},
  volume = {29},
  number = {2},
  pages = {617--630},
  issn = {1941-0107},
  doi = {10.1109/TPEL.2013.2258684},
  eventtitle = {{{IEEE Transactions}} on {{Power Electronics}}}
}

@article{johnsonSynthesizingVirtualOscillators2016,
  title = {Synthesizing {{Virtual Oscillators}} to {{Control Islanded Inverters}}},
  author = {Johnson, Brian B. and Sinha, Mohit and Ainsworth, Nathan G. and Dörfler, Florian and Dhople, Sairaj V.},
  date = {2016-08},
  journaltitle = {IEEE Transactions on Power Electronics},
  volume = {31},
  number = {8},
  pages = {6002--6015},
  issn = {1941-0107},
  doi = {10.1109/TPEL.2015.2497217},
  url = {https://ieeexplore.ieee.org/document/7317584/},
  urldate = {2025-10-28}
}

@article{harneforsUniversalControllerGridConnected2021,
  title = {A {{Universal Controller}} for {{Grid-Connected Voltage-Source Converters}}},
  author = {Harnefors, Lennart and Kukkola, Jarno and Routimo, Mikko and Hinkkanen, Marko and Wang, Xiongfei},
  date = {2021-10},
  journaltitle = {IEEE Journal of Emerging and Selected Topics in Power Electronics},
  volume = {9},
  number = {5},
  pages = {5761--5770},
  issn = {2168-6785},
  doi = {10.1109/JESTPE.2020.3039407},
  url = {https://ieeexplore.ieee.org/document/9264177/},
  urldate = {2025-10-23}
}

@article{liuAdaptiveHybridGridForming2024,
  title = {Adaptive {{Hybrid Grid-Forming}} and {{Grid-Following Control}} of {{IBRs With Enhanced Small-Signal Stability Under Varying SCRs}}},
  author = {Liu, Pengyin and Xie, Xiaorong and Shair, Jan},
  date = {2024-06},
  journaltitle = {IEEE Transactions on Power Electronics},
  volume = {39},
  number = {6},
  pages = {6603--6607},
  issn = {1941-0107},
  doi = {10.1109/TPEL.2024.3373659},
  url = {https://ieeexplore.ieee.org/document/10460137/},
  urldate = {2025-10-23}
}

@article{chenGeneralizedMultivariableGridForming2022,
  title = {Generalized {{Multivariable Grid-Forming Control Design}} for {{Power Converters}}},
  author = {Chen, Meng and Zhou, Dao and Tayyebi, Ali and Prieto-Araujo, Eduardo and Dörfler, Florian and Blaabjerg, Frede},
  date = {2022-07},
  journaltitle = {IEEE Transactions on Smart Grid},
  volume = {13},
  number = {4},
  pages = {2873--2885},
  issn = {1949-3061},
  doi = {10.1109/TSG.2022.3161608},
  eventtitle = {{{IEEE Transactions}} on {{Smart Grid}}}
}

@article{zhangModelPredictiveDualControlLoop2025,
  title = {Model-{{Predictive Dual-Control Loop With Improved Current-Limiting Capability}} for {{Grid-Forming Inverter Under Grid Faults}}},
  author = {Zhang, Han and Xue, Cheng and Liu, Rui and Li, Yunwei},
  date = {2025-01},
  journaltitle = {IEEE Transactions on Power Electronics},
  volume = {40},
  number = {1},
  pages = {813--827},
  issn = {1941-0107},
  doi = {10.1109/TPEL.2024.3483177},
  url = {https://ieeexplore.ieee.org/document/10721289/},
  urldate = {2025-10-28}
}

@inproceedings{seoDispatchableVirtualOscillator2018,
  title = {Dispatchable {{Virtual Oscillator Control}} for {{Decentralized Inverter-dominated Power Systems}}: {{Analysis}} and {{Experiments}}},
  shorttitle = {Dispatchable {{Virtual Oscillator Control}} for {{Decentralized Inverter-dominated Power Systems}}},
  author = {Seo, Gab-Su and Colombino, Marcello and Subotić, Irina and Johnson, Brian and Groß, Dominic and Dörfler, Florian},
  date = {2018-11-21},
  eprint = {1811.08842},
  eprinttype = {arXiv},
  eprintclass = {math},
  publisher = {arXiv},
  doi = {10.48550/arXiv.1811.08842},
  url = {http://arxiv.org/abs/1811.08842},
  urldate = {2025-10-29}
}

@article{carnieluttiFixedSwitchingFrequency2025,
  title = {Fixed {{Switching Frequency Model Predictive Control}} for {{Grid-Forming Inverters}}},
  author = {Carnielutti, Fernanda and Busarello, Tiago Davi Curi and Resende, Ênio Costa and Ullah, Qudrat and Simões, Marcelo Godoy},
  date = {2025-07},
  journaltitle = {IEEE Transactions on Power Electronics},
  volume = {40},
  number = {7},
  pages = {9080--9089},
  issn = {1941-0107},
  doi = {10.1109/TPEL.2025.3544823},
  url = {https://ieeexplore.ieee.org/document/10899871/},
  urldate = {2025-10-29}
}

@report{ENTSOE2013-RfG-IGD,
  title = {Implementation Guideline for Network Code “Requirements for Generators”},
  author = {{ENTSO-E}},
  date = {2013-10-16},
  institution = {European Network of Transmission System Operators for Electricity (ENTSO-E)},
  url = {https://www.entsoe.eu/fileadmin/user_upload/_library/resources/RfG/131016_-_NC_RfG_implementation_guideline.pdf},
  urldate = {2025-11-21}
}

@article{ahmedDynamicGridStability2023,
  title = {Dynamic Grid Stability in Low Carbon Power Systems with Minimum Inertia},
  author = {Ahmed, Faraedoon and Al Kez, Dlzar and McLoone, Seán and Best, Robert James and Cameron, Ché and Foley, Aoife},
  date = {2023-07},
  journaltitle = {Renewable Energy},
  shortjournal = {Renewable Energy},
  volume = {210},
  pages = {486--506},
  issn = {09601481},
  doi = {10.1016/j.renene.2023.03.082},
  url = {https://linkinghub.elsevier.com/retrieve/pii/S0960148123003774},
  urldate = {2026-01-06},
  langid = {english}
}

@inproceedings{alferinkRobustModelPredictive2024,
  title = {Robust {{Model Predictive Control-Based Energy Management System}} for the {{Island Power System}} of {{Suðuroy}}, {{Faroe Islands}}},
  booktitle = {2024 {{IEEE}} 63rd {{Conference}} on {{Decision}} and {{Control}} ({{CDC}})},
  author = {Alferink, Marco and Reus, Lucas and Hofmann, Lutz and Michels, Kai},
  date = {2024-12},
  pages = {6519--6524},
  issn = {2576-2370},
  doi = {10.1109/CDC56724.2024.10886828},
  url = {https://ieeexplore-ieee-org.cardiff.idm.oclc.org/document/10886828},
  urldate = {2026-01-06},
  eventtitle = {2024 {{IEEE}} 63rd {{Conference}} on {{Decision}} and {{Control}} ({{CDC}})}
}

@article{katsaprakakisFaroeIslands1002019,
  title = {Faroe {{Islands}}: {{Towards}} 100\% {{R}}.{{E}}.{{S}}. Penetration},
  shorttitle = {Faroe {{Islands}}},
  author = {Katsaprakakis, Dimitris Al and Thomsen, Bjarti and Dakanali, Irini and Tzirakis, Kostas},
  date = {2019-05},
  journaltitle = {Renewable Energy},
  shortjournal = {Renewable Energy},
  volume = {135},
  pages = {473--484},
  issn = {09601481},
  doi = {10.1016/j.renene.2018.12.042},
  url = {https://linkinghub.elsevier.com/retrieve/pii/S0960148118314733},
  urldate = {2026-01-06},
  langid = {english}
}

@inproceedings{raufIntegratingRenewablesInsights2025,
  title = {Integrating {{Renewables}}: {{Insights}} from {{Global Pumped Hydro Storage Case Studies}}},
  shorttitle = {Integrating {{Renewables}}},
  booktitle = {2025 9th {{International Conference}} on {{Green Energy}} and {{Applications}} ({{ICGEA}})},
  author = {Rauf, Qasim and Yanpin, Li and Ashraf, Anam},
  date = {2025-03},
  pages = {1--10},
  issn = {2694-3298},
  doi = {10.1109/ICGEA64602.2025.11009747},
  url = {https://ieeexplore.ieee.org/document/11009747/},
  urldate = {2026-01-06},
  eventtitle = {2025 9th {{International Conference}} on {{Green Energy}} and {{Applications}} ({{ICGEA}})}
}

@inproceedings{reusFrequencyRestorationControl2024,
  title = {Frequency {{Restoration Control}} and {{Its Impact}} on {{Model Predictive Control}} of {{Island Power Systems}}},
  booktitle = {2024 {{IEEE International Conference}} on {{Power System Technology}} ({{PowerCon}})},
  author = {Reus, Lucas and Hofmann, Lutz and Alferink, Marco and Michels, Kai},
  date = {2024-11},
  pages = {1--5},
  doi = {10.1109/PowerCon60995.2024.10870546},
  url = {https://ieeexplore.ieee.org/document/10870546/},
  urldate = {2026-01-06},
  eventtitle = {2024 {{IEEE International Conference}} on {{Power System Technology}} ({{PowerCon}})}
}

@article{trondheim100SustainableElectricity2021,
  title = {100\% {{Sustainable Electricity}} in the {{Faroe Islands}}: {{Expansion Planning Through Economic Optimization}}},
  shorttitle = {100\% {{Sustainable Electricity}} in the {{Faroe Islands}}},
  author = {Tróndheim, Helma Maria and Niclasen, Bár{\dh}r A. and Nielsen, Terji and Silva, Filipe Faria Da and Bak, Claus Leth},
  date = {2021},
  journaltitle = {IEEE Open Access Journal of Power and Energy},
  volume = {8},
  pages = {23--34},
  issn = {2687-7910},
  doi = {10.1109/OAJPE.2021.3051917},
  url = {https://ieeexplore.ieee.org/document/9323058/},
  urldate = {2026-01-06}
}

@inproceedings{trondheim2021frequency,
  title={Frequency and voltage analysis of the hybrid power system in Su{\dh}uroy, Faroe Islands},
  author={Trondheim, Helma Maria and Hofmann, Lutz and Gartmann, Pascal and Quitmann, Eckard and da Silva, Filipe Faria and Bak, Claus Leth and Nielsen, Terji and Niclasen, B{\'a}r{\dh}ur A},
  booktitle={Proceedings of the virtual 5th international hybrid power systems workshop},
  year={2021},
  organization={Energynautics GmbH}
}

@misc{eem_ce4eui_isolated_services_2023,
  author       = {{Empresa de Electricidade da Madeira (EEM)}},
  title        = {System/grid services in isolated electric systems},
  howpublished = {Presentation slides, Clean Energy for EU Islands (CE4EUI), Workshop 5: 100\% Decarbonization},
  year         = {2023},
  note         = {[Online]. Available: \url{https://clean-energy-islands.ec.europa.eu/system/files/2023-12/CE4EUI_100\%25\%20RES_EEM-Madeira-final.pdf}. Accessed: 2026-01-09}
}

@article{beiresNeedSynchronousInertia2018,
  title = {The {{Need}} of {{Synchronous Inertia}} in {{Autonomous Power Systems}} with {{Increasing Shares}} of {{Renewables}}},
  author = {Beires, Pedro P. and Moreira, C. L. and Lopes, João Peças and Figueira, Agostinho G.},
  date = {2018},
  journaltitle = {Natural gas},
  volume = {54},
  pages = {14--6},
  url = {https://hybridpowersystems.org/wp-content/uploads/sites/9/2018/05/7A_2_TENE18_027_paper_Moreira_Carlos.pdf},
  urldate = {2026-01-08}
}

@article{fotopoulouReviewEnergyStorage2024,
  title = {A {{Review}} of the {{Energy Storage Systems}} of {{Non-Interconnected European Islands}}},
  author = {Fotopoulou, Maria and Pediaditis, Panagiotis and Skopetou, Niki and Rakopoulos, Dimitrios and Christopoulos, Sotirios and Kartalidis, Avraam and Fotopoulou, Maria and Pediaditis, Panagiotis and Skopetou, Niki and Rakopoulos, Dimitrios and Christopoulos, Sotirios and Kartalidis, Avraam},
  date = {2024-02-12},
  journaltitle = {Sustainability},
  volume = {16},
  number = {4},
  publisher = {publisher},
  issn = {2071-1050},
  doi = {10.3390/su16041572},
  url = {https://www.mdpi.com/2071-1050/16/4/1572},
  urldate = {2026-01-08},
  langid = {english}
}

@article{miguelEnergyStorageRenewable2017,
  title = {Energy Storage for Renewable Energy Integration: The Case of {{Madeira Island}}, {{Portugal}}},
  shorttitle = {Energy Storage for Renewable Energy Integration},
  author = {Miguel, Miguel and Nogueira, Teresa and Martins, Florinda},
  date = {2017-10},
  journaltitle = {Energy Procedia},
  shortjournal = {Energy Procedia},
  volume = {136},
  pages = {251--257},
  issn = {18766102},
  doi = {10.1016/j.egypro.2017.10.277},
  url = {https://linkinghub.elsevier.com/retrieve/pii/S187661021735227X},
  urldate = {2026-01-08},
  langid = {english}
}

@article{ntomalisDynamicModelingSimulation2020,
  title = {Dynamic {{Modeling}} and {{Simulation}} of {{Non-Interconnected Systems}} under {{High-RES Penetration}}: {{The Madeira Island Case}}},
  shorttitle = {Dynamic {{Modeling}} and {{Simulation}} of {{Non-Interconnected Systems}} under {{High-RES Penetration}}},
  author = {Ntomalis, Stefanos and Iliadis, Petros and Atsonios, Konstantinos and Nesiadis, Athanasios and Nikolopoulos, Nikos and Grammelis, Panagiotis and Ntomalis, Stefanos and Iliadis, Petros and Atsonios, Konstantinos and Nesiadis, Athanasios and Nikolopoulos, Nikos and Grammelis, Panagiotis},
  date = {2020-11-04},
  journaltitle = {Energies},
  volume = {13},
  number = {21},
  publisher = {publisher},
  issn = {1996-1073},
  doi = {10.3390/en13215786},
  url = {https://www.mdpi.com/1996-1073/13/21/5786},
  urldate = {2026-01-08},
  langid = {english}
}

@article{paivaEffectsBatteryEnergy2024,
  title = {Effects of {{Battery Energy Storage Systems}} on the {{Frequency Stability}} of {{Weak Grids}} with a {{High-Share}} of {{Grid-Connected Converters}}},
  author = {Paiva, Pedro and Castro, Rui and Paiva, Pedro and Castro, Rui},
  date = {2024-03-13},
  journaltitle = {Electronics},
  volume = {13},
  number = {6},
  publisher = {publisher},
  issn = {2079-9292},
  doi = {10.3390/electronics13061083},
  url = {https://www.mdpi.com/2079-9292/13/6/1083},
  urldate = {2026-01-08},
  langid = {english}
}

@article{vasconcelosAdvancedControlSolutions2015,
  title = {Advanced {{Control Solutions}} for {{Operating Isolated Power Systems}}: {{Examining}} the {{Portuguese}} Islands},
  shorttitle = {Advanced {{Control Solutions}} for {{Operating Isolated Power Systems}}},
  author = {Vasconcelos, Helena and Moreira, Carlos and Madureira, Andre and Lopes, Joao Pecas and Miranda, Vladimiro},
  date = {2015-03},
  journaltitle = {IEEE Electrification Magazine},
  volume = {3},
  number = {1},
  pages = {25--35},
  issn = {2325-5889},
  doi = {10.1109/MELE.2014.2380131},
  url = {https://ieeexplore.ieee.org/document/7050401/},
  urldate = {2026-01-08}
}

@misc{renault_porto_santo_smart_island_2018,
  author       = {{Renault Group Press Office}},
  title        = {Groupe Renault and {EEM} create first ``smart island'' in Porto Santo},
  year         = {2018},
  month        = feb,
  note         = {Press release, published Feb. 21, 2018. Accessed: 2026-01-09},
  url          = {https://media.renaultgroup.com/groupe-renault-and-eem-create-first-smart-island-in-porto-santo/}
}

@techreport{martins_madeira_replication_action_plan_2024,
  author       = {Martins, Rui Miguel F. and Correia, Henrique Pinto and Vasconcelos, Diogo and Zepter, Jan Martin and Marinelli, Mattia},
  title        = {Madeira Large-scale Replication Action Plan},
  institution  = {European Climate, Infrastructure and Environment Executive Agency},
  year         = {2024},
  type         = {Report (INSULAE deliverable D7.5)},
  note         = {Accessed: 2026-01-09},
  url          = {https://orbit.dtu.dk/en/publications/madeira-large-scale-replication-action-plan/}
}

@article{darcoEquivalenceVirtualSynchronous2014,
  title = {Equivalence of {{Virtual Synchronous Machines}} and {{Frequency-Droops}} for {{Converter-Based MicroGrids}}},
  author = {D'Arco, Salvatore and Suul, Jon Are},
  date = {2014-01},
  journaltitle = {IEEE Transactions on Smart Grid},
  volume = {5},
  number = {1},
  pages = {394--395},
  issn = {1949-3061},
  doi = {10.1109/TSG.2013.2288000},
  url = {https://ieeexplore.ieee.org/document/6683080/},
  urldate = {2025-10-20}
}

@article{darcoVirtualSynchronousMachine2015,
  title = {A {{Virtual Synchronous Machine}} Implementation for Distributed Control of Power Converters in {{SmartGrids}}},
  author = {D’Arco, Salvatore and Suul, Jon Are and Fosso, Olav B.},
  date = {2015-05-01},
  journaltitle = {Electric Power Systems Research},
  shortjournal = {Electric Power Systems Research},
  volume = {122},
  pages = {180--197},
  issn = {0378-7796},
  doi = {10.1016/j.epsr.2015.01.001},
  url = {https://www.sciencedirect.com/science/article/pii/S0378779615000024},
  urldate = {2026-01-26}
}

@article{qoriaPLLFreeGridFormingControl2020a,
  title = {A {{PLL-Free Grid-Forming Control With Decoupled Functionalities}} for {{High-Power Transmission System Applications}}},
  author = {Qoria, Taoufik and Rokrok, Ebrahim and Bruyere, Antoine and François, Bruno and Guillaud, Xavier},
  date = {2020},
  journaltitle = {IEEE Access},
  volume = {8},
  pages = {197363--197378},
  issn = {2169-3536},
  doi = {10.1109/ACCESS.2020.3034149},
  url = {https://ieeexplore.ieee.org/abstract/document/9240975},
  urldate = {2026-01-26}
}

@article{Lee2020,
  author = {Lee, T. and Glick, M. B. and Lee, J.-H.},
  title = {Island energy transition: Assessing Hawaii’s multi-level, policy-driven approach},
  journal = {Renewable and Sustainable Energy Reviews},
  volume = {118},
  pages = {109500},
  year = {2020}
}

@article{HawaiiSolarIntegration,
  author = {Eber, K. and Corbus, D.},
  title = {Hawaii Solar Integration Study: Executive Summary},
  institution = {NREL},
  year = {2013}
}

@article{SurveyGridIncidents,
  author = {Wachter, J. and Gröll, L. and Hagenmeyer, V.},
  title = {Survey of real-world grid incidents – opportunities, arising challenges and lessons learned for the future converter dominated power system},
  journal = {IEEE Open Journal of Power Electronics},
  year = {2023}
}

@article{Technologies2025,
  author = {Ochoa-Correa, D. and Arévalo, P. and Martinez, S.},
  title = {Pathways to 100\% Renewable Energy in Island Systems: A Systematic Review of Challenges, Solutions Strategies, and Success Cases},
  journal = {Technologies},
  volume = {13},
  number = {5},
  pages = {180},
  year = {2025}
}

@article{Acha_2016,
  title = {{On the Provision of Frequency Regulation in Low Inertia AC Grids Using HVDC Systems}},
  author = {Castro, L M and Acha, E},
  year = {2016},
  journaltitle = {IEEE Transactions on Smart Grid},
  volume = {7},
  number = {6},
  pages = {2680 -- 2690},
}

@misc{mathworks_2026,
  author       = {{Mathworks}},
  title        = {www.mathworks.com (accessed 27-1-2026)},
  year         = {2026},
}

@article{Cheah_2026,
  title = {{A New Paradigm for Small-Signal Stability Analysis in Modern Power Systems: Challenges, Models, and Methods in Power Systems Rich in Power Electronics}},
  author = {Cheah-Mane, Marc and Arevalo-Soler, Josep and Moutevelis, Dionysios and Mateu-Barriendos, Elia and Prieto-Araujo, Eduardo and Gomis-Bellmunt, Oriol and Renedo, Javier and Martin-Almenta, Macarena and Nu{\~n}o-Martinez, Edgar and Martinez-Villanueva, Sergio and Jung, JooYong and Kim, Namkyu and Kwon, YoungJin},
  year = {2026},
  journaltitle = {IEEE Power and Energy Magazine},
  volume = {24},
  number = {1},
  pages = {67 -- 81},
}

@article{Echavarren_part1_2011,
  title = {{Formulation, computation and improvement of steady state security margins in power systems. Part I: Theoretical framework}},
  author = {Echavarren, F M and Lobato, E and Rouco, L and Gomez, T},
  year = {2011},
  journaltitle = {International Journal of Electrical Power and Energy Systems},
  volume = {33},
  number = {2},
  pages = {340 -- 346},
}

@article{Echavarren_part2_2011,
  title = {{Formulation, computation and improvement of steady state security margins in power systems. Part II: Results}},
  author = {Echavarren, F M and Lobato, E and Rouco, L and Gomez, T},
  year = {2011},
  journaltitle = {International Journal of Electrical Power and Energy Systems},
  volume = {33},
  number = {2},
  pages = {347 -- 358},
}

@book{chiang2011direct,
  author    = {H.-D. Chiang},
  title     = {Direct Methods for Stability Analysis of Electric Power Systems},
  publisher = {Wiley},
  year      = {2011}
}

@book{vancutsem1998voltage,
  author    = {Thierry Van Cutsem and Costas Vournas},
  title     = {Voltage Stability of Electric Power Systems},
  publisher = {Springer},
  year      = {1998}
}

@article{Egido_2009,
  title = {{Maximum frequency deviation calculation in small isolated power systems}},
  author = {Egido, I and Fernandez-Bernal, F and Centeno, P and Rouco, L},
  year = {2009},
  journaltitle = {IEEE Transactions on Power Systems},
  volume = {24},
  number = {4},
  pages = {1731 -- 1738},
}

@article{Sigrist_2013,
  title = {{Energy storage systems providing primary reserve and peak shaving in small isolated power systems: an economic assessment}},
  author = {Sigrist, L and Lobato, E and Rouco, L},
  year = {2013},
  journaltitle = {International Journal of Electrical Power and Energy Systems},
  volume = {53},
  pages = {675 -- 683},
}

@article{Sigrist_2011,
  title = {{A method for the design of UFLS schemes of small isolated power systems}},
  author = {Sigrist, L and Egido, I and Rouco, L},
  year = {2011},
  journaltitle = {IEEE Transactions on Power Systems},
  volume = {72},
  number = {2},
  pages = {951 -- 958},
}

@article{Sigrist_2012,
  title = {{Principles of a centralized UFLS scheme for small isolated power systems}},
  author = {Sigrist, L and Egido, I and Rouco, L},
  year = {2012},
  journaltitle = {IEEE Transactions on Power Systems},
  volume = {28},
  number = {2},
  pages = {1779 -- 1786},
}

@inproceedings{Rouco_2024,
  title = {{Synchronous compensators: Old devices for new problems}},
  booktitle = {Proc. 3th International Conference on Future Electric Power Systems and the Energy Transition (FEPSET), Champéry (Switzerland), 04-09 Feb},
  author = {Rouco, L},
  Year = {2024},
  pages = {1--8},
}

@misc{redlectrica_planif_2021,
  author       = {{Red Eléctrica}},
  title        = {Plan de Desarrollo de la Red de Transporte de Energía Eléctrica: 
Período 2021-2026, https://www.planificacionelectrica.es/},
  year         = {2021},
}
\sectionbreak

\end{document}